\documentclass[12pt]{article}
\pdfoutput=1

\usepackage{amsthm}
\usepackage{putex}
\usepackage{pgf,interval}
\usepackage{graphicx}
\usepackage{caption}
\usepackage{amsmath}
\usepackage{array}
\usepackage{subcaption}
\usepackage{epstopdf}
\usepackage{enumerate}
\usepackage{cite}
\usepackage{youngtab}
\usepackage{tensor}
\usepackage{slashed}
\usepackage[aligntableaux=center]{ytableau}
\usepackage[utf8]{inputenc}
\usepackage{rotating}
\usepackage{bigfoot}
\usepackage{multirow}
\usepackage[
colorlinks=true,
linkcolor=blue,
urlcolor=blue,
filecolor=black,
citecolor=red,
]{hyperref}

\usepackage{tikz,varwidth}
\usetikzlibrary{decorations.pathreplacing,positioning}
\usetikzlibrary{calc}

\newcommand{\abs}[1]{\left\lvert #1 \right\rvert}

\newcommand {\be} {\begin {equation}}
\newcommand {\ee} {\end {equation}}

\newcommand {\bes} {\begin {equation*}}
\newcommand {\ees} {\end {equation*}}

\newcommand{\es}[2] {\begin{equation} \label{#1} \begin{split} #2 \end{split} \end{equation}}

\newcommand{\Z}{\mathbb{Z}}

\newcommand{\R}{\mathbb{R}}

\newcommand{\cA}{{\mathcal A}}
\newcommand{\cB}{{\mathcal B}}

\newcommand{\cE}{{\mathcal E}}
\newcommand{\cF}{{\mathcal F}}
\newcommand{\cG}{{\mathcal G}}
\newcommand{\cH}{{\mathcal H}}
\newcommand{\cJ}{{\mathcal J}}

\newcommand{\cN}{{\mathcal N}}

\newcommand{\cS}{{\mathcal S}}
\newcommand{\cT}{{\mathcal T}}

\newcommand{\cM}{{\mathcal M}}

\newcommand{\cZ}{{\mathcal Z}}

\newcommand{\beq}{\begin{equation}}
\newcommand{\eeq}{\end{equation}}

\newcommand\ep{\epsilon}

\newcommand\ga{{\ensuremath{{\gamma}}}}

\def\ie{\begin{equation}\begin{aligned}}
\def\fe{\end{aligned}\end{equation}}

\newcommand{\la}{\langle}
\newcommand{\ra}{\rangle}

\newcommand{\A}{{\alpha}}
\newcommand{\B}{{\beta}}
\newcommand{\D}{{\delta}}

\newcommand{\mZ}{{\mathbb Z}}
\newcommand{\mR}{{\mathbb R}}

\newcommand{\mf}{\mathfrak }

\newcommand{\Mands}{{\bf s}}
\newcommand{\Mandt}{{\bf t}}
\newcommand{\Mandu}{{\bf u}}

\numberwithin{equation}{section}

\def\<{\langle}
\def\>{\rangle}

\def\half{{\scriptstyle \frac 12}}

\def\sevenh{{\scriptstyle \frac 72}}
\def\threeh{{\scriptstyle \frac 32}}
\def\fiveh{{\scriptstyle \frac 52}}

\begin{document}
	
	\preprint{PUPT-2620\\ QMUL-PH-20-16}

	\institution{Weizmann}{Department of Particle Physics and Astrophysics, Weizmann Institute of Science, Rehovot, Israel}
	\institution{QueenMary}{School of Physics and Astronomy, Queen Mary University of London, London, E1 4NS, UK}
	\institution{DAMTP}{DAMTP, Wilberforce Road, Cambridge CB3 0WA, UK}
	\institution{PU}{Joseph Henry Laboratories, Princeton University, Princeton, NJ 08544, USA}
	\institution{CMSA}{Center of Mathematical Sciences and Applications, Harvard University, Cambridge, MA 02138, USA}
	\institution{HU}{Jefferson Physical Laboratory, Harvard University,
		Cambridge, MA 02138, USA}
	%\institution{Harvard}{Jefferson Physical Laboratory, Harvard University, Cambridge, MA 02138, USA}

	\title{New Modular Invariants in $\mathcal{N}=4$ Super-Yang-Mills Theory
	}

	\authors{Shai M.~Chester,\worksat{\Weizmann} Michael B.~Green,\worksat{\QueenMary, \DAMTP} Silviu S.~Pufu,\worksat{\PU} Yifan Wang,\worksat{\CMSA,\HU}\\[10 pt] and Congkao Wen\worksat{\QueenMary}}

	\abstract{
		We study modular invariants arising in the four-point functions of the stress tensor multiplet operators of the ${\cal N} = 4$ $SU(N)$ super-Yang-Mills theory, in the limit where $N$ is taken to be large while the complexified Yang-Mills coupling $\tau$ is held fixed.  The specific four-point functions we consider are integrated correlators obtained by taking various combinations of four derivatives of the squashed sphere partition function of the ${\cal N} = 2^*$ theory with respect to the squashing parameter $b$ and mass parameter $m$, evaluated at the values $b=1$ and $m=0$ that correspond to the ${\cal N} = 4$ theory on a round sphere.  At each order in the $1/N$ expansion, these fourth derivatives are modular invariant functions of $(\tau, \bar \tau)$.  We present evidence that at half-integer orders in $1/N$, these modular invariants are linear combinations of non-holomorphic Eisenstein series, while at integer orders in $1/N$, they are certain ``generalized Eisenstein series'' which satisfy inhomogeneous Laplace eigenvalue equations on the hyperbolic plane.  These results reproduce  known features of the low-energy expansion of the four-graviton amplitude in  type IIB superstring theory in ten-dimensional flat space and  have interesting implications for the structure of the analogous expansion in $AdS_5\times S^5$.}

	\date{August 2020}

	\maketitle

	\tableofcontents

	\pagebreak
	
	\section{Introduction}
	\label{introduction}
	
	One of the most intriguing features of four-dimensional gauge theories is the possibility of a mysterious duality that exchanges elementary quarks and magnetic monopoles, while relating physics at strong and weak gauge couplings. First conjectured by Montonen and Olive in \cite{Montonen:1977sn} following the work of Goddard, Nuyts, and Olive (GNO) \cite{Goddard:1976qe} as a direct generalization of the  electric-magnetic duality in Maxwell theories, it is commonly known as the Montonen-Olive duality or S-duality. 
	It was soon realized that such a duality is more likely to hold in a supersymmetric gauge theory rather than in QCD, because supersymmetry provides more control over the spectrum of solitons \cite{Witten:1978mh}. In the case of the maximally supersymmetric $\cN=4$ super-Yang-Mills (SYM) theory \cite{Osborn:1979tq}, an overwhelming amount of compelling evidence for S-duality has been provided by analyzing the dyon-monopole bound states \cite{Sen:1994yi}, from certain topologically twisted partition functions on four manifolds \cite{Vafa:1994tf}, and from embedding this model into type IIB string theory.
	
	Under S-duality, the $\cN=4$ SYM theory with gauge group $SU(N)$ and complexified gauge coupling
	\ie
	\tau\equiv \tau_1+i\tau_2 ={\theta\over 2\pi} + {4\pi i\over g_{_{\rm YM}}^2}
	\fe
	is equivalent to the SYM theory with gauge group $SU(N) / \Z_N$ and gauge coupling $\tau^\vee=-\frac{1}{\tau} $.  While the distinction between the gauge group being $SU(N)$ and $SU(N) / \Z_N$ is important for studying non-local operators, it does not affect local operators, which are the subject of this work.    
	The S-duality transformation combined with the T-transformation $\tau \to \tau +1$ from the periodic identification of the $\theta$-angle gives rise to an $SL(2,\mZ)$  duality that acts on the complexified coupling as
	\es{SL2ZTransf}{
		\tau \to {a \tau +b \over c\tau +d}\,,
	}
	with $a,b,c,d\in \mZ$ and $ad-bc=1$.\footnote{In general, under S-duality, ${\cal N} = 4$ SYM with the gauge group $G$ is mapped to ${\cal N} = 4$ SYM with gauge group $G^\vee$ given by the GNO dual of $G$, and the gauge coupling $\tau$ is mapped to $\tau^\vee=-1/( n_G \tau )$, where $n_G=1$ when $G$ is simply-laced, $n_G=2$ for $B_r,C_r,F_4$, and $n_G=3$ for $G_2$.  Consequently, for non-simply-laced $G$, the combination of S-duality and T-duality yields dualities by (extensions of) congruence subgroups of $SL(2,\mZ)$ \cite{Argyres:2006qr}. }
	The Coulomb branch of the SYM theory supports an infinite tower of massive BPS particles from W-bosons, monopoles, and their bound states that transform nontrivially under the duality group, while the mass spectrum stays invariant~\cite{Sen:1994yi}.  Correlation functions of local operators have definite $SL(2,\mZ)$ transformation properties, and correlation functions of half-BPS operators are invariant \cite{Intriligator:1998ig,Intriligator:1999ff}. 
	
	Embedding the $\cN=4$  SYM theory  into string theory provides an elegant picture of this non-perturbative duality. The $SU(N)$ $\cN=4$ SYM theory is the low energy theory on a stack of $N$ coincident  D3-branes immersed in asymptotically flat ten-dimensional spacetime. The gauge coupling $\tau$ is identified with the axion-dilaton background $\tau_s= \chi +  i e^{-\phi}$ ($\chi$ being the axion and $\phi$ being the dilaton), and the $SL(2, \Z)$ duality of the gauge theory is a direct consequence of the $SL(2,\mZ)$ duality in type IIB string theory \cite{Hull:1994ys}.

	While the string theory perspective is conceptually useful, it is more satisfying to directly investigate the S-duality properties of $\cN=4$ SYM using field theory methods, and this will be our approach here.  In fact, one may argue that the field theory methods provide nontrivial support for the duality structures in the quantum gravity theory.  Over the past twenty or so years, there have been steady developments on investigating the S-duality properties of $\cN=4$ SYM using field theory methods, including numerous sophisticated checks based on supersymmetric partition functions \cite{Vafa:1994tf,Alday:2009aq,Alday:2010vg,Billo:2014bja,Dabholkar:2020fde},  extensions that incorporate supersymmetric defects   \cite{Kapustin:2005py,Kapustin:2006pk,Gukov:2006jk,Alday:2009fs,Gomis:2009ir,Giombi:2009ek}, as well as refinements of the duality by keeping track of global structures of the gauge group and topological couplings in the theory \cite{Witten:1998wy,Aharony:2013hda,Garcia-Etxebarria:2019cnb}. 
	
	The goal of this paper is to continue the study began in \cite{Chester:2020dja} of the $SL(2, \Z)$ duality properties of certain  correlation functions of the $SU(N)$ ${\cal N} = 4$ SYM theory in the $1/N$ expansion.  In particular, we focus on all possible integrated four-point correlation functions that can be related to fourth derivatives of the partition function $Z(b, m, \tau, \bar \tau)$ of the ${\cal N} = 2$-preserving mass deformation of the ${\cal N} = 4$ SYM theory (also known as the ${\cal N} = 2^*$ theory) placed on a squashed four-sphere.  Here, $m$ is the mass parameter and $b$ is the squashing parameter, defined such that $(b, m) = (1, 0)$ corresponds to the (massless) ${\cal N} = 4$ SYM theory on a round sphere.  The main reason for focusing on these derivatives of $Z(b, m, \tau, \bar \tau)$ is that $Z(b, m, \tau, \bar \tau)$ itself can be computed exactly at any $N$ and any coupling $(\tau, \bar \tau)$ using supersymmetric localization \cite{Pestun:2007rz,Hama:2012bg} (see also \cite{Russo:2013qaa,Russo:2013sba,Russo:2013kea,Russo:2019lgq,Buchel:2013id,Bobev:2013cja}).  Each of the following combinations of derivatives, evaluated at $(b, m) = (1, 0)$, provides, in principle, a different $SL(2, \Z)$-invariant integrated four-point function in $\cN = 4$ SYM:\footnote{As we will discuss, the partition function $Z(b, m, \tau, \bar \tau)$ suffers from scheme-dependent ambiguities, but the combinations of derivatives in \eqref{Derivatives} are scheme-independent.  In particular, the subtraction of $15 \partial_b^2 Z$ in the third quantity is needed for removing such an ambiguity. We will discuss these scheme-dependent ambiguities in Appendix~\ref{app:scheme}.}
	\es{Derivatives}{
		&\tau^2_2\partial_\tau\partial_{\bar\tau} \partial_m^2 
		\log Z  \Bigr|_{\substack{m=0\\b=1}} \,, \qquad 
		\tau_2^2 \partial_\tau\partial_{\bar\tau} \partial_b^2 \log Z \Bigr|_{\substack{m=0\\b=1}}  \,, \qquad
		(\partial_b^4 - 15 \partial_b^2)  \log Z \Bigr|_{\substack{m=0\\b=1}} \,, \\
		&\partial_m^4 \log Z  \Bigr|_{\substack{m=0\\b=1}} \,, \qquad  \qquad \quad
		\partial_m^2\partial_b^2 \log Z \Bigr|_{\substack{m=0\\b=1}}
		\,.
	}
	Because both $m$ and $\delta b \equiv b-1$ couple in the action to integrated operators that belong to the ${\cal N} = 4$ stress tensor multiplet, it should be possible to express all quantities in \eqref{Derivatives} in terms of integrated four-point functions of stress tensor multiplet operators.  Of course, it is plausible that not all such integrated correlators are independent, because there may be relations between them that are implied by the $\cN = 4$ superconformal symmetry.  In fact, as we will discuss in Section~\ref{sec:instantons}, one of our main results is a derivation\footnote{For one of these relations, we do not have a full proof, but amass significant evidence in the case where the gauge group is $SU(N)$. In Appendix~\ref{app:relgenG}, we will make comments about these relations in $\cN=4$ SYM with a general gauge group.} of three linear relations between the quantities in \eqref{Derivatives}, as well as the conformal anomaly $c$, based on the supersymmetric localization results of \cite{Pestun:2007rz,Hama:2012bg}. 
	
	Taking into account the three linear relations mentioned above, one can take the independent quantities in \eqref{Derivatives} to be 
	\es{IndepDeriv}{
		\tau^2_2\partial_\tau\partial_{\bar\tau} \partial_m^2  \log Z  \Bigr|_{\substack{m=0\\b=1}} \,, 
		\qquad 
		\partial_m^4 \log Z  \Bigr|_{\substack{m=0\\b=1}} \,.
	}  
	The precise relation of these two quantities to integrated correlation functions was explained in \cite{Binder:2019jwn} and \cite{Chester:2020dja}, respectively.  In slightly more detail, due to the fact that the stress tensor multiplet of ${\cal N} = 4$ SYM is a $1/2$-BPS multiplet, it can be shown that supersymmetry requires the correlators of any four operators from this multiplet to be {\em algebraically} related to a single function $\cT(U, V)$ of the conformally-invariant cross ratios $U$ and $V$ \cite{Belitsky:2014zha}.  Thus, the two independent quantities in \eqref{IndepDeriv} should be expressible in terms of integrals $\cT(U, V)$ with potentially different integration measures.  It is these explicit expressions in terms of integrals of $\cT(U, V)$ that were given in \cite{Binder:2019jwn} and in \cite{Chester:2020dja}, respectively.
	
	The main question we ask in this work is what modular invariants\footnote{We emphasize here that while all correlators of half-BPS operators are $SL(2,\mZ)$ invariant, the correlators that involve their superconformal descendants may not be. In particular they would violate the $U(1)_Y$ bonus symmetry for five- and higher-point functions \cite{Intriligator:1998ig,Intriligator:1999ff}. For four-point functions, it was conjectured in \cite{Intriligator:1998ig,Intriligator:1999ff} that the $U(1)_Y$ bonus symmetry and consequently $SL(2,\mZ)$ invariance hold  for half-BPS operators and their descendants.  For four-point functions of stress-tensor multiplet operators, which are of interest here, the $U(1)_Y$ invariance follows from the fact that the superconformal Ward identities impose coupling-independent algebraic relations between any four-point function of stress tensor multiplet operators and the $U(1)_Y$-invariant four-point function of the half-BPS superconformal primary of this multiplet.  Therefore, while the fourth derivatives in \eqref{Derivatives} and \eqref{IndepDeriv} are expected to be $SL(2,\mZ)$ invariant, the modular properties of the higher derivatives will be more complicated.}  appear in the $1/N$ expansion of the quantities in \eqref{IndepDeriv} for the large $N$ $\cN=4$ SYM?   For $ \tau^2_2\partial_\tau\partial_{\bar\tau} \partial_m^2  \log Z  \Bigr|_{\substack{m=0\\b=1}} $, this question was answered in \cite{Chester:2019jas}, building on the work of \cite{Binder:2019jwn,Chester:2019pvm} where only the perturbative terms in $g_\text{YM}$ were studied.  Ref.~\cite{Chester:2019jas} found that this quantity has an expansion in half-integer\footnote{By a half-integer we mean a number in the set $\Z + {1\over 2}$.} powers of $1/N$, and that at each order in the expansion the answer can be written as a finite sum of non-holomorphic Eisenstein series
	\es{EisensteinDef}{
		E(s,\tau,\bar\tau)=\sum_{(m,n)\ne (0,0)} \frac{\tau_2^s}{|m + n\tau|^{2s}}
	} 
	for various half-integer values of $s$.    The perturbative terms in the second quantity in \eqref{IndepDeriv} were studied in \cite{Chester:2020dja}, and the non-perturbative contributions will be studied here.  As we will show, we find strong evidence that the $1/N$ expansion of this quantity involves not only the non-holomorphic Eisenstein series but also another class of modular-invariant functions that generalize the non-holomorphic Eisenstein series in the following sense. The Eisenstein series \eqref{EisensteinDef} satisfies the homogeneous Laplace eigenvalue equation
	\es{LaplaceEq}{
		\left( 4\tau_2^2\partial_\tau \partial_{\bar \tau} - s(s-1) \right) E(s,\tau,\bar\tau) = 0 \,.
	}
	The new modular functions we encounter are solutions to similar Laplace eigenvalue equations but with a source term given by a product of two Eisenstein series:\footnote{The Laplace equation \eqref{laplacewsource} and $SL(2,\mZ)$ invariance do not completely fix  $\cE(r,s_1,s_2,\tau,\bar \tau)$.  In particular, the solution to \eqref{laplacewsource} is ambiguous up to a shift by the Eisenstein series $E(r+1,\tau,\bar \tau)$. Later we will fix this ambiguity by specifying the cusp behavior $\cE(r,s_1,s_2,\tau,\bar \tau)$ as $\tau_2 \to \infty$. See also Appendix~\ref{lapsol} for more details.} 
	\es{laplacewsource}{
		\left( 4\tau_2^2\partial_\tau \partial_{\bar \tau} - r(r+1) \right) \cE(r, s_1, s_2, \tau,\bar\tau)= - E(s_1,\tau,\bar\tau)E(s_2,\tau,\bar\tau) \,.
	} 
	In particular, we find that at half-integer orders in $1/N$, the second quantity in \eqref{IndepDeriv} is still written in terms of the Eisenstein series \eqref{EisensteinDef}, while at integer orders in $1/N$ the expansion is in terms of $\cE(r, s_1, s_2, \tau, \bar \tau)$ for various values of $r$, $s_1$, and $s_2$.\footnote{In particular, we will see that the order ${1\over N^{p-2}}$ contributions to the SYM free energy $F=-\log Z$ with $p \in \mZ_{\geq 0}+{1\over 2}$ are given by the Eisenstein series $E(s,\tau,\bar \tau)$ with $s=p,p-2,\dots,{3\over 2}$. On the other hand, the order ${1\over N^{q}}$ contributions with $q \in \mZ_{>0}$ are controlled by the general modular functions $\cE(r, s_1, s_2,  \tau, \bar \tau)$ for  $s_1,s_2\in \mZ_{>0}+{1\over 2}$ with $s_1+s_2=q+2,q,\dots$.}
	
	At low orders in the $1/N$ expansion, these findings are perhaps not entirely surprising, because, as we will explain in Section~\ref{sec:applyto4pf}, at these orders one can establish a precise connection between the integrated correlators \eqref{IndepDeriv}, expanded in $1/N$, and 10d type IIB superstring scattering amplitudes of gravitons and their superpartners, expanded at low momentum, as a consequence of the AdS/CFT correspondence \cite{Maldacena:1997re,Witten:1998qj,Gubser:1998bc}.  At leading orders in the momentum expansion, the latter quantity contains certain supersymmetric terms that are purely analytic in momentum and whose coefficients are modular functions such as the ones encountered above. Most notably, the S-matrix contributions from $R^4$, $D^4 R^4$, and $D^6 R^4$ vertices are suppressed relative to the tree-level supergravity contribution by six, ten, and twelve orders, respectively, in the small momentum expansion, and they are proportional to the modular functions $E(\threeh, \tau, \bar \tau)$, $E(\fiveh, \tau, \bar \tau)$, and $\cE(3, \threeh, \threeh, \tau, \bar \tau)$, respectively \cite{Green:1997tv,Green:1997as,Green:1998by,Green:1999pu,Green:2005ba}.\footnote{See \cite{Wang:2015aua} for another perspective on the differential equation \eqref{laplacewsource} as coming from constraints of IIB supersymmetry. Analogous arguments have also been applied to higher-point interactions which violate the $U(1)$ symmetry \cite{Green:2019rhz}.}${}^,$\footnote{ Functions satisfying \eqref{laplacewsource} with various values of $s_1$ and $s_2$  were discussed in \cite{Green:2008bf}, where they arose in the context of higher-order terms in the low energy expansion of flat-space type II superstring amplitudes.}
	
	The connection between the superstring scattering amplitudes and the integrated correlators \eqref{Derivatives} is based on two facts.  The first is that for CFTs with weakly-coupled holographic duals, the CFT correlators in Mellin space represent the AdS analogs of scattering amplitudes, and, moreover, from the asymptotic form of the `Mellin amplitudes' in the limit of large Mellin space variables one can recover the scattering amplitudes in flat space \cite{Polchinski:1999ry,Susskind:1998vk,Giddings:1999jq,Penedones:2010ue,Fitzpatrick:2011hu,Fitzpatrick:2011jn}.  Conversely, if the flat space amplitude is known, it can be used to determine the leading term in the Mellin amplitude.  The second fact is that order by order in the $1/N$ expansion, analytic bootstrap conditions (meaning analyticity, crossing symmetry, and supersymmetry) can be used to write the separated point correlation functions of the stress tensor multiplet operators, encoded in the function $\cT(U, V)$ mentioned above, as a finite sum of specific functions of $(U, V)$ with a priori arbitrary coefficients  \cite{Alday:2013opa, Rastelli:2016nze, Aharony:2016dwx, Rastelli:2017ymc,Rastelli:2017udc, Caron-Huot:2018kta,Alday:2014qfa,Alday:2018pdi,Alday:2017xua, Aprile:2017bgs, Aprile:2017qoy, Alday:2017vkk, Aprile:2018efk, Aprile:2019rep, Alday:2019nin, Drummond:2019hel, Giusto:2018ovt, Rastelli:2019gtj, Giusto:2019pxc, Giusto:2020neo,Alday:2020tgi}.  The number of coefficients that are not determined by the bootstrap approach grows with the order in the expansion.  In particular, the Mellin amplitude corresponding to $\cT(U, V)$ is
	\es{mellinintro}{
		\cM(s, t) =& \frac{8}{(s - 2) (t - 2) (u - 2)} \frac 1c 
		+ \frac{\alpha}{c^{7/4}} + \frac { \cM_{\text{1-loop}}(s, t)}{c^2}
		\\
		&+ \frac{\beta_2 (s^2 + t^2 + u^2) + \beta_1}{c^{9/4}} + \frac{\gamma_3 stu+\gamma_2 (s^2 + t^2 + u^2) + \gamma_1}{c^{5/2}} + O(c^{-11/4}) \,,
	}
	where $s$, $t$ are the Mellin space variables with $u = 4 - s - t $, and where we have re-expressed the ${1\over N}$ expansion as a $1\over c$ expansion in terms of the conformal anomaly coefficient $c={N^2-1\over 4}$ of the $\cN=4$ SYM theory, which is more natural from the CFT perspective.  In \eqref{mellinintro}, the coefficients $\A, \B_i,\ga_i$ depend on $(\tau, \bar \tau)$.\footnote{Here $\cM_{\rm 1-loop}$ is a meromorphic term that corresponds to the regularized supergravity one-loop amplitude in the holographic dual, and it also contains a coefficient that is not determined from the bootstrap approach.  We will not discuss this term, but we note that the coefficient mentioned above was determined in \cite{Chester:2019pvm} from supersymmetric localization.} To determine these coefficients, the approach proposed in \cite{Binder:2019jwn, Chester:2019jas,Chester:2020dja}\footnote{ See also \cite{Binder:2018yvd,Binder:2019mpb} for similar computations in 3d CFTs. The general method of computing higher derivative corrections using non-trivial CFT was initiated in \cite{Chester:2018aca} in 3d, where a certain OPE coefficient computed using localization was used to the fix the $R^4$ correction. A similar approach was also taken to compute $R^4$ in 6d in \cite{Chester:2018dga}, where the nontrivial OPE coefficient was now computed using the 2d chiral algebra subsector of \cite{Beem:2014kka}.} was to use the integrated correlators that are calculable from supersymmetric localization as well as the flat space limit of the Mellin amplitude.  In this case, as we explain in Section~\ref{sec:applyto4pf}, the constraints coming from the integrated correlators \eqref{IndepDeriv}, expanded in $1/N$, are sufficient to determine all the constants $\A, \B_i,\ga_i$ in \eqref{mellinintro}.\footnote{It is possible that not all relations that reduce the integrated correlators from the list in \eqref{Derivatives} to that in \eqref{IndepDeriv} follow from $\cN=4$ supeconformal symmetry, and if this is the case, then one would be able to determine $\A, \B_i,\ga_i$ purely from the integrated correlators without the need for the flat space limit.}  As a preview, one finds
	\es{abg}{
		\A &= {15\over 4\sqrt{2 \pi^3}} E(\threeh,\tau,\bar \tau)\,, \qquad \B_1 = -{1\over 3}\B_2={315\over 128 \sqrt{2\pi^5}} E(\fiveh,\tau,\bar \tau) \,, \\
		\ga_3 &= -4 \gamma_2 = - \frac{1}{4} \gamma_1 = \frac{315\cE(3,\threeh,\threeh, \tau, \bar \tau)}{64\pi^3} \,.
	}
	
	We should note that developing the $1/N$ expansion of the quantities \eqref{IndepDeriv} is an onerous task.  The supersymmetric localization results of  \cite{Pestun:2007rz,Hama:2012bg} cast \eqref{IndepDeriv} as $(N-1)$-dimensional integrals over the zero modes of certain scalars in the ${\cal N} = 4$ vector multiplet.  The integrand contains a product of two factors coming from fluctuations localized at the poles of the sphere, where each factor takes the form of a Nekrasov instanton partition function \cite{Nekrasov:2002qd,Nekrasov:2003rj,Pestun:2007rz}.  These factors are the ones responsible for the non-perturbative effects we study in this work that are crucial for obtaining the modular functions mentioned above.

	The rest of the paper is organized as follows. In Section~\ref{sec:modularfcn}, we start with a general discussion of the large $N$ expansion of the integrated correlators and introduce the new modular functions  $\cE(r, s_1, s_2, \tau, \bar \tau)$ that generalize the non-holomorphic Eisenstein series. In Section~\ref{sec:instantons}, we study in detail the localization constraints coming from \eqref{IndepDeriv} keeping track of the instanton effects.  We apply these results in Section~\ref{sec:applyto4pf} to correlation functions at separated points. In Section~\ref{sec:conclusion}, we end with a brief summary and discuss future directions. Various technical details are contained in the Appendices.  In particular, Appendix~\ref{lapsol} contains some  details concerning solutions of \eqref{laplacewsource}  that are important in guiding the analysis in Section~\ref{sec:instantons}.

	\section{Overview of modular invariants and integrated correlators at large $N$}
	\label{sec:modularfcn}
	
	As mentioned in the Introduction, our interest is in the integrated four-point functions  \eqref{Derivatives} of the $\cN=4$ SYM expanded at large~$N$. Before delving into detailed calculations, let us provide an overview of these expansions, review previous results, and state our main results.
	
	\subsection{Sphere partition function from supersymmetric localization}
	
	The partition function of the mass-deformed $SU(N)$ $\cN=4$ SYM theory placed on a squashed four-sphere parameterized by squashing parameter $b$ is, up to an overall normalization constant that is independent of $b$ and the mass $m$, given by \cite{Pestun:2007rz,Hama:2012bg}:\footnote{Note that the factors $\Upsilon'_b(0)$  were missing in the localization formula of \cite{Hama:2012bg}. They come from the (regularized) one-loop determinant of the $\cN=2$ vector multiplet associated to each Cartan generator, namely
		\ie
		\Upsilon'_b(0)=\prod_ {m,n \geq 0,\, (m,n)\neq 0 } (m b+ n/b+Q)(m b+ n/b)\,,
		\fe
		where $\Upsilon'_b(0)$ denotes the $z$-derivative of $\Upsilon_b(z)$ at $z=0$. One can also rewrite $\Upsilon'_b(0)$ as
		\ie
		\Upsilon'_b(0)= {2\pi \over \Gamma_b(Q)^2}
		\fe
		in terms of  the Barnes double Gamma function $\Gamma_b(x)$  (see e.g. \cite{Drukker:2009id,Tachikawa:2016kfc} for properties of these special functions). We emphasize that these squashing-dependent factors are crucial for producing the correct CFT correlators by taking derivatives of the SYM partition function with respect to $b$.
	}
	\es{N2starMatrixModel}{
		Z
		= {1\over N!}\int d^{N-1} a\,  \abs{Z_\text{inst}(m,\tau,b,a_{ij})}^2  e^{-\frac{8 \pi^2 N }{\lambda} \sum_i a_i^2} \frac{\Upsilon'_b(0)^{N-1}           \prod_{i < j}\Upsilon_b(ia_{ij})\Upsilon_b(-ia_{ij})}{\Upsilon_b(im+\frac{Q}{2})^{N-1}\prod_{i \neq j}\Upsilon_b(im+\frac{Q}{2}+ia_{ij})} \,,
	}
	where $Q\equiv b+{1\over b}$ and $a_{ij}\equiv a_i-a_j$. The integration is over $N$ real variables $a_i$, $i = 1, \ldots, N$, subject to the constraint $\sum_i a_i = 0$.\footnote{The constrained integral over the $a_i$ can be implemented, for instance, by an integral over $N$ unconstrained $a_i$'s with a $\delta(\sum_i a_i)$ insertion. 
		%We will only consider derivatives of $\log Z$ for which this constraint will not matter.
	} In the integrand of  \eqref{N2starMatrixModel}, $\abs{Z_\text{inst}}^2$ captures the contribution from instantons localized at opposite poles of $S^4$ that we will come to shortly \cite{Nekrasov:2002qd,Nekrasov:2003rj,Pestun:2007rz}, and the one-loop determinants of SYM fields are written in terms of  $\Upsilon_b(x)$ which has the following convenient integral definition \cite{Nakayama:2004vk}
	\es{upsilon}{
		\log\Upsilon_b(x)=\int_0^\infty \frac{d\omega}{\omega}
		\left[ e^{-2 \omega }\left( {Q\over 2}-x\right)^2 - \frac{\sinh ^2\left(\omega  \left( {Q\over 2}-x\right) \right)}{ \sinh \frac{\omega}{b} \sinh (b \omega) } \right]\,.
	}
	The bare sphere free energy, $F_{\rm bare} = - \log Z_{\rm bare}$, of the deformed $\cN=4$ SYM theory has logarithmic divergences in addition to power-law divergences, as is the case for all (deformed) sphere free energies in even dimensions. Consequently, the regularized expression in \eqref{N2starMatrixModel} has an ambiguity of the form \cite{Bobev:2013cja,Crossley:2014oea}
	\es{weyl}{
		\log Z \to \log Z + \kappa_1 + \kappa_2 m^2+\kappa_3(b+b^{-1})^2\,,
	}
	where the coefficients $\kappa_i(\tau,\bar\tau)$ satisfying $\partial_\tau \partial_{\bar \tau}\kappa_i=0$ depend on the regularization scheme (see Appendix~\ref{app:scheme} for details).
	We must therefore be careful to only consider derivatives of $\log Z$ such that regularization-dependent terms cancel. For instance, the only well-defined two-derivative term with respect to $(b,m,\tau,\bar\tau)$ is $\partial_\tau\partial_{\bar\tau}\log Z\big\vert_{m=0,b=1}$, and it can be seen from the supersymmetric localization result \cite{Pestun:2007rz} quoted above that it equals
	\es{2der}{
		\partial_\tau\partial_{\bar\tau}\log Z\big\vert_{m=0,b=1}=\frac{c}{2(\Im\tau)^2}\,,
	}
	where $c=\frac{N^2-1}{4}$ is the conformal anomaly for $SU(N)$ SYM\@.  For three derivatives, we have the non-ambiguous terms
	\es{3der}{
		(\partial_b^3+3\partial_b^2)\log Z\big\vert_{m=0,b=1}=\partial_m^3\log Z\big\vert_{m=0,b=1}=0\,,
	}
	which vanish identically. The vanishing of the latter term follows immediately from the fact that $Z$ is an even function of $m$, while the vanishing of the former comes from the invariance of $Z$ under $b\leftrightarrow 1/b$ (see Appendix~\ref{app:relgenG} for details). At four derivatives, we consider the nontrivial quantities listed in \eqref{Derivatives}, which we argue should satisfy the three relations 
	\es{relation1}{
		0&=(\partial_\tau\partial_{\bar\tau}\partial_m^2-\partial_\tau\partial_{\bar\tau}\partial_b^2)\log Z\big\vert_{m=0,b=1}\,,\\
		0&=(-6\partial_b^2\partial_m^2+\partial_m^4+\partial_b^4-15\partial_b^2)\log Z\big\vert_{m=0,b=1}\,,\\
		-16c&=(3\partial_b^2\partial_m^2-\partial_m^4-16\tau_2^2\partial_\tau\partial_{\bar\tau}\partial_m^2)\log Z\big\vert_{m=0,b=1}\,.
	}
	We will prove the first two statements, and amass significant evidence for the third.\footnote{While we focus on the $SU(N)$ case in the main text, we expect these relations to hold for $\cN=4$ SYM with general gauge groups. See Appendix~\ref{app:relgenG} for related discussions.} These relations imply that the five quantities in \eqref{Derivatives} can all be written in terms of the two quantities in \eqref{IndepDeriv}. 
	
	\subsection{Large $N$ expansion of integrated correlators}

	As mentioned in the Introduction, the first quantity in \eqref{IndepDeriv} was previously expanded in $1/N$ in  \cite{Chester:2019jas}.  The expansion took the form:
	\es{oldpaper}{
		& \tau_2^2 \partial_\tau\partial_{\bar\tau} \partial_m^2 \log Z\big\vert_{m=0,b=1} =\frac{ N^2}{4} -\frac{3\sqrt{N}}{2^4 \,\pi^{\frac32}} E( {\scriptstyle {3 \over 2}},\tau,\bar\tau)+\frac{45}{2^8 \sqrt{N}\pi^{\frac52}}E( {\scriptstyle {5 \over 2}},\tau,\bar\tau)\\
		&\quad+\frac{1}{{N}^{\frac32}}\left[-\frac{39}{2^{13} \pi^{\frac32}}E( {\scriptstyle {3 \over 2}},\tau,\bar\tau)+\frac{4725}{2^{15} \pi^{\frac72}}E( {\scriptstyle {7 \over 2}},\tau,\bar\tau)\right]+\frac{1}{{N}^{\frac52}}\left[-\frac{1125}{2^{16} \pi^{\frac52}}E( {\scriptstyle {5 \over 2}},\tau,\bar\tau)+\frac{99225}{2^{18} \pi^{\frac92}}E( {\scriptstyle {9 \over 2}},\tau,\bar\tau)\right]\\
		&\quad+\frac{1}{{N}^{\frac72}}\left[\frac{4599}{2^{22} \pi^{\frac32}}E( {\scriptstyle {3 \over 2}},\tau,\bar\tau)-\frac{2811375}{2^{25} \pi^{\frac72}}E( {\scriptstyle {7 \over 2}},\tau,\bar\tau)+\frac{245581875}{2^{27} \pi^{\frac{11}{2}}}E( {\scriptstyle {11 \over 2}},\tau,\bar\tau)\right]+O(N^{-\frac92})\,,
	}
	where the non-holomorphic Eisenstein series $E(s,\tau,\bar\tau)$ were defined in our normalization in \eqref{EisensteinDef}.  In particular, $E(s, \tau, \bar \tau)$ is the unique $SL(2,\Z)$-invariant solution of moderate growth (i.e.~behaving as $\tau_2^a$, $a\in \R$, as $\tau_2\to \infty$) of the Laplace eigenvalue equation \eqref{LaplaceEq}.  The differential operator $\Delta_\tau\equiv 4\tau_2^2\partial_\tau\partial_{\bar\tau} = \tau_2^2\, (\partial_{\tau_1}^2+\partial_{\tau_2}^2)$ appearing in this equation is the hyperbolic Laplacian.   As a periodic function of $\tau_1$ with unit period, the Eisenstein series has a Fourier expansion in the form
	\es{eisdef}{
		E(s, \tau, \bar \tau)
		&= 2 \zeta(2 s)\tau_2^s + 2 \sqrt{\pi} \tau_2^{1-s} \frac{\Gamma(s - \frac 12)}{\Gamma(s)} \zeta(2s-1) \\
		&{}+ \frac{4 \pi^s\sqrt{\tau_2}}{\Gamma(s)} \sum_{k\ne 0} \abs{k}^{s-\half}
		\sigma_{1-2s}(|k|) \, 
		K_{s - \frac 12} (2 \pi \tau_2 \abs{k}) \, e^{2 \pi i k \tau_1} \, ,
	}
	where the divisor sum $\sigma_p(k)$ is defined by $\sigma_p(k)=\sum_{d>0,{d|k}}  d^p$, and $K_{s - \frac 12}$ is a  Bessel function of second kind.   Notably, the constant term in $\tau_1$ (the zero Fourier mode) has only two power-behaved (i.e.~perturbative) terms in $1/\tau_2$.  The non-zero Fourier modes are interpreted as the contributions of  D-instantons, since in the weak coupling regime: $\tau_2 \gg 1$, the Bessel functions have expansions of the form  $2 (|k| \tau_2)^\half\, K_{s-\half}(2\pi |k| \tau_2) = e^{-2\pi |k| \tau_2} \left(1+ O(\tau_2^{-1})\right)$.  From \eqref{eisdef}, we see that for $\tau_2 \gg 1$  the $k$th and $(-k)$th  modes ($k>0$) contribute a term proportional to $\left(e^{2\pi i k \tau}+ e^{-2\pi i k \bar\tau}\right)$, which is the sum of a charge-$k$ D-instanton and a charge-$(-k)$ anti-D-instanton contributions.  While the expression \eqref{oldpaper} was derived in  \cite{Chester:2019jas} by computing the coefficients of $e^{2 \pi i n \tau}$ in a power series in $1/\tau_2$, in Appendix~\ref{oldnew} we will provide an alternative derivation in which it is not necessary to expand these coefficients in $1/\tau_2$.

	In the next section we will determine the first few terms of the large-$N$ expansion of the second quantity in \eqref{IndepDeriv}, $\partial_m^4 \log Z\big\vert_{m=0,b=1}$.  As we will see, this quantity has an expansion in integer powers of $N^{-\half}$ of the general form
	\es{mderiv}{ 
		\partial_m^4 \log Z\big\vert_{m=0,b=1}= 6 N^2 + \sum_{p=0}^\infty a_{p}\, N^{-p+\half} \cG(p,\tau,\bar \tau)  
		+ \sum_{q=1}^\infty b_{q}\, N^{-q} \cH(q,\tau, \bar \tau)  \,,
	}
	where $a_q$ and $b_q$ are rational numbers multiplying powers of $\pi$. The functions   $\cG(p,\tau,\bar \tau)$  and $\cH(q,\tau,\bar \tau)$ are modular functions  (i.e.~functions that are invariant under the $SL(2,\Z)$ transformations \eqref{SL2ZTransf}).  Whereas the  large-$N$ expansion of the correlation function \eqref{oldpaper} was an expansion in half-integer powers of $1/N$, the expansion \eqref{mderiv} contains both even and odd powers of $N^{-\half}$, as was demonstrated in the analysis of the terms that are perturbative in $\tau_2^{-1}$ in \cite{Chester:2020dja}.
	
	According to the AdS/CFT correspondence, when $(\tau, \bar \tau)$ is fixed, the expansion in powers of $N^{-\half}$ corresponds to an expansion in powers of $\alpha^\prime/L^2$ and $\alpha^\prime D^2$, where $L$ is the curvature radius of anti de-Sitter space, and $D$ denotes, schematically, a space-time derivative. The large-$N$ expansion \eqref{mderiv} is therefore interpreted as a small curvature and low momentum expansion in the bulk superstring theory.   The leading term in  \eqref{mderiv} is proportional to $N^2$  and corresponds to the contribution of classical IIB supergravity, which is of order $(\alpha^\prime)^{-4}$ in our conventions.

	\subsubsection*{Half-integer powers of $1/N$}
	As we will see, the coefficients of the half-integer powers of $N$---the functions $\cG(p,\tau,\bar \tau)$---bear a strong similarity  to the analogous coefficients in \eqref{oldpaper}.  They are rational sums of non-holomorphic Eisenstein series $E(s,\tau,\bar\tau)$ with half-integer indices ${3\over 2}\leq s \leq p+{3\over 2}$.   As in \eqref{oldpaper}, the first term with a half-integer power of $N$ that appears in the large-$N$ expansion is proportional to $N^\half\, E(\threeh,\tau,\bar\tau)$.  Setting $s=\threeh$ in \eqref{eisdef} we see that this has perturbative contributions proportional to $N^\half \tau_2^\threeh \sim N^2 \lambda^{-\threeh}$ and $N^\half \tau_2^{-\half} \sim N^0 \lambda^{-\half}$, where $\lambda=4\pi\tau_2^{-1} N$ is the 't Hooft coupling. This is the order corresponding to the $(\alpha^\prime)^{-1}R^4$ interaction in the flat-space type IIB superstring effective  action.  The next term with a half-integer power of $N$ is proportional to $N^{-\half}\, E(\fiveh,\tau,\bar\tau)$ and corresponds to an $\alpha^\prime D^4 R^4$ interaction in the flat-space type IIB superstring effective  action.

	\subsubsection*{Integer powers of $1/N$}
	The coefficients of the  terms in \eqref{mderiv} that have integer powers of $1/N$---the functions $\cH(q,\tau,\bar\tau)$---are modular functions that are linear combinations of the generalized Eisenstein series $\cE(r,s_1,s_2,\tau,\bar\tau)$ obeying the inhomogeneous Laplace eigenvalue equations \eqref{laplacewsource}.  These equations generalize the equation satisfied by the coefficient of the $D^6R^4$ interaction in the flat-space type IIB superstring S-matrix, which has $r = 3$ and $s_1 = s_2 = 3/2$ \cite{Green:2005ba,Wang:2015aua}.  As can be seen from the equation \eqref{laplacewsource}, the argument $r$ is an integer that labels  the eigenvalue $r(r+1)$, while $s_1$, $s_2$  are indices of the Eisenstein series in the source term subject to the condition $s_1+s_2\leq q+2$ and $s_1,s_2\geq {3\over 2}$.\footnote{The function $\cE(3,\threeh,\threeh,\tau,\bar\tau)$ was denoted $\cE_{\threeh,\threeh} (\tau,\bar\tau)$ in \cite{Green:2005ba} and by $\cE_{0} (\tau,\bar\tau)$ in more recent literature---see, for instance, \cite{Green:2019rhz}.} As previously mentioned, the solution to Laplace equation \eqref{laplacewsource} is ambiguous. Here we fix the ambiguity by requiring $\cE(r,s_1,s_2,\tau,\bar\tau)\sim  \tau_2^{s_1+s_2}$ in the limit $\tau_2\to \infty$. In the case that corresponds to the $D^6 R^4$ interaction discussed above, this condition is required by consistency of the string perturbation expansion \cite{Green:2014yxa}. More generally, we demand $\cH(q,\tau,\bar\tau)\sim \tau_2^{q+2}$ as $\tau_2 \to \infty$, as expected for the genus zero string amplitude  corresponding to the $D^{4q+2} R^4$ interaction (after transformation to the Einstein frame).
	
	General equations of the form \eqref{laplacewsource} were considered in  \cite{Green:2008bf} where the  method for extracting the perturbative terms (power-behaved in $\tau_2$) in the zero Fourier mode (in $\tau_1$) of the functions $\cE(r,s_1,s_2,\tau,\bar\tau)$ was presented.    These terms are summarized as:
	\es{power}{
		\cE(r,s_1,s_2,\tau,\bar\tau)\vert_{\text{power}} = a_1 \tau_2^{s_1+s_2}+ a_2 \tau_2^{1+s_1-s_2}+ a_3 \tau_2^{1+s_2-s_1}+ a_4 \tau_2^{2-s_1-s_2} + \beta_r \tau_2^{-r} \,,
	} 
	where the coefficients $a_r$ are easy to determine since the corresponding powers of $\tau_2$ arise from the zero Fourier modes of  
	the source, $E(s_1,\tau,\bar\tau)E( s_2,\tau,\bar\tau)$. Equating the power-behaved terms on the left-hand and right-hand sides of \eqref{laplacewsource}  determines the values of  $a_r$.   The term proportional to $\tau_2^{-r}$ does not arise in the source term, and furthermore it satisfies the homogeneous equation since it is in the kernel of $(\Delta_{\tau}-r(r+1))$, so its contribution is zero on both the left-hand and right-hand sides of \eqref{laplacewsource}.\footnote{ Note that there is another homogeneous solution to \eqref{laplacewsource} given by $\tau_2^{r+1}$. For the cases we consider here $r\geq s_1+s_2$ (see \eqref{newpaper}) and thus such term is forbidden by the boundary condition that $\cE(r,s_1,s_2,\tau,\bar\tau)\sim  \tau_2^{s_1+s_2}$ as $\tau_2 \to \infty$.}   The value of its coefficient $\beta_r$ may be  determined by the procedure in  \cite{Green:2005ba,Green:2008bf}, which involves multiplying \eqref{laplacewsource} by $E(r+1,\tau,\bar\tau)$ and integrating over a cut-off fundamental domain of $SL(2,\mathbb{Z} )$.  The integral over the left-hand side reduces to a boundary term evaluated at the cutoff, while the integral over the product of  three Eisenstein series on the right-hand side can be evaluated by the Rankin-Selberg method.  In Appendix~\ref{lapsol} we review the details of this procedure and present explicit results for the perturbative terms appearing in \eqref{power}.
	
	The functions $\cE(r,s_1,s_2,\tau,\bar\tau)$ have a rich structure of D-instanton and anti-D-instanton contributions.  In particular, this structure includes D-instanton/anti-D-instanton pairs, unlike for the ordinary Eisenstein series $E(s,\tau,\bar\tau)$.  Indeed, the zero mode consists of the sum of the power-behaved terms in \eqref{power} together with an infinite series of D-instanton/anti D-instanton pairs with zero net instanton charge.
	This non-perturbative structure will also be studied in Appendix~\ref{lapsol}\@.  
	The cases that will be considered in this paper are the following:
	\begin{itemize}
		\item  The $1/N$ contribution, which is the order where the flat-space interaction   $(\alpha^\prime)^2 D ^6R^4$ appears.  This is a $1/8$-BPS interaction of dimension 14 with coefficient the generalized Eisenstein series $ \cE(3,\threeh, \threeh,\tau,\bar\tau)$ that was determined in detail in \cite{Green:2014yxa}.  Although we will not perform a complete analysis here, we will find strong evidence that the fourth mass derivative $\partial_m^4 \log Z \big \vert_{m=0,b=1}$ as computed from supersymmetric localization is also proportional to this generalized Eisenstein series.
		
		\item The $1/N^2$ contribution, which is the order where the flat-space interaction is $(\alpha^\prime)^4 D^{10}R^4$.  In this case we will see the function $\cH(2,\tau,\bar\tau)$ contains a rational sum of two new modular functions, $ \cE(4,\threeh, \fiveh,\tau,\bar\tau)$ and $ \cE(6,\threeh, \fiveh,\tau,\bar\tau)$.  The power-behaved terms correspond in the flat-space limit to perturbative string theory contributions ranging from genus zero to genus five. Although details of  the perturbative sector of these functions are well-understood, we will not discuss the complete expressions for the instanton contributions in the Fourier expansion. Many details of these perturbative terms as well as the $k$-instanton/anti-instanton pairs have been determined and are presented in Appendix~\ref{lapsol}.  This data (and that from other instanton sectors) will be compared with terms arising in the analysis of the localization formula for the SYM free energy, and provides compelling evidence that the $1/N^2$ coefficient is proportional to a particular rational  linear combination of the two modular invariants mentioned above.

		\item  The $1/N^3$ contribution, which is the order where the flat-space interaction is $(\alpha^\prime)^6 D^{14}R^4$.  In this case we will see that the function $\cH(3,\tau,\bar\tau)$ contains a rational sum of ten modular functions, These consist of the nine functions  $\cE(r,\threeh, \threeh,\tau,\bar\tau)$, $\cE(r,\fiveh, \fiveh,\tau,\bar\tau)$ and $ \cE(r,\threeh, \sevenh,\tau,\bar\tau)$, where $r=5,7,9$, together with the function  $ \cE(3,\threeh, \threeh,\tau,\bar\tau)$.   The sum of these ten terms contains power-behaved  (perturbative) contributions ranging from tree-level to genus-seven. Once again, these perturbative contributions of these functions are well understood but we have not completely analyzed the D-instanton contributions.  However, we have obtained sufficient information of particular single D-instanton contributions, as well as the contributions from instanton/anti D-instanton pairs to  compare with the corresponding terms that are obtained from   supersymmetric localization.  This again provides compelling evidence that the $1/N^3$ coefficient is proportional to a particular rational linear combination of the ten modular-invariant functions mentioned above.
	\end{itemize} 
	
	\subsubsection*{Explicit formula}
	
	To be concrete, the explicit formula that we find with qualitative features described above is
	\es{newpaper}{
		& \partial_m^4 \log Z\big\vert_{m=0,b=1} =6 N^2+\frac{6\sqrt{N}}{\pi^{\frac32}} E( {\scriptstyle {3 \over 2}},\tau,\bar\tau)+C_0-\frac{9}{2\sqrt{N}\pi^{\frac52}}E( {\scriptstyle {5 \over 2}},\tau,\bar\tau)-\frac{27}{2^3\pi^3N}\cE(3,{\scriptstyle {3 \over 2}},{\scriptstyle {3 \over 2}},\tau,\bar\tau)\\
		&\quad+\frac{1}{{N}^{\frac32}}\left[\frac{117}{2^8 \pi^{\frac32}}E( {\scriptstyle {3 \over 2}},\tau,\bar\tau)-\frac{3375}{2^{10} \pi^{\frac72}}E( {\scriptstyle {7 \over 2}},\tau,\bar\tau)\right]+\frac{1}{N^2}\left[C_1+\frac{14175}{704\pi^4}\cE(6,{\scriptstyle {5 \over 2}},{\scriptstyle {3 \over 2}},\tau,\bar\tau)-\frac{1215}{88\pi^4}\cE(4,{\scriptstyle {5 \over 2}},{\scriptstyle {3 \over 2}},\tau,\bar\tau)\right]\\
		&\quad+\frac{1}{{N}^{\frac52}}\left[\frac{675}{2^{10} \pi^{\frac52}}E( {\scriptstyle {5 \over 2}},\tau,\bar\tau)-\frac{33075}{2^{12} \pi^{\frac92}}E( {\scriptstyle {9 \over 2}},\tau,\bar\tau)\right]+\frac{1}{N^3}\Big[\alpha_3\cE(3,{\scriptstyle {3 \over 2}},{\scriptstyle {3 \over 2}},\tau,\bar\tau)\\
		&\qquad\qquad+\sum_{r=5,7,9}[\alpha_r\cE(r,{\scriptstyle {3 \over 2}},{\scriptstyle {3 \over 2}},\tau,\bar\tau)+\beta_r\cE(r,{\scriptstyle {5 \over 2}},{\scriptstyle {5 \over 2}},\tau,\bar\tau)+\gamma_r\cE(r,{\scriptstyle {7 \over 2}},{\scriptstyle {3 \over 2}},\tau,\bar\tau)]\Big]+O(N^{-\frac72})\,,\\
	}
	where $C_0, C_1$ are numerical constants that we will not determine here, while $\alpha_r,\beta_r,\gamma_r$ are
	\es{alphas}{
		\alpha_3 = {1161 \over 1144 \pi^3 }   \, , \quad\alpha_5 &=-\frac{135}{52 \pi ^3}  \, , \quad \alpha_7 = \frac{17364375}{1244672 \pi ^3}  \, , \quad  \alpha_9 = -\frac{7203735}{452608 \pi ^3} \, , \cr
		\beta_5 &=-\frac{30375}{832 \pi ^5} \, , \quad \beta_7 = \frac{6251175}{56576 \pi ^5} \, , \quad    \beta_9 = -\frac{2679075}{34816 \pi ^5} \, , \cr 
		\gamma_5&=-\frac{42525}{832 \pi ^5} \, , \quad  \gamma_7 = \frac{28704375}{226304 \pi ^5} \, , \quad     \gamma_9 = -\frac{9823275}{139264 \pi ^5}  \, .
	}
We emphasize that we do not have a complete proof of \eqref{newpaper}.  Instead, in the next section, we will provide abundant evidence for this expression by considering various limits of the terms in the $1/N$ expansion.  In particular, we will study perturbative contributions in the zero-instanton sector, as well as perturbative expansions around certain $(n_1, n_2)$ (anti)instanton-pair backgrounds for $|n_1|, |n_2| \leq 3$.  We include  a summary of the terms we have computed in Table~\ref{ChecksTable}.
	\begin{table}[h]
		\begin{center}
			\begin{tabular}{l|c|c|c|c|c}
				\multirow{2}{*}{order} & perturbative & \multicolumn{4}{c}{$(n_1, n_2)$ instanton terms $\sim e^{2 \pi i (n_1 \tau - n_2 \bar \tau)}$}  \\
				\cline{3-6}
				& terms & $n_1 =1, n_2 = 0$ & $n_1 =2, n_2 = 0$ & $n_1 =1, n_2 = -1$
				& $1 \leq \abs{n_{1, 2}} \leq 3$ \\
				\hline
				$N^2$ & $1$ (all)  & $-$ & $-$ & $-$ & $-$ \\
				$N^{1/2}$ & $g_{_{\rm YM}}^{-3}$, $g_{_{\rm YM}}$ (all) & $1$, $g_{_{\rm YM}} ^2$, $g_{_{\rm YM}} ^4$ & $1$ & $-$ & $-$\\
				$N^0$ & not computed & $-$ & $-$ & $-$ & $-$\\
				$N^{-1/2}$ & $g_{_{\rm YM}} ^{-5}$, $g_{_{\rm YM}} ^3$ (all) & $1$, $g_{_{\rm YM}} ^2$, $g_{_{\rm YM}} ^4$ & $1$ & $-$ & $-$\\
				$N^{-1}$ & $g_{_{\rm YM}} ^{-6}$ & $g_{_{\rm YM}} ^{-1}$, $g_{_{\rm YM}} $, $g_{_{\rm YM}} ^3$ & $g_{_{\rm YM}} ^{-1}$, $1$, $g_{_{\rm YM}} $, $g_{_{\rm YM}} ^3$ & finite $g_{_{\rm YM}} $ (all) & $g_{_{\rm YM}} ^4$, $g_{_{\rm YM}} ^6$ \\
				$N^{-3/2}$ & not computed & $1$, $g_{_{\rm YM}} ^2$, $g_{_{\rm YM}} ^4$ & $1$ & $-$ & $-$\\
				$N^{-2}$ & $g_{_{\rm YM}} ^{-8}$ & $g_{_{\rm YM}} ^{-3}$, $g_{_{\rm YM}} ^{-1}$, $g_{_{\rm YM}} $ & $g_{_{\rm YM}} ^{-3}$, $g_{_{\rm YM}} ^{-1}$, $1$, $g_{_{\rm YM}} $ & finite $g_{_{\rm YM}} $ (all) & $g_{_{\rm YM}} ^4$, $g_{_{\rm YM}} ^6$ \\
				$N^{-5/2}$ & not computed & $1$, $g_{_{\rm YM}} ^2$ & $1$ & $-$ & $-$ \\
				$N^{-3}$ & $g_{_{\rm YM}} ^{-10}$ & $g_{_{\rm YM}} ^{-5}$, $g_{_{\rm YM}} ^{-3}$, $g_{_{\rm YM}} ^{-1}$ & $g_{_{\rm YM}}^{-5}$, $g_{_{\rm YM}} ^{-3}$, $g_{_{\rm YM}} ^{-1}$, $1$ & finite $g_{_{\rm YM}} $ (all) & $g_{_{\rm YM}} ^4$, $g_{_{\rm YM}} ^6$
			\end{tabular}
		\end{center}
		\caption{A schematic summary of the evidence we have collected for \eqref{newpaper}.  An entry $g_{_{\rm YM}} ^{a} e^{2 \pi i (n_1 \tau - n_2 \bar \tau)}$ in this table means we have computed the coefficient of a term proportional to $g_{_{\rm YM}} ^a e^{2 \pi i (n_1 \tau - n_2 \bar \tau)}$ in the expansion of \eqref{newpaper}.  (The perturbative terms correspond to $n_1 = n_2 = 0$.)   A dash means  such terms are absent in the expansion.  The cases where we have computed all non-zero terms are marked with ``all.'' }  
		\label{ChecksTable}
	\end{table}%

	\section{Derivatives of deformed $S^4$ partition function}
	\label{sec:instantons}

	In this section, we will give evidence for the relations \eqref{relation1} and the large $N$ expansion \eqref{newpaper} by explicit evaluation of the various derivatives of \eqref{N2starMatrixModel}, which take the form of expectation values in the familiar Hermitian Gaussian matrix model that describes the $m=0,b=1$ partition function
	\es{Zfree}{
		Z\big \vert_{m=0,b=1}
		=  \int d^{N-1} a\, e^{-\frac{8 \pi^2  }{g_{_{\rm YM}} ^2} \sum_i a_i^2}  \prod_{i < j} a_{ij}^2\,.
	}
	We will compute these expectation values using topological recursion \cite{Eynard:2004mh,Eynard:2008we}. Since the application of this method to the $\cN=4$ partition function was already explained in detail in \cite{Chester:2019pvm,Chester:2020dja}, we will relegate the explicit calculations to Appendix \ref{app:top} and only present the results below.

	\subsection{Perturbative sector}
	\label{pert}
	
	We begin by considering the perturbative part $Z^\text{pert}$ of \eqref{N2starMatrixModel} obtained by setting $Z_\text{inst}=1$. Taking derivatives in $m,b$ and using \eqref{upsilon}, we find
	\es{pertmain}{
		\partial_\tau\partial_{\bar\tau}\partial_m^2 \log Z\big\vert^\text{pert}_{m=0,b=1}&=\partial_\tau\partial_{\bar\tau}\partial_b^2 \log Z\big\vert^\text{pert}_{m=0,b=1}=-\partial_\tau\partial_{\bar\tau}\int_0^\infty d\omega\frac{2 \omega  \mathcal{I}(\omega)}{\sinh^2\omega} \,,\\
		\partial_b^2\partial_m^2 \log Z\big\vert^\text{pert}_{m=0,b=1}&=4-12\zeta(3)+\int_0^\infty d\omega\frac{2 \omega ^2 \mathcal{I}(\omega)}{\sinh^4\omega} {(\sinh (2 \omega )-2 \omega )} \\
		&{}+\int_0^\infty d\omega\int_0^\infty dw\frac{4w\omega \mathcal{J}(\omega,w)}{\sinh^2w\sinh^2\omega}\,,\\
		\partial_m^4 \log Z\big\vert^\text{pert}_{m=0,b=1}&=-12\zeta(3)+\int_0^\infty d\omega\frac{ 8 \omega ^3\mathcal{I}(\omega)}{\sinh^2\omega}  +\int_0^\infty d\omega\int_0^\infty dw\frac{12w\omega \mathcal{J}(\omega,w)}{\sinh^2w\sinh^2\omega}\,,\\
		[\partial_b^4-15\partial_b^2] \log Z\big\vert^\text{pert}_{m=0,b=1}&=24-50\zeta(3)+\int_0^\infty d\omega\frac{ 4\omega^2\mathcal{I}(\omega)}{ \sinh^4\omega} (3\sinh(2\omega)-5\omega-\omega\cosh(2\omega))\\
		&\qquad+\int_0^\infty d\omega\int_0^\infty dw\frac{12w\omega \mathcal{J}(\omega,w)}{\sinh^2w\sinh^2\omega}\,,\\
	}
	where, as in \cite{Chester:2020dja}, we define the two- and four-body expectation values
	\es{exp2}{
		\mathcal{I}(\omega)\equiv \sum_{i, j}\langle e^{2i\omega a_{ij}}\rangle\,,\qquad\mathcal{J}(\omega,w) =\sum_{i,j,k,l}\left[\langle e^{2i\omega a_{ij}}e^{2iw a_{kl}}\rangle-\langle e^{2i\omega a_{ij}}\rangle\langle e^{2iw a_{kl}}\rangle\right]\,,
	}
	which are taken with respect to the $(b, m) = (1, 0)$ partition function \eqref{Zfree}. The terms that do not involve ${\cal I}$ and $\cJ$ come from the factor $ \frac{\Upsilon'_b(0) }{\Upsilon_b(im+\frac{Q}{2})}$ in \eqref{N2starMatrixModel} that does not depend on the integration variables $a_i$. The $\cJ$ terms in \eqref{pertmain} take the same form for each expression, up to a factor of 3, so the nontrivial difference between the expressions in \eqref{pertmain} comes from the ${\cal I}$ terms, which are also easier to evaluate. 
	
	From these expressions we see that the first and second relations in \eqref{relation1} are identically satisfied even before computing the expectation values, while the third relation follows from integration by parts after taking the expectation value.  To justify the latter statement, note that from \eqref{pertmain} we can write the perturbative contributions to the RHS of the third relation in \eqref{relation1} as
	\es{RHS2}{
		&4+\int_0^\infty \frac{d\omega}{\sinh^2 \omega} \left[- \frac{12 \omega ^3}{\sinh^2 \omega }-8 \omega ^3 +12 \omega ^2 \coth \omega 
		+\frac{8 \omega }{ g_{_{\rm YM}} ^4}\partial^2_{ g_{_{\rm YM}} ^{-2}}\right]\mathcal{I}(\omega)\\
		&=4- \frac{4 \omega ^3 \cosh\omega}{\sinh^3 \omega} \mathcal{I}(\omega)\bigg\vert_{\omega=0}+\int_0^\infty \frac{d\omega}{\sinh^2 \omega} \left[ -2   \partial_\omega \omega ^3\partial_\omega +\frac{8 \omega }{g_{_{\rm YM}} ^4 }\partial^2_{g_{_{\rm YM}} ^{-2}}\right]\mathcal{I}(\omega) \\
		&=4(1-N^2)\,.
	}
	Here, the first equality comes from performing integration by parts twice, where only one boundary term is non-vanishing, and the $\tau,\bar\tau$ derivatives in \eqref{relation1} can be written as $g_{_{\rm YM}}$ derivatives because the perturbative terms do not depend on $\theta$. The second equality comes from evaluating this boundary term, which is what yields the $N^2$, while the integrand in the second line vanishes from the explicit expression for $\mathcal{I}(\omega)$ in \eqref{exp2} and \eqref{Zfree}.

	Now that we have established the identities \eqref{relation1} at the perturbative level, let us examine the perturbative contributions to $\partial_m^4 \log Z\big\vert^\text{pert}_{m=0,b=1}$, which include both a two- and four-body term (proportional to ${\cal I}$ and $\cJ$, respectively). We can evaluate the two-body term in a large $N$ and large $\lambda$ expansion as in \cite{Chester:2019pvm}, which we then translate to large $N$ and finite $g_{_{\rm YM}}$ by simply setting $\lambda=N g_{_{\rm YM}}^2$. It is harder to determine the large $\lambda$ expansion of the four-body term, since as explained in \cite{Chester:2020dja}, the dependence on the Fourier variables $(w,\omega)$ does not factorize and thus the analytic method in Appendix D of \cite{Binder:2019jwn} cannot be easily applied. Instead, these terms were evaluated numerically to high precision in \cite{Chester:2020dja}, and from a numerical fit it was possible to extract the first few terms in the $1/\lambda$ expansion.  In Appendix \ref{app:pert}, we furthermore show that all terms of the form $N^2\lambda^{-\text{integer}}$ can in fact be computed analytically, since the Fourier variables factorize for these terms. After combining these results and converting $\lambda=g_{_{\rm YM}}^2N$, we get 
	\es{treeExpect2}{
		\partial^4_m\log Z\big\vert^\text{pert}_{m=0,b=1}&=6N^2+\sqrt{N}\left[\frac{96\zeta(3)}{g_{_{\rm YM}} ^{3}}+2g_{_{\rm YM}}  \right]+f_0(g_{_{\rm YM}} )-\frac{1}{\sqrt{N}}\left[\frac{288\zeta(5)}{g_{_{\rm YM}} ^5}+\frac{g_{_{\rm YM}} ^3}{60}\right]\\
		&-\frac1N\left[\frac{144\zeta(3)^2}{g_{_{\rm YM}} ^6}+O(g_{_{\rm YM}} ^{-4})\right]+\frac{f_{\frac32}(g_{_{\rm YM}} )}{N^{\frac32}}+\frac{1}{N^2}\left[-\frac{1080\zeta(3)\zeta(5)}{g_{_{\rm YM}} ^8}+O(g_{_{\rm YM}} ^{-6})\right]\\
		&+\frac{f_{\frac52}(g_{_{\rm YM}} )}{N^{\frac52}}+\frac{1}{N^3}\left[-\frac{6885\zeta(5)^2}{g_{_{\rm YM}} ^{10}}-\frac{42525\zeta(3)\zeta(7)}{4g_{_{\rm YM}} ^{10}}+O(g_{_{\rm YM}} ^{-8})\right]+O(N^{-\frac72})\,,
	}
	where the $f_i(g_{_{\rm YM}})$ denote functions that we have not yet been able to compute due to the aforementioned technical difficulties, while the leading small $g_{_{\rm YM}}$ term at any $N^{-\text{integer}}$ order can be computed analytically. The terms shown in \eqref{treeExpect2} match with the expectation \eqref{newpaper}, after expanding the Eisenstein series using \eqref{eisdef} and extracting the perturbative parts of the other modular functions shown in \eqref{power}. Note that we have not computed enough perturbative terms to unambiguously fix the modular functions in \eqref{newpaper}, so we need to look at the instanton sector, which we will discuss next.
	
	\subsection{Instanton sector}
	\label{instanton}
	
	We now consider the other parts of $ \log Z$ in \eqref{N2starMatrixModel} that involve $Z_\text{inst}$.  In particular, we define the non-perturbative contributions to $\log Z$ by $\log Z^\text{NP} \equiv \log (Z/Z^\text{pert})$. Taking derivatives in $(m,b)$ we find
	\es{explicitDers}{
		\partial_b^2\partial_m^2 \log Z\big\vert^\text{NP}_{m=0,b=1}&=\big[\cZ+\langle\partial_b^2\partial_m^2(Z_\text{inst}+\bar Z_\text{inst})\rangle\big]_{m=0,b=1}\,,\\
		\partial_m^4 \log Z\big\vert^\text{NP}_{m=0,b=1}&=\big[3\cZ+\langle\partial_m^4(Z_\text{inst}+\bar Z_\text{inst})\rangle\big]_{m=0,b=1}\,,\\
		(\partial_b^4-15\partial_b^2) \log Z\big\vert^\text{NP}_{m=0,b=1}&=\big[3\cZ+\langle(\partial_b^4-15\partial_b^2) (Z_\text{inst}+\bar Z_\text{inst})\rangle\big]_{m=0,b=1}\,,
	}
	where
	\es{ZZ}{
		\cZ &\equiv 2\langle  \partial_m^2Z_\text{inst}  \partial_m^2\bar Z_\text{inst}\rangle- 2\langle  \partial_m^2Z_\text{inst} \rangle\langle \partial_m^2\bar Z_\text{inst}\rangle-\langle \partial_m^2Z_\text{inst}\rangle^2-\langle  \partial_m^2\bar Z_\text{inst}\rangle^2 \\
		&{}-\int_0^\infty \frac{4\omega \,d\omega}{\sinh^2\omega}\sum_{i, j}\left[\langle e^{2\omega a_{ij}}({  \partial_m^2Z_\text{inst}}+{  \partial_m^2\bar Z_\text{inst}})\rangle-\langle e^{2\omega a_{ij}}\rangle\langle({  \partial_m^2Z_\text{inst}}+{  \partial_m^2\bar Z_\text{inst}})\rangle\right]
	}
contains expectation values of $n>2$-body terms.  As was the case with the $\cJ$ contribution in \eqref{pertmain}, $\cZ$ appears in each expression with a coefficient proportional to the coefficient of the $\cJ$ term in \eqref{pertmain}.  In deriving \eqref{ZZ}, we have used the fact that $\partial_b^2 Z_\text{inst}\big\vert_{m=0,b=1}=\partial_m^2Z_\text{inst}\big\vert_{m=0,b=1}$ (which we will explain momentarily from the structure of $Z_\text{inst}$ in \eqref{Zkfactor}) to write $\cZ$ purely in terms of derivatives with respect to $m$. The second line of \eqref{ZZ} comes from a mixed term involving derivatives of both $Z_\text{inst}$ and the $\Upsilon$ factors in \eqref{N2starMatrixModel}.
	
	The explicit Nekrasov partition function $Z_\text{inst}(m,\tau,b,a_{ij})$ was computed in \cite{Nekrasov:2002qd,Nekrasov:2003rj}.  It can be written in terms of a sum over $k$-instanton sectors $Z_\text{inst}^{(k)} (m, b,a_{ij})$ as
	\es{ZInstSum}{
		Z_\text{inst}(m, \tau,  b,a_{ij}) = \sum_{k=0}^\infty e^{2 \pi i k \tau} Z_\text{inst}^{(k)} (m, b,a_{ij}) \,,
	} 
	which is normalized so that  $Z_\text{inst}^{(0)} (m, b, a_{ij}) = 1$.\footnote{We emphasize here that $Z_\text{inst}$ is the $U(N)$ instanton partition function, which differs from the $SU(N)$ answer by the $U(1)$ instanton contribution $Z_\text{inst}^{U(1)}$  \cite{Nekrasov:2003rj,Alday:2009aq,Wyllard:2009hg,Alday:2010vg},
		\ie
		Z_\text{inst}^{U(1)}=\left [  \prod_{i=1}^\infty (1-q^i)\right]^{N\left ({Q^2\over 4}-m^2 \right )-1}\,.
		\fe
		Since $Z_\text{inst}^{U(1)}$ can be completely absorbed by the counter-term ambiguities in \eqref{weyl}, this difference  does not affect the physical observables we compute.}  Notably, $Z_\text{inst}(0, 1, a_{ij}) = 1$ \cite{Pestun:2007rz} so the instantons do not contribute to the sphere partition function at the conformal point. Inserting this expansion into \eqref{explicitDers}, we see that the two-body terms are sums over single instantons only, while $\cZ$ involves pairs of instantons in the first line and single instantons in the second line. The explicit form of $Z_\text{inst}^{(k)} (m, b, a_{ij}) $ can be found in Appendix B of \cite{Chester:2019jas} and satisfies\footnote{The instanton partition function $Z^{(k)}_{\rm inst}$ receives contributions from each $N$-vector of Young diagrams $\vec Y$ whose total size is $k$, in the form of a rational expression in $a_{ij}$, $m$ and squashing parameters $\ep_1=b,\ep_2={1\over b}$ that is determined by the shapes of the Young diagrams in $\vec Y$. The  factor $(m^2+{1\over 4}(b-1/b)^2)$ in \eqref{Zkfactor} comes from a universal part of this rational expression for each $\vec Y$ that is due to the outer-corner entries of the non-empty Young diagrams. See Appendix~B of \cite{Chester:2019jas} for details.} 
	\ie
	Z^{(k)}_{\rm inst}(m,b,a)= \left[m^2+{1\over 4}(b-1/b)^2 \right] G^{(k)}(m,b,a)\,,
	\label{Zkfactor}
	\fe
	where $G^{(k)}(m,b,a)$ is analytic in $(m,b)\in \mR^2$ around an open neighborhood of $(m,b)=(0,1)$  for generic $a_i \in \mR$. Furthermore, $G^{(k)}(m,b,a)$ is a symmetric function under $b\to 1/b$ and under $m\to -m$ separately. From these properties, we deduce
	\ie
	&(\partial_m^2 -\partial_b^2) \left.   Z^{(k)}_{\rm inst}(m,b,a) \right |_{m=0,b=1}=0\,,
\\
	&(-6\partial_b^2\partial_m^2+\partial_m^4+\partial_b^4-15\partial_b^2) \left.  
	Z^{(k)}_{\rm inst}(m,b,a)  \right |_{m=0,b=1}=-24  \partial_b\left.  
	G^{(k)}(m,b,a)  \right |_{m=0,b=1}=0 \,,
	\fe
 and consequently the first two relations in \eqref{relation1} hold identically before taking expectation values. Combined with the derivation of the perturbative terms in the previous subsection, this completes the proof of these first two relations for $SU(N)$ SYM at finite $N$ and gauge coupling $\tau$.\footnote{For SYM with general gauge groups, see Appendix~\ref{app:relgenG} for relevant discussions.} For the third relation in \eqref{relation1}, as well as for computing $\partial_m^4 \log Z\big\vert^\text{pert}_{m=0,b=1}$, we require the explicit expressions for $G^{(k)}(m,b,a)$. These are in general very complicated, so in the next few subsections we will consider just the lowest few values of $k$. We will compute the expectation values arising from these terms using both the small $g_{_{\rm YM}}$ expansion introduced in \cite{Chester:2019jas}, as well as a more powerful finite $g_{_{\rm YM}}$ method that uses the full power of topological recursion. 
	
	\subsubsection{One-instanton sector}
	\label{1inst}
	
	We begin by considering the one-instanton sector of \eqref{explicitDers}, which consists of terms proportional to $e^{2\pi i\tau}$, and denote the overall coefficient by $\log Z^{\text{NP},(1)}$ with the corresponding derivatives in $(b,m)$.\footnote{Similarly, the one-anti-instanton sector consists of terms proportional to $e^{-2\pi i\bar\tau}$, and the calculation is identical to the one in this section.} For the two-body (i.e.~non-$\cZ$) terms in \eqref{explicitDers}, we can simply set $Z_\text{inst}\to Z^{(1)}_\text{inst}$. For $\cZ$, we can replace $\cZ \to \cZ^{(1)}$, with
	\es{ZZ1}{
		\cZ^{(1)}=   &-\int_0^\infty \frac{4\omega \,d\omega}{\sinh^2\omega}\sum_{i j}\left[\langle e^{2\omega a_{ij}}{  \partial_m^2Z^{(1)}_\text{inst}}\rangle-\langle e^{2\omega a_{ij}}\rangle\langle{  \partial_m^2Z^{(1)}_\text{inst}}\rangle\right]\,,
	}
	since the other terms in \eqref{ZZ} can only contribute to higher instanton sectors. We use the explicit expression for $Z_\text{inst}^{(1)} (m, b, a_{ij}) $ in \cite{Nekrasov:2002qd,Nekrasov:2003rj}:
	\es{I1ExpressionM}{
		Z_\text{inst}^{(1)}(m,b,a_{ij})& =-{(m^2+\frac14{( b-1/b)^2 })}  \sum_{l=1}^N \prod_{l\neq i}
		{((a_{li}-i {b+1/b\over 2})^2-m^2)     \over a_{li}  (a_{li}- i(b+1/b)) } \,,
	}
	which satisfies the general form in \eqref{Zkfactor}.
	
	In order to check the third relation in \eqref{relation1}, we consider the two-body term $(3\partial_b^2\partial_m^2-\partial_m^4) Z\big\vert^\text{NP}_{m=0,b=1}$, for which contributions from $\cZ$ cancel.  We can evaluate this expectation value in a large $N$ expansion at finite $g_{_{\rm YM}}$ using topological recursion as shown in Appendix \ref{app:top}, and find
	\es{top1mmbb}{
		&(3\partial_b^2\partial_m^2-\partial_m^4) \log Z\big\vert^{\text{NP},(1)}_{m=0,b=1}=e^{\frac{8\pi^2}{g_{_{\rm YM}} }}\Bigg[-\frac{48 \sqrt{N}
			K_1 \left(\frac{8 \pi ^2}{g_{_{\rm YM}} ^2} \right)}{g_{_{\rm YM}}  }+\frac{30 K_2 \left(\frac{8 \pi ^2}{g_{_{\rm YM}} ^2} \right)}{g_{_{\rm YM}} 
			\sqrt{N}}\\
		&+\frac{{315 K_3\Big(\frac{8 \pi ^2}{g_{_{\rm YM}} ^2}\Big)}-{39 K_1\Big(\frac{8 \pi ^2}{g_{_{\rm YM}} ^2}\Big)} }{32 g_{_{\rm YM}}  N^{\frac32}}+\frac{{945 K_4\Big(\frac{8 \pi ^2}{g_{_{\rm YM}} ^2}\Big)}-{375 K_2\Big(\frac{8 \pi
				^2}{g_{_{\rm YM}} ^2}\Big)}}{128 g_{_{\rm YM}}  N^{\frac52}}\Bigg]+O(N^{-\frac72})\,,
	}
	which satisfies the third relation in \eqref{relation1} after comparing to $\tau_2^2\partial_\tau\partial_{\bar \tau}\partial_m^2\log Z\big\vert_{m=0,b=1}$ in \eqref{oldpaper} and extracting the 1-instanton term in the Eisenstein series \eqref{eisdef}.
	
	We can then include $\cZ$ by considering $\partial_m^4  Z\big\vert^\text{NP}_{m=0,b=1}$, which includes both the two-body term $\partial_m^4 Z_\text{inst}^{(1)}\big\vert_{m=0,b=1}$ and the $n>2$-body term $\cZ^{(1)}\big\vert_{m=0,b=1}$. It is harder to evaluate $\cZ^{(1)}\big\vert_{m=0,b=1}$ at finite $g_{_{\rm YM}}$, so instead we will compute it in a small $g_{_{\rm YM}}$ expansion by expanding $\partial_m^2 Z^{(1)}_\text{inst}\big\vert_{m=0,b=1}$ for small eigenvalue $a_{ij}$, computing the resulting $n$-body expectation values using topological recursion, and then performing the $\omega$ integral using a large $\lambda$ expansion, as detailed in Appendix \ref{app:top}. Finally, we add the contribution of the small $g_{_{\rm YM}}$ expansion of $\partial_m^4 Z_\text{inst}^{(1)}$, which we also compute  at finite $g_{_{\rm YM}}$ in the Appendix, to get the full result
	\es{top1Fin}{
		&\partial_m^4  \log Z\big\vert^{\text{NP},(1)}_{m=0,b=1}=\sqrt{N}\left[\frac{24 }{\sqrt{\pi }}+\frac{9 g_{_{\rm YM}} ^2 }{8 \pi ^{5/2}}-\frac{45 g_{_{\rm YM}} ^4}{1024 \pi
			^{9/2}}+O(g_{_{\rm YM}} ^6)\right] \\
		&{}+\frac{1}{\sqrt{N}}\left[-\frac{12}{\sqrt{\pi }}-\frac{45 g_{_{\rm YM}} ^2}{16 \pi ^{5/2}}-\frac{315 g_{_{\rm YM}} ^4}{2048 \pi
			^{9/2}}+O(g_{_{\rm YM}} ^6)\right]\\
		&+\frac{1}{N}\left[-\frac{54 \zeta (3)}{\pi ^{5/2} g_{_{\rm YM}}  }-\frac{2565 g_{_{\rm YM}}  \zeta (3)}{32 \pi ^{9/2}}+\frac{3 g_{_{\rm YM}} ^3 (512 \pi
			^4-51345 \zeta (3))}{4096 \pi ^{13/2}}+O(g_{_{\rm YM}} ^5)\right]\\
		&+\frac{1}{N^{\frac32}}\left[-\frac{27}{16 \sqrt{\pi }}-\frac{1881 g_{_{\rm YM}} ^2}{1024 \pi ^{5/2}}-\frac{53595 g_{_{\rm YM}} ^4}{131072 \pi
			^{9/2}}+O(g_{_{\rm YM}} ^6)\right]\\
		&+\frac{1}{N^2}\left[\frac{135 \zeta (5)}{2 \pi ^{5/2} g_{_{\rm YM}} ^3}+\frac{135 (32 \pi ^2 \zeta (3)-705 \zeta (5))}{128
			\pi ^{9/2} g_{_{\rm YM}} }+\frac{945 g_{_{\rm YM}}  (6464 \pi ^2 \zeta (3)-83985 \zeta (5))}{16384 \pi
			^{13/2}}+O(g_{_{\rm YM}} ^3)\right]\\
		&+\frac{1}{N^{\frac52}}\left[-\frac{45}{64 \sqrt{\pi }}-\frac{8235 g_{_{\rm YM}} ^2}{4096 \pi ^{5/2}}+O(g_{_{\rm YM}} ^4)\right]\\
		&+\frac{1}{N^3}\Bigg[\frac{8505 \zeta (7)}{64 \pi ^{5/2} g_{_{\rm YM}} ^5}-\frac{675 (256 \pi ^2 \zeta (5)-5229 \zeta
			(7))}{4096 \pi ^{9/2} g_{_{\rm YM}} ^3}\\
		&-\frac{27 (886784 \pi ^4 \zeta (3)-62630400 \pi ^2 \zeta
			(5)+555626925 \zeta (7))}{524288 \pi ^{13/2} g_{_{\rm YM}} }+O(g_{_{\rm YM}})\Bigg]+O(N^{-\frac{7}{2}})\,,
	}
	which matches the 1-instanton sector of the expected combinations of modular invariants in \eqref{newpaper}, which for the Eisenstein series is given in  \eqref{eisdef} and for the other modular functions is given in Appendix~\ref{lapsol}.

	\subsubsection{Two-instanton sector}
	\label{2inst}
	
	The two-instanton sector  of \eqref{explicitDers} consists of terms proportional to $e^{4\pi i\tau}$ whose coefficients we denote by $\log Z^{\text{NP},(2)}$ with the corresponding $(b,m)$ derivatives. For the two-body terms, we can simply set $Z_\text{inst}\to Z^{(2)}_\text{inst}$, while for $\cZ$ in \eqref{ZZ} we have
	\es{ZZ2}{
		\cZ^{(2)}=   &-\langle \partial_m^2Z^{(1)}_\text{inst}\rangle^2-\int_0^\infty \frac{4\omega \,d\omega}{\sinh^2\omega}\sum_{i j}\left[\langle e^{2\omega a_{ij}}{  \partial_m^2Z^{(2)}_\text{inst}}\rangle-\langle e^{2\omega a_{ij}}\rangle\langle{  \partial_m^2Z^{(2)}_\text{inst}}\rangle\right]\,,
	}
	where the first term is new relative to the one-instanton case \eqref{ZZ1}. We use the explicit expression for $Z_\text{inst}^{(2)} (m, b, a_{ij}) $ in \cite{Nekrasov:2002qd,Nekrasov:2003rj}:
	\es{I2ExpressionM}{
		Z^{(2)}_{\rm inst}(m,b,a_{ij})
		=Z_{\rm inst}^{\tiny\yng(2)}+Z_{\rm inst}^{\tiny\yng(1,1)}+Z_{\rm inst}^{\tiny\yng(1),\tiny\yng(1)}\,,
	}
	where the three terms correspond to the three vectors of Young diagrams $\vec Y$ given by the superscript (we have only listed the non-empty Young diagrams) and have the following explicit forms
	\es{Y1}{
		&Z_{\rm inst}^{\tiny\yng(2)}(m,b,a_{ij})= Z_{\rm inst}^{\tiny\yng(1,1)}(m,1/b,a_{ij})
		\\
		=&{ (m^2+{(b-{1/ b})^2\over 4})(m^2+{(3b-{1/ b})^2\over 4})\over  -2(b^2-1)}
		\sum_{i=1}^N \prod_{j\neq i} {((a_{ji}- i{b+{1/ b} \over 2})^2-m^2)((a_{ji}- i{3b+{1/b} \over 2})^2-m^2)
			\over
			a_{ji} (a_{ji}-i b)(a_{ji}-i(b+1/b))(a_{ji}-i(2b+1/b))   }\,,
	}
	and
	\es{Y2}{
		&Z_{\rm inst}^{\tiny\yng(1),\tiny\yng(1)}(m,b,a_{ij})
		={1\over 2}{(m^2+{( b-1/b)^2\over 4})^2}  \sum_{i=1}^N \sum_{j\neq i}
		{(a_{ij}^2+(i m+{b-1/b\over 2})^2)(a_{ij}^2+(i m-{b-1/b\over 2})^2) \over (a_{ij}^2+b^2)(a_{ij}^2+1/b^2)}
		\\
		&\times 
		\prod_{ k\neq i,j}  {((a_{ki}-i {b+1/b\over 2})^2-m^2)  ((a_{kj}-i {b+1/b\over 2})^2-m^2)   \over a_{ki}  (a_{ki}- i(b+1/b))a_{kj}  (a_{kj}- i (b+1/b))}\,.
	}
	which again satisfies the general form in \eqref{Zkfactor}.
	
	Since the two-instanton expression is much more complicated than the one-instanton expression, we will only perform perturbative in $g_{_{\rm YM}}$ calculations. As shown in Appendix~\ref{app:top}, we find 
	\es{top2mmbb}{
		(3\partial_b^2\partial_m^2-\partial_m^4) \log Z\big\vert^{\text{NP},(2)}_{m=0,b=1}=&\bigg[-15 \sqrt{\frac{2N}{\pi }}+\frac{255 }{8 \sqrt{2 \pi N}}+\frac{19695
		}{1024 \sqrt{2 \pi }N^{\frac32}}\\
		&+\frac{217365
		}{8192 \sqrt{2 \pi }N^{\frac52}}+O(N^{-\frac72})\bigg]+O(g_{_{\rm YM}}^2)\,,
	}
	which satisfies the two-instanton sector of the second relation in \eqref{relation1}, and 
	\es{top2Fin}{
		&\partial_m^4  \log Z\big\vert^{\text{NP},(2)}_{m=0,b=1}=\sqrt{N}\left[30 \sqrt{\frac{2}{\pi }}+O(g_{_{\rm YM}} ^2)\right]+\frac{1}{\sqrt{N}}\left[-\frac{51}{\sqrt{2 \pi }}+O(g_{_{\rm YM}} ^2)\right]\\
		&\!\!\!\!\! +\frac{1}{N}\left[ -\frac{135 \zeta (3)}{2 \sqrt{2} \pi ^{5/2} g_{_{\rm YM}}  } -\frac{18}{\pi }-\frac{12825 g_{_{\rm YM}} 
			\zeta (3)}{256 \sqrt{2} \pi ^{9/2}}+\mathfrak{k} g_{_{\rm YM}} ^2+\frac{15 g_{_{\rm YM}} ^3 \left(2048 \pi ^4-51345 \zeta (3)\right)}{65536 \sqrt{2} \pi ^{13/2}}+O(g_{_{\rm YM}} ^{4})\right]\\
		&\!\!\!\!\! +\frac{1}{N^{\frac32}}\left[-\frac{12285}{512 \sqrt{2 \pi }}+O(g_{_{\rm YM}} ^2)\right]\\
		&\!\!\!\!\! +\frac{1}{N^2}\left[\frac{675 \zeta (5)}{8 \sqrt{2} \pi ^{5/2} g_{_{\rm YM}} ^3}+\frac{135 \left(544 \pi ^2 \zeta (3)-3525 \zeta
			(5)\right)}{1024 \sqrt{2} \pi ^{9/2} g_{_{\rm YM}}  }-\frac{63}{2 \pi }+\frac{945 g_{_{\rm YM}}  \left(109888 \pi ^2 \zeta
			(3)-419925 \zeta (5)\right)}{262144 \sqrt{2} \pi ^{13/2}}+O(g_{_{\rm YM}} ^3)\right]\\
		& \!\!\!\!\! +\frac{1}{N^{\frac52}}\left[-\frac{65655}{2048 \sqrt{2 \pi }}+O(g_{_{\rm YM}} ^2)\right]\\
		& \!\!\!\!\! +\frac{1}{N^3}\Bigg[\frac{42525 \zeta (7)}{256 \sqrt{2} \pi ^{5/2} g_{_{\rm YM}} ^5}-\frac{675 \left(4352 \pi ^2 \zeta (5)-26145 \zeta
			(7)\right)}{32768 \left(\sqrt{2} \pi ^{9/2}\right) g_{_{\rm YM}} ^3}\\
		& \!\!\!\!\! -\frac{135 \left(1611776 \pi ^4 \zeta
			(3)-212943360 \pi ^2 \zeta (5)+555626925 \zeta (7)\right)}{8388608 \left(\sqrt{2} \pi
			^{13/2}\right) g_{_{\rm YM}}  }-\frac{279}{4 \pi }+O(g_{_{\rm YM}} )\Bigg]+O(N^{-\frac{7}{2}})\,,
	}
	which matches the two-instanton sector of the expected modular function \eqref{newpaper}.  Here, $\mathfrak{k}$ denotes the term at order $g_{_{\rm YM}} ^2/N$ that we have not yet computed.\footnote{As explained in  Appendix \ref{app:top}, the $g_{_{\rm YM}} ^n$ for even/odd $n$ are computed from different terms in the localization expression. In particular, the even $n$ terms come from the two body terms $\langle \partial_m^2Z_\text{inst}^{(1)}\rangle^2\big\vert_{m=0,b=1}$ and $\langle\partial_m^4Z_\text{inst}^{(2)}\rangle\big\vert_{m=0,b=1}$, while the odd $n$ terms come from $\mathcal{Z}^{(2)}$, and it turns out the latter is easier to compute to higher order in $g_{_{\rm YM}} $.} Note that the $N^{-1}$, $N^{-2}$, and $N^{-3}$ contributions at zeroth order in $g_{_{\rm YM}}$ are new types of terms that did not appear in the one-instanton expression \eqref{top1Fin}. The only class of terms that we have not checked so far are the instanton/anti-instanton pairs, which we will consider in the next subsection.

	\subsubsection{Instanton/anti-instanton sector}
	\label{mixedinst}
	
	The mixed $p$ instanton and $q$ anti-instanton sector  of \eqref{explicitDers} consists of terms proportional to $e^{2\pi i(p\tau-q\bar \tau)}$ with coefficients denoted by the corresponding derivatives of $\log Z^{\text{NP},(p,-q)}$, and only receives contributions from the first two terms in $\cZ$ in Eq.~\eqref{ZZ}:
	\es{pqinst}{
		\cZ^{(p,-q)}=2\langle  \partial_m^2Z^{(p)}_\text{inst}  \partial_m^2\bar Z^{(q)}_\text{inst}\rangle- 2\langle  \partial_m^2Z^{(p)}_\text{inst} \rangle\langle \partial_m^2\bar Z^{(q)}_\text{inst}\rangle\,.
	}
	These terms therefore take the same form for all the $(m,b)$ derivatives we consider, so they trivially satisfy the relations in \eqref{relation1}, but they can be used to nontrivially check the formula for $\partial_m^4\log Z\big\vert_{m=0,b=1}$ in \eqref{newpaper}. Conveniently, we can use the expressions for $ \partial_m^2Z^{(p)}_\text{inst} \big\vert_{m=0,b=1}$ that were already computed for any $p$ in \cite{Chester:2019jas}. For the $(1,-1)$ sector, we compute the resulting expectation values in Appendix \ref{app:top} in a large $N$ expansion at finite $g_{_{\rm YM}} $. The answer takes the form of a complicated integral that we write explicitly in Appendix~\ref{app:top}\@. We can evaluate this integral for any value of $g_{_{\rm YM}} $, and find that it matches the relevant term in \eqref{newpaper}. We have also computed contributions from the $(p,-q)$ sectors for $p,q\leq3$ in a small $g_{_{\rm YM}} $ expansion as shown in Appendix \ref{app:top}. For instance, the $(2,-2)$ term is 
	\es{pq2-2}{
		\partial_m^4  \log Z\big\vert^{\text{NP},(2,-2)}_{m=0,b=1}&=\frac{1}{N}\left[\frac{675 g_{_{\rm YM}} ^4}{2048 \pi ^5}-\frac{3375 g_{_{\rm YM}} ^6}{131072 \pi ^7}+\frac{111375 g_{_{\rm YM}} ^8}{16777216 \pi ^9}+O(g_{_{\rm YM}} ^{10})\right]\\
		&-\frac{1}{N^2}\left[\frac{11475 g_{_{\rm YM}} ^4}{16384 \pi ^5} +\frac{11475 g_{_{\rm YM}} ^6}{1048576 \pi ^7}+\frac{7952175 g_{_{\rm YM}} ^8}{134217728 \pi ^9} +O(g_{_{\rm YM}} ^{10})\right]\\
		&+\frac{1}{N^3}\left[\frac{585225
			g_{_{\rm YM}} ^4}{2097152 \pi ^5}-\frac{8304525 g_{_{\rm YM}} ^6}{134217728 \pi ^7}+\frac{6173214525 g_{_{\rm YM}} ^8}{17179869184 \pi ^9}+O(g_{_{\rm YM}} ^{10})\right]+O(N^{-4})\,,
		\\
	}
	while the other terms take a similar form and are given in Appendix \ref{app:top}\@. All these terms agree perfectly with \eqref{newpaper}, which completes the check of that formula.

	\section{Four-point function in $\cN=4$ SYM}
	\label{sec:applyto4pf}
	
	We will now apply the localization results of the previous section to constrain the four-point function $\langle SSSS\rangle$ of the stress tensor superconformal primary, which can also be constrained from its relation to the 10d IIB flat space graviton S-matrix. The superprimary $S$ transforms in the ${\bf 20}'$ of the $SO(6)_R$ R-symmetry, and it can be represented as a traceless symmetric tensor $S_{IJ}(\vec{x})$ with $I, J = 1, \ldots, 6$ as $SO(6)_R$ fundamental indices.  In order to avoid a proliferation of indices, it is customary to contract them with null polarization vectors $Y^I$ satisfying $Y \cdot Y \equiv \sum_{I=1}^6 Y^I Y^I = 0$.  Superconformal symmetry implies that the four-point function of the operator $S(\vec{x}, Y) \equiv S_{IJ}(\vec{x}) Y^I Y^J$ takes the form \cite{Eden:2000bk, Nirschl:2004pa}
	\es{FourPoint}{
		\langle S(\vec{x}_1, Y_1) \cdots S(\vec{x}_4, Y_4) \rangle 
		= \frac{1}{\vec{x}_{12}^4 \vec{x}_{34}^4}
		\left[ 
		\vec{\cS}_\text{free} + {\cal T}(U, V) \vec{\Theta} 
		\right] \cdot \vec{\cB} \,,
	}
	where $\vec{x}_{ij}  \equiv \vec{x}_i - \vec{x}_j$, and 
	\es{SThetaB}{
		\vec{\cS}_\text{free} &\equiv \begin{pmatrix}
			1 & U^2 & \frac{U^2}{V^2} & \frac{1}{c} \frac{U^2}V & \frac 1c \frac UV  & \frac 1c U
		\end{pmatrix} \,, \\
		\vec{\Theta} &\equiv \begin{pmatrix}
			V & UV & U & U(U- V - 1) & 1 - U - V & V (V - U - 1) 
		\end{pmatrix} \,, \\
		{\cal B} & \equiv \begin{pmatrix}
			Y_{12}^2 Y_{34}^2 & Y_{13}^2 Y_{24}^2 & Y_{14}^2 Y_{23}^2
			& Y_{13} Y_{14} Y_{23} Y_{24} & Y_{12} Y_{14} Y_{23} Y_{34}
			& Y_{12} Y_{13} Y_{24} Y_{34}
		\end{pmatrix} \,.
	}
	Here, $U \equiv \frac{ \vec{x}_{12}^2 \vec{x}_{34}^2}{ \vec{x}_{13}^2 \vec{x}_{24}^2}$ and $V\equiv \frac{ \vec{x}_{14}^2 \vec{x}_{23}^2}{ \vec{x}_{13}^2 \vec{x}_{24}^2}$ are the usual conformal invariant cross-ratios,  and $Y_{ij} \equiv Y_i \cdot Y_j$ are $SO(6)_R$ invariants.  Importantly, the only non-trivial information in the correlator \eqref{FourPoint} is encoded in a single function of the conformal cross-ratios, $\cT(U,V)$.
	
	We would like to study $\langle SSSS\rangle$ in the large $c$ expansion at finite $\tau$, which is related to the small momentum expansion of the IIB S-matrix at finite complexified string coupling $\tau_s = \chi_s + i / g_s$. In this limit, it is convenient to use the Mellin transform \cite{Mack:2009gy,Mack:2009mi} $\cM$ of $\cT$, which is defined as \cite{Rastelli:2017udc}:
	\es{MellinDef}{
		\cT(U, V)
		= \int_{-i \infty}^{i \infty} \frac{ds\, dt}{(4 \pi i)^2} U^{\frac s2} V^{\frac u2 - 2}
		\Gamma \left[2 - \frac s2 \right]^2 \Gamma \left[2 - \frac t2 \right]^2 \Gamma \left[2 - \frac u2 \right]^2
		\cM(s, t) \,,
	} 
	where $u \equiv 4 - s - t$.  Crossing symmetry $\cM(s, t) = \cM(t, s) = \cM(s, u)$, the conformal Ward identity, as well as the analytic properties of the Mellin amplitude (for a detailed description see \cite{Binder:2019jwn}), restrict $\cM(s, t)$ to have the following $1/c$ expansion at fixed Yang-Mills coupling:
	\es{MExpansion}{
		\cM(s, t) =& \frac{8}{(s - 2) (t - 2) (u - 2)} \frac 1c 
		+ \frac{\alpha}{c^{7/4}} + \frac { \cM_{\text{1-loop}}(s, t)}{c^2}
		\\
		&+ \frac{\beta_2 (s^2 + t^2 + u^2) + \beta_1}{c^{9/4}} + \frac{\gamma_3 stu+\gamma_2 (s^2 + t^2 + u^2) + \gamma_1}{c^{5/2}} + O(c^{-11/4}) \,,
	} 
	where the coefficients $\alpha$, $\beta_i$, $\gamma_i$, etc.~are potentially non-trivial functions of $(\tau, \bar \tau)$.  The first term corresponds to tree-level supergravity, while $\cM_\text{1-loop}$ is the regularized supergravity one-loop amplitude that can be found in \cite{Chester:2019pvm} and will not be discussed here. We will instead focus mostly on the $1/c^{7/4}$, $1/c^{9/4}$, and $1/c^{5/2}$ terms, which correspond to the $R^4$, $D^4 R^4$, and $D^6R^4$ interaction vertices in type IIB string theory, respectively.  At each order in $1/c$, one can impose constraints on the coefficients $\alpha$, $\beta_i$, $\gamma_i$, etc.~by either comparing with the (super)graviton four-point scattering amplitude in type IIB string theory in the flat space limit or using the quantities \eqref{oldpaper} and \eqref{newpaper} (or other similar quantities) derived from supersymmetric localization.  Let us first discuss the constraints from the flat space scattering amplitude, and then those from supersymmetric localization.
	
	\newpage
	
	\subsection{Constraints from the flat space limit}

	The IIB four-point scattering amplitude of 10d gravitons and superpartners are restricted by supersymmetry to be proportional to a single function $f(\Mands, \Mandt)$
	\es{ScattAmp}{
		{\cal A}(\Mands, \Mandt) = \cA_\text{SG tree} (\Mands, \Mandt) f(\Mands, \Mandt) \,,
	}
	where $\cA_\text{SG tree}$ is the tree-level four-point supergravity amplitude,\footnote{This is given by $\D^{16}(Q)\over {\bf stu}$ in the superamplitude notation where $Q$ denotes the 16-component super-momentum variable.  See, for instance, \cite{Boels:2012ie, Wang:2015jna}. In particular, the component corresponding to the four-graviton scattering  is given by ${R^4 \over \Mands \Mandt \Mandu}$, where $R$  denotes the linearized Riemann curvature tensor.} $\Mands$ and $\Mandt$ are the Mandelstam invariants.  We will also define $\Mandu \equiv - \Mands - \Mandt$.  
	In turn, this function has an expansion at small momentum (more correctly, the expansion is for small values of the dimensionless product between momentum and the string length $\ell_s$) of the form
	\es{fDef}{
		f(\Mands, \Mandt) \equiv
		1 + f_{R^4}(\Mands, \Mandt) \ell_s^6 + f_\text{1-loop}(\Mands, \Mandt) \ell_s^8 + f_{D^4 R^4}(\Mands, \Mandt) \ell_s^{10} + f_{D^6 R^4}(\Mands, \Mandt) \ell_s^{12} + \cdots \,,
	} 
	where the coefficient function that appears at each order in the expansion may be a non-trivial function of the complexified string coupling $\tau_s $.   The functions $f_{R^4}$, $f_{D^4 R^4}$, and $f_{D^6 R^4}$ can be written in terms of the modular functions introduced in Section \ref{sec:modularfcn} as \cite{Green:2005ba,Green:1997as,Green:1998by,Green:1999pu,Green:2014yxa}
	\es{fEisenstein}{
		f_{R^4} &= \frac{ \Mands \Mandt \Mandu}{64} g_s^{\frac 32} E(\threeh, \tau_s, \bar \tau_s) \,, \\
		f_{D^4 R^4} &= \frac{\Mands \Mandt \Mandu (\Mands^2 + \Mandt^2 + \Mandu^2)}{2^{11}} g_s^{\frac 52} E(\fiveh, \tau_s, \bar \tau_s) \, ,\\
		f_{D^6 R^4} &= \frac{3(\Mands \Mandt \Mandu)^2 }{2^{12}} g_s^{3} \cE(3,\threeh,\threeh, \tau_s, \bar \tau_s) \, ,
	}
	where the non-holomorphic Eisenstein series was defined in \eqref{eisdef} and the other modular function was defined as the $SL(2, \Z)$-invariant solution of the inhomogeneous equation \eqref{laplacewsource}.
	
	The relation between the function $f(\Mands, \Mandt)$ in \eqref{ScattAmp} and the Mellin amplitude \eqref{MExpansion} is given by the flat space limit formula \cite{Binder:2019jwn}
	\es{FlatLimit}{
		f(\Mands, \Mandt) = \frac{\Mands \Mandt \Mandu}{2^{11} \pi^2 g_s^2 \ell_s^8}
		\lim_{L / \ell_s \to \infty} L^{14}
		\int_{\kappa - i \infty}^{\kappa + i \infty}
		\frac{d \alpha}{2 \pi i }
		e^\alpha \alpha^{-6} \cM\left( \frac{L^2}{2 \alpha} \Mands, \frac{L^2}{2 \alpha} \Mandt \right)  \,,
	}
	where $\kappa > 0$.\footnote{When evaluating this integral, it is useful to note that $\int_{\kappa-i \infty}^{\kappa + i \infty} \frac{d \alpha}{2 \pi i} e^{\alpha} \alpha^{-n} = \frac{1}{\Gamma(n)}$.}  This relation, as well as the AdS/CFT dictionary 
	\es{AdSCFTDict}{
		\tau_s = \tau \,, \qquad
		\frac{L^4}{\ell_s^4} = \lambda=g_\text{YM}^2\sqrt{4c+1}
	}
	allow us to fix the leading $s,t$ terms in \eqref{MExpansion}, such as
	\es{betagammaFlat}{
		\alpha = \frac{15E(\threeh, \tau, \bar \tau) }{4 \sqrt{2 \pi^3}}  \,, \qquad
		\beta_2 = \frac{315E(\fiveh, \tau, \bar \tau)}{128 \sqrt{2 \pi^5}}  \,, \qquad
		\gamma_3 = \frac{945\cE(3,\threeh,\threeh, \tau, \bar \tau)}{64\pi^3}\,.
	}
	
	\subsection{Constraints from supersymmetric localization}
	
	As explained in \cite{Binder:2019jwn,Chester:2020dja}, the localization quantities $\tau_2^2\partial_\tau\partial_{\bar\tau}\partial_m^2\log Z\big\vert_{m=0,b=1}$ and $\partial_m^4\log Z\big\vert_{m=0,b=1}$ impose constraints on $\langle SSSS\rangle$ integrated over $S^4$. In the large $c$ expansion, these constraints take the form
	\es{IntConstraint}{
		\frac{   \partial_\tau \partial_{\bar \tau} \partial_m^2 \log Z}
		{\partial_\tau \partial_{\bar \tau}  \log Z} \Bigg|_{m=0,b=1}
		&= 2 -\frac{ \alpha}{5 c^{3/4}} + \frac{C_\text{1-loop}}{c} 
		-  \frac{7\beta_1 + 16 \beta_2}{35 c^{5/4}} -\frac{7\gamma_1+16\gamma_2+32\gamma_3}{35c^{3/2}}+ \cdots \,,\\
		c^{-1} {    \partial_m^4 \log Z}\big|_{m=0,b=1}
		&= 24 +\frac{ 16\alpha}{5 c^{3/4}} + \frac{C'_\text{1-loop}}{c} 
		+  \frac{112\beta_1 + 272 \beta_2}{35 c^{5/4}} +\frac{112\gamma_1+272\gamma_2+512\gamma_3}{35c^{3/2}}+ \cdots \,,\\
	}
	where $C_\text{1-loop},C'_\text{1-loop}$ are constants that depends on the precise form of the $\cM_\text{1-loop}$ amplitude that we will not study here, and $\partial_\tau  \partial_{\bar \tau}  \log Z \Big|_{m=0,b=1}$ was given in \eqref{2der}.   The right-hand sides of \eqref{IntConstraint} are obtained by integrating the Mellin amplitude \eqref{MExpansion} with certain integration measures that produce the integrated correlators which are accessible by localization. For $\partial_\tau \partial_{\bar \tau} \partial_m^2 \log Z$, the integration measure was first obtained in \cite{Binder:2019jwn}, and for ${    \partial_m^4 \log Z}\big|_{m=0,b=1}$ the measure was derived in \cite{Chester:2020dja}. Explicitly, the two integration measures are given in equations (2.15) and (2.16), respectively, of \cite{Chester:2020dja}.  As for the left-hand sides of \eqref{IntConstraint}, we use the explicit localization results in \eqref{oldpaper} and \eqref{newpaper}. After converting the expressions in \eqref{oldpaper} and \eqref{newpaper} into the $1/c$ expansion using $c=(N^2-1)/4$, the constraints \eqref{IntConstraint} fix the coefficients to be
	\es{locRes}{
		\alpha = \frac{15E(\threeh, \tau, \bar \tau) }{4 \sqrt{2 \pi^3}}  \,, \qquad
		-\frac{\beta_1}{3}=&\beta_2 = \frac{315E(\fiveh, \tau, \bar \tau)}{128 \sqrt{2 \pi^5}}  \,, \\
		8\gamma_3 +\frac{7\gamma_1}{4}=&-4\gamma_2= \frac{945\cE(3,\threeh,\threeh, \tau, \bar \tau)}{64\pi^3}\,.
	}
	Note that the values of $\alpha$ and $\beta_2$ match those computed from the flat space limit in \eqref{betagammaFlat}, which is a non-perturbative in $\tau$ check of AdS/CFT to this order in $1/c$. We can then combine the flat space limit and localization constraints to fix all the coefficients shown in \eqref{MExpansion} and obtain
	\es{MExpansion2}{
		\cM(s, t) &= \frac{8}{(s - 2) (t - 2) (u - 2)} \frac 1c 
		+  \frac{15E(\threeh, \tau, \bar \tau) }{4 \sqrt{2 \pi^3}c^{7/4}} + \frac { \cM_{\text{1-loop}}(s, t)}{c^2}\\
		&{}+ \frac{315E(\fiveh, \tau, \bar \tau)}{128 \sqrt{2 \pi^5}c^{9/4}}  \left[ (s^2 + t^2 + u^2) -3\right]
		\\
		&{}+  \frac{945\cE(3,\threeh,\threeh, \tau, \bar \tau)}{64\pi^3c^{5/2}}\left[stu-\frac14 (s^2 + t^2 + u^2) -4\right] + O(c^{-11/4}) \,,
	} 
	which is one of our main results. Note that we could not yet make use of the localization quantities involving derivatives of squashed parameter $b$, since the integrated constraints for those have not yet been derived.
	
	Note that the this Mellin amplitude takes the same form as the strong coupling expression in \cite{Chester:2020dja}, except that the coefficients of each term are promoted to their natural $SL(2, \mathbb{Z})$ completion as conjectured in  \cite{Binder:2019jwn},  for example $\zeta(3) \rightarrow g^3_{_{\rm YM}}/(16\pi^3) E(\threeh, \tau, \bar \tau)$. The CFT data can be extracted from this Mellin amplitude as done in \cite{Binder:2019jwn, Chester:2020dja}, and again takes the same form with the appropriate replacements. In particular, the Konishi operator does not receive corrections in the large $N$ limit in either the strong or very strong coupling expansions, since the only single trace operator that appear in the correlator we consider in this limit are those in the stress tensor multiplet.

	\section{Conclusion}
	\label{sec:conclusion}
	
	Let us start with a summary of our results, and afterwards discuss several future directions.  In this paper, we have studied  integrated correlators of four operators from the stress tensor multiplet of the $SU(N)$ $\mathcal{N}=4$ SYM theory, as defined by the various fourth derivatives of the $\cN=2^*$ partition function $Z(b, m)$ on a squashed four-sphere, evaluated at the conformal point $(b, m) = (1, 0)$.  In order to exhibit the $SL(2, \Z)$ modular invariance of the integrated correlators {\em and} to interpret these correlators in terms of the string theory derivative expansion around $AdS_5 \times S^5$, we considered the $1/N$ expansion of these quantities at fixed $(\tau, \bar \tau)$.  From the supersymmetric localization work of \cite{Pestun:2007rz,Hama:2012bg}, it is known that $Z(b, m)$ can be expressed as an $(N-1)$-dimensional integral, with the integrand being a product of classical, one-loop, and instanton contributions, so our main task was to expand this quantity in $1/N$ at fixed $(\tau, \bar \tau)$.  However, this expansion is quite difficult to perform in general, and thus, at the technical level, the bulk of our paper consisted in expanding various contributions to (the derivatives of) $Z(b, m)$ in $1/N$ and providing evidence that each term in the $1/N$ expansion can be written as sums of non-holomorphic Eisenstein series and generalizations thereof.
	
	Our first result was that among five possible (nontrivial) combinations of derivatives of $Z$ with respect to $(m, b,\tau, \bar \tau)$, only two are independent and can be taken to be $\tau_2^2\partial_{\tau} \partial_{\bar \tau} \partial^2_m \log Z\big\vert_{m=0,b=1}$ and $\partial^4_m \log Z\big\vert_{m=0,b=1}$.  The former was studied in \cite{Chester:2019jas} where strong evidence was presented that, beyond the leading term that scales as $N^2$, this quantity has an expansion only in half-integer powers of $1/N$ whose coefficients are linear combinations of non-holomorphic Eisenstein series.  (In Appendix~\ref{oldnew}, we presented an alternative method that improved on the one in \cite{Chester:2019jas}.). In this paper, we focussed on the other integrated correlator, $\partial^4_m \log Z\big\vert_{m=0,b=1}$, for which the $1/N$ expansion contains both half-integral and integral powers of $1/N$. Based on our computations, we conjectured that the coefficients of the half-integer powers of $1/N$ are again linear combinations of non-holomorphic Eisenstein series, while the coefficients of integer powers of $1/N$  are generalized Eisenstein series, which obey inhomogeneous  Laplace eigenvalue  equations of the form \eqref{laplacewsource}.   In particular,  the modular invariant coefficient at order $1/N$ is proportional to  $\cE(3,\threeh,\threeh,\tau,\bar\tau)$,  the well-known coefficient of $D^6R^4$ in the low-energy expansion of the flat-space type IIB superstring  amplitude.  The terms at order $1/N^2$ and $1/N^3$ have coefficients that are linear combinations of  generalized Eisenstein series with rational coefficients.   See Table~\ref{ChecksTable} where we summarized our evidence for our conjectures for the specific modular functions appearing in the $1/N$ expansion.

	Lastly, in Section~\ref{sec:applyto4pf} we discussed the relation between the integrated correlators we computed and the type IIB superstring low-energy effective action on $AdS_5\times S^5$ which encodes scattering amplitudes of bulk (super)gravitons.  As shown in \cite{Binder:2019jwn}, the separated point correlation functions of the same operators can be determined by general consistency conditions up to one, two, and three undetermined coefficients at orders $N^{\half}$, $N^{-\half}$ and $N^{-1}$, respectively.  At these orders, the correlators are determined from contact $R^4$ (at order $N^{\half}$),  $\frac{1}{L^4} R^4$ and $D^4 R^4$ (at order $N^{-\half}$), and $\frac{1}{L^6} R^4$, $\frac{1}{L^2} D^4 R^4$, and $D^6 R^4$ (at order $1/N$) interaction vertices, and in Mellin space they asymptote to the flat space scattering amplitudes corresponding, respectively, to the $R^4$, $D^4 R^4$, and $D^6 R^4$ contact interactions in type IIB string theory.   This information from the flat space limit combined with the integrated correlators $\partial_{\tau} \partial_{\bar \tau} \partial^2_m \log Z\big\vert_{m=0,b=1}$ or $\partial^4_m \log Z\big\vert_{m=0,b=1}$ allows us to uniquely determine the separated point correlators at orders $N^{\half}$, $N^{-\half}$, and $N^{-1}$.  As we discussed, it is plausible that one of the relations \eqref{relation1} does not follow from superconformal symmetry and that it thus imposes an additional non-trivial constraint on the separated-point correlation function.  If this is the case, then one would be able to determine the separated-point correlator at orders up to order $1/N$, and from it {\em derive} the flat space scattering amplitude corresponding to the $R^4$, $D^4 R^4$, and $D^6 R^4$ contact interactions.    These are precisely the terms that are also determined by supersymmetry in flat space.

	The structure of the integrated correlators beyond order $1/N$ is worth highlighting.  At these orders, the integrated correlators that can be computed using supersymmetric localization do not provide enough constraints to determine the separated-point correlators, so the Eisenstein and generalized Eisenstein series that we find do not completely characterize superstring scattering in AdS or in flat space.    Nevertheless, they do represent supersymmetry-protected interactions in $AdS_5 \times S^5$.  It is interesting to analyze their perturbative structure by examining the powers of $ g_{_{\rm YM}} $ that appear in the zero-instanton expansion of the Eisenstein series in Eqs.~\eqref{oldpaper} and \eqref{newpaper}.   For terms of order $N^{(1-2n)/2}$ for $n=0,1,2,\dots$, which correspond to $D^{4n}R^4$ vertices in the bulk, the Eisenstein series have perturbative terms that can arise from up to genus-$(n+1)$ string worldsheets.  For terms of order $N^{-n}$ for $n=1,2,3,\dots$, which correspond to $D^{2(2n+1)}R^4$ vertices in the bulk, we find contributions of up to genus-$(2n+1)$.  Interestingly, these features match previous observations about the type IIB S-matrix on flat space in \cite{Berkovits:2006vc, Green:2006gt}.  Indeed, in \cite{Green:2006gt} it was argued that if the duality between M-theory and string theory is naively assumed to hold exactly for all terms in the effective action, then one would conclude that the $D^{2k} R^4$ interaction vertices in type IIB string theory receive contributions from up to genus $k$ worldsheets, in agreement with the observation we made above about the integrated four-point functions.  It is believed, however, that the relations implied by the M-theory/string theory duality hold only for supersymmetry-protected interactions and are violated otherwise.\footnote{ We emphasize that this instance of M-theory/string theory duality (see \cite{Green:2006gt} for details) assumes that the 11d supergravity description continues to be valid  from large radius in the 11th dimension to small radius (to make contact with type IIA string theory) or from a large two-torus to small two-torus (to make contact with type IIB string theory). While this turns out to be true for BPS interactions protected by supersymmetry, in general non-perturbative M-theory effects (e.g. from M2 and M5 branes) that become large in these continuations cannot be ignored.
	} Thus, one expects that for the non-supersymmetric $D^{2k} R^4$ interactions with $k>3$, both in flat space as well as their corresponding AdS Mellin amplitudes, there should be no restriction on the genus of the string worldsheets that contribute.  However, the integrated correlators we study here are much simpler quantities than the full interaction vertex or the full Mellin amplitude, and, as mentioned above, these integrated correlators are supersymmetric.  It is thus not entirely surprising that the arguments based on the M-theory/string theory duality seem to apply to them and restrict the genus of the worldsheet contributions in a manner consistent with our explicit computations in ${\cal N} = 4$ SYM theory.  Whether this observation is a coincidence or whether it can be made more precise are questions that we leave for future work.

	Another future direction is the study of modular functions that appear in higher-point correlators.\footnote{The five-point function of the stress tensor multiplet superconformal primary was considered in \cite{Goncalves:2019znr} in the supergravity approximation.}  Whereas the four-point correlators studied in this paper conserve the bonus $U(1)_Y$ symmetry of \cite{Intriligator:1998ig,Intriligator:1999ff}, $n$-point correlators may violate $U(1)_Y$ by a maximum of $2(n-4)$ units.   Maximal $U(1)_Y$-violating $n$-point correlators of operators in the stress tensor multiplet are holographically dual to type IIB superstring $n$-particle amplitudes that violate the $U(1)$ R-symmetry maximally.   The coefficients of terms in the large-$N$ expansion of these correlators transform as  $SL(2,\Z)$ modular forms with modular weights related to their $U(1)$ charges.   In \cite{Green:2020eyj} these correlators are determined up to order $1/N$ by a recursion relation analogous to the soft dilaton relations of flat-space superstring amplitudes.  These relate the higher-point correlators to the four-point correlators determined in this paper and in \cite{Chester:2019jas}, and make contact with the results in  \cite{Green:2019rhz} concerning flat-space type IIB maximal $U(1)$-violating superstring amplitudes.

It would be interesting to construct $n$-point correlators that violate the bonus $U(1)_Y$ symmetry directly from the localization procedure by generalizing the analysis of this paper to cases where one takes more than four derivatives of the ${\cal N} = 2^*$ partition function.  In this manner we would hope to determine expressions for the modular form coefficients to any order in the large-$N$ expansion of the integrated $n$-point  correlators.

	An important loose end of our work is the proof of the last relation in \eqref{relation1} as well as determining whether or not any of these relations are consequences of supersymmetry (see Appendix~\ref{app:relgenG} for evidence for these relations for general gauge groups).  The Ward identities relating the four-point functions of various operators in the stress tensor multiplet were solved in \cite{Belitsky:2014zha}.\footnote{The first relation of \eqref{relation1} can be derived from the simpler superconformal Ward identities that relate two-point functions of the stress tensor multiplet. We comment on this near the end of Appendix~\ref{app:relgenG}.} The derivatives in the relations \eqref{relation1} involving squashing are directly related to correlators of operators with spin in the stress tensor multiplet (namely the stress tensor, the R-symmetry current, as well as a rank-two anti-symmetric tensor operator), and one would have to use the Ward identity solution in \cite{Belitsky:2014zha} to relate such correlators to those of the stress tensor multiplet superconformal primary.  It would be very valuable to perform this analysis, because it could have applications beyond $1/N$ perturbation theory, for instance in numerical bootstrap studies.

	Lastly, let us point out that the large $N$ expansion in this paper is asymptotic, as can be seen already from the all orders in $1/N$ expressions for $\partial_\tau \partial_{\bar \tau} \partial_m^2 Z\big\vert_{m=0,b=1}$ in \cite{Chester:2019pvm}, and so is expected to have exponentially small in $N$ corrections.  In the bulk, we speculate that these exponentially-suppressed corrections can be interpreted as boundary-anchored strings and branes in $AdS_5 \times S^5$.  It would be interesting to understand these contributions as well as their dependence on $(\tau, \bar \tau)$.  More generally, one might hope that our $1/N$ expansion supplemented by these exponential corrections can be resummed into a finite-$N$ modular function.  For the perturbative terms in $g_\text{YM}$, the finite-$N$ integrated correlators were computed using the method of orthogonal polynomials for both $\partial_\tau \partial_{\bar \tau} \partial_m^2 Z\big\vert_{m=0,b=1}$ \cite{Chester:2019pvm} and $\partial_m^4 Z\big\vert_{m=0,b=1}$ \cite{Chester:2020dja}, but such an analysis would be more challenging for the instanton terms.  We nevertheless hope to come back to these issues in the near future.
	
	\section*{Acknowledgments}
	
	We thank David Simmons-Duffin for useful discussions.  SMC is supported in part by a Zuckerman STEM Leadership Fellowship.   MBG has been partially supported by STFC consolidated grant ST/L000385/1.   The work of SSP was supported in part by the US NSF under Grant No.~PHY-1820651 and by the Simons Foundation Grant No.~488653.  The work of YW is  supported in part by the Center for Mathematical Sciences and Applications and the Center for the Fundamental Laws of Nature at Harvard University. CW is supported by a Royal Society University Research Fellowship No. UF160350.

	\appendix
	
	%%%%%%%%%
	%%%%%%%
	
	\section{Scheme dependence and supersymmetric  counter-terms}
	\label{app:scheme}
	The supersymmetric free energy $\log Z(b,m)$  of the $\cN=2^*$ SYM theory is subject to finite regularization ambiguities as in \eqref{weyl}.  These ambiguitites can be understood from $\cN=2$ supersymmetric counter-terms of the form \cite{Butter:2013lta,Gomis:2014woa}
	\ie
	\int d^4 x d^4 \theta \,\cE  \cF\,,
	\label{n2ct}
	\fe
	as an integral over the $\cN=2$ chiral superspace, and similarly for its complex-conjugate anti-chiral version.  Here $\cE$ is the Berezinian (superdeterminant) of the chiral superspace vielbein and $\cF$ is a (composite) background chiral superfield of Weyl weight 2 and  chiral weight $-2$.  Here, the relevant background supergravity fields consist of the Weyl (chiral) superfield $W_{\A\B}$ and the vector (chiral) superfield $\Phi$ both having Weyl weight 1 and chiral weight $-1$, as well as a chiral superfield $\cA$ of vanishing Weyl and chiral weights. They couple to squashing, mass, and marginal deformations of the theory, respectively. In particular, the bottom component of $\Phi$ parameterizes the $\cN=2$ mass parameter $m$, and that of $\cA$ parametrizes the marginal coupling $\tau$.
	
	As shown in \cite{Butter:2013lta}, with the chiral superfields $W_{\A\B},\Phi$, and $\cA$,
	there are three classes of composite fields $\cF$ of chiral weight $-2$ and Weyl weight 2,
	\ie
	\cF= f_1(\cA)\Phi^2,\quad   f_2(\cA)\mathbb{T}(\log\bar\Phi'),\quad   f_3(\cA)W_{\A \B }W^{\A\B}\,,
	\label{Ffield}
	\fe
	corresponding to three counter-terms in \eqref{n2ct}. Here $\Phi'$ is an auxiliary chiral superfield that can be identified with the compensating vector multiplet in $\cN=2$ supergravity \cite{Gomis:2014woa}, and $\mathbb{T} \propto \bar D^4$ is an antisymmetric combination of the four anti-chiral superspace covariant derivatives \cite{Butter:2013lta}.
Importantly the corresponding counter-term does not depend on the value of $\Phi'$ when $\cA$ take constant values \cite{Butter:2013lta} which is the case for the deformations considered here.

	When evaluated on the supersymmetric mass deformed background \cite{Gomis:2014woa}, the first term in \eqref{Ffield} gives $ f_2(\tau) m^2$ which (along with its complex conjugate) explains the $m$-dependent ambiguity in \eqref{weyl}. The $m$-independent ambiguities in \eqref{weyl} are related to the conformal anomaly \cite{Gomis:2015yaa} and explained by (combinations of)   the last two terms in \eqref{Ffield} (to show this explicitly requires evaluating the counter-term on the supersymmetric squashing background \cite{Hama:2012bg} which we do not pursue here).

	\section{Comments on relations between fourth derivatives of $\log Z(b,m)$ for general gauge groups}
	\label{app:relgenG}
	Here we provide some further evidence for  the three relations \eqref{relation1} between various fourth-derivatives of the SYM free energy with respect to the mass and squashing deformations as well as the complexified gauge coupling $(\tau,\bar\tau)$ for $\cN=4$ SYM with general gauge groups.  We will also argue for the first relation in \eqref{relation1} based on superconformal Ward identities.

	The $\cN=2^*$ partition function of SYM with a general gauge group $G$ on a squashed sphere  is given by\footnote{We emphasize again that the factors of $ \Upsilon'_b(0)$ that have been missing the previous works (e.g. \cite{Hama:2012bg}) carry nontrivial $b$ dependence, \ie
		\partial_b^2 \left. \Upsilon'_b(0) \right |_{b=1}=-2 (1+\gamma ),\quad 
		\partial_b^4 \left. \Upsilon'_b(0) \right |_{b=1}=12 \zeta (3)-30 \gamma -38\,,
		\fe
		where $\gamma$ is the Euler's constant,
		and are thus crucial to produce the correct CFT free energy in the presence of squashing deformations.
	} 
	\ie
	Z(b,m)={1\over |W|}\int [d^r a]  \abs{Z_\text{inst}(m,\tau,b, a)}^2 \,e^{- 2\pi \tau_2 (a,a)} {\Upsilon'_b(0)^r\over  \Upsilon_b(im+{Q\over 2})^r} \prod_{\A \in \Delta}{ \Upsilon_b(i \A( a))\over   \Upsilon_b(i \A( a)+im+{Q\over 2})}
	\label{genZ}
	\fe
	where $r$ denotes the rank of $G$, $W$ is the Weyl group, $\Delta$ is the set of roots and $(\cdot,\cdot)$ defines the standard Killing form on the Lie algebra $\mf{g}$. The instanton contributions at the two poles of $S^4$ are captured by the factor $Z_\text{inst}(m,\tau,b, a)=1+\sum_{k\geq 1}q^k Z_{\rm inst}^{(k)}(m,b,a)$ and its conjugate respectively, whose explicit forms are available for $G=SU(N)$ and are used extensively in the following  sections. For more general classical Lie groups of $BCD$ types,  $Z_{\rm inst}^{(k)}(m,b,a)$ admits a contour integral expression at each instanton number $k$ thanks to the ADHM construction of the instanton moduli space and an equivariant localization procedure thereof  \cite{Nekrasov:2002qd,Nekrasov:2003rj,Nekrasov:2004vw,Marino:2004cn,Nakamura:2014nha}. The instanton contribution to $Z$ for  exceptional Lie groups is still an open question, though by the AGT correspondence they are related to torus one-point blocks of the corresponding W-algebras (with twist for the non-simply laced cases) \cite{Alday:2009aq,Keller:2011ek}. 
	
	We note the following simple properties of $Z$
	\ie
	Z(b,m)=Z(b,-m)=Z(1/b,m)\,.
	\fe
	The first equality is due to $m$ being the mass parameter for an $SU(2)$ flavor symmetry of the SYM theory, which flips sign under an $SU(2)$ Weyl reflection. The second equality is a consequence of the fact that $b$ and $1/b$ parametrizes the identical supersymmetric squashed sphere background up to a relabelling of coordinates \cite{Hama:2012bg}. For cases where the complete integral form of  $Z(b,m)$ is known, it is easy to check that these equalities hold separately for the perturbative and instanton pieces in the integrand of \eqref{genZ}.\footnote{To verify this for the perturbative contributions,  the following identities of the Upsilon function is useful,
		\ie
		\Upsilon_b(x)=\Upsilon_{1/b}(x)=\Upsilon_b(b+1/b-x)\,.
		\label{Upsilonsym}
		\fe}
	Consequently the single derivatives of $Z$  (and separately for the perturbative and instanton contributions in \eqref{genZ} before integration) with respect to $b$ or $m$ vanishes at the symmetric values $b=1,m=0$.\footnote{Physically the vanishing of the single derivatives at $b=1,m=0$ corresponds to the vanishing of one-point functions in the CFT.} This implies that
	\ie
	&\partial_m^2 \left. \log  Z(b,m) \right |_{m=0,b=1}=\left. {\partial_m^2  Z(b,m)\over Z(b,m)} \right |_{m=0,b=1}\, ,
	\\
	&\partial_b^2 \left. \log  Z(b,m) \right |_{m=0,b=1}=\left. {\partial_b^2  Z(b,m)\over Z(b,m)} \right |_{m=0,b=1} \, ,
	\\
	&(-6\partial_b^2\partial_m^2+\partial_m^4+\partial_b^4) \left. \log Z(b,m) \right |_{m=0,b=1}
	\\
	=&
	\left. {(-6\partial_b^2\partial_m^2+\partial_m^4+\partial_b^4)  Z(b,m)\over Z(b,m)} 
	\right |_{m=0,b=1}
	-3\left. \left({(\partial_b^2 -\partial_m^2)  Z(b,m)\over Z(b,m)} \right)^2
	\right |_{m=0,b=1} \, ,
	\label{logwo}
	\fe
	and we can perform  the derivatives inside the matrix integral in \eqref{genZ} to study relations of the form \eqref{relation1}.
	
	Let us define the one-loop contribution in \eqref{genZ} from a pair of root vector $\A$  and its Weyl reflection $-\A$ as
	\ie
	H(b,m,z)\equiv { \Upsilon_b(iz)\Upsilon_b(-iz)\over   \Upsilon_b(iz+im+{b+1/b\over 2})\Upsilon_b(-iz+im+{b+1/b\over 2})} \, ,
	\fe
	with $z=\A(a) \in \mR$. Then by using the integral expression of the Upsilon function \eqref{upsilon}, one finds that $H(b,m,z)$ satisfies\footnote{Note that $H(b,m,z)$ is invariant under $b \to 1/b$ or $m \to -m$ thanks to \eqref{Upsilonsym}. Thus the equalities below hold also with $\log H(b,m,z)$ replaced by $H(b,m,z)$.}
	\ie
	&(\partial_m^2 -\partial_b^2) \left. \log  H(b,m,z) \right |_{m=0,b=1}=0\,,
	\\
	&(-6\partial_b^2\partial_m^2+\partial_m^4+\partial_b^4-15\partial_b^2) \left. \log H(b,m,z) \right |_{m=0,b=1}=0\,.
	\label{relationoneloop}
	\fe
	Note that the Upsilon function $\Upsilon_{b}(iz)$ has a simple zero at $z=0$.  Consequently, the relations \eqref{relationoneloop} continue to hold with $H(b,m,z)$ replaced by
	\ie
	{ \Upsilon_b'(0)\over   \Upsilon_b(im+{b+1/b\over 2})}\,.
	\fe
	Putting them together, we conclude
	\ie
	&(\partial_m^2 -\partial_b^2) \left. \log  Z_{\rm pert} \right |_{m=0,b=1}=0\,,
	\\
	&(-6\partial_b^2\partial_m^2+\partial_m^4+\partial_b^4-15\partial_b^2) \left.  
	\log  Z_{\rm pert}  \right |_{m=0,b=1}=0\,,
	\fe
	and thus we verify the first two relations in \eqref{relation1} for the perturbative contributions in \eqref{genZ} (before $\int [d^r a]$ integral).\footnote{Said differently, in a weak coupling expansion, $Z_{\rm pert}$ captures  the perturbative contributions to the full SYM partition function $Z$. Thus we have verified the first two relations in \eqref{relation1} up to instanton effects. } 
	In the main text, we have further proved the first two relations of \eqref{relation1} non-perturbatively for $G=SU(N)$ at finite $N$ by using the explicit form of the instanton partition function which can be found in Appendix B of \cite{Chester:2019jas}. 
	
	Concerning the last relation of \eqref{relation1}, we have verified it perturbatively for $G=SU(N)$ at finite $N$ using integration by parts as in \eqref{RHS2}. The derivation extends trivially to general gauge group $G$ after replacing ${\cal I}(w)$ by
	\ie
	{\cal I}^G(w) \equiv r+\sum_{\A \in \Delta} \la e^{2iw \A(a)} \ra \,.
	\fe 
	Furthermore, we have provided evidence for this relation of  \eqref{relation1} in the main text at the non-perturbative level for $G=SU(N)$. 
	
	Let us now comment on \eqref{relation1} in relation to $\cN=4$ superconformal Ward identities. As explained in the Introduction and in the Conclusion sections, the four point functions of operators in the $\cN=4$ stress tensor multiplet are related by the superconformal Ward identities, which upon integration over the positions, could lead to relations between integrated correlators that appear in \eqref{relation1}. Here we will provide an argument for the first relation of \eqref{relation1} as a  consequence of the superconformal Ward identity and leave the rest to future investigation.
	
	Despite its look, the first relation in \eqref{relation1} can be understood as a consequence of the supersymmetric Ward identity that relates two-point functions of operators in the stress-tensor multiplet. Indeed, such two point functions are completed fixed  up to a common normalization factor (which may depend on $(\tau,\bar\tau)$). Since both $m$ and $b$ parameterize supersymmetric background configurations that couple to the $\cN=4$ stress-tensor multiplet, second derivatives of $Z$ with respect to $(m,b)$ naturally produces these two-point functions, up to potential harmonic ambiguities \eqref{weyl} in $(\tau,\bar\tau)$ that can be removed by taking $\partial_\tau \partial_{\bar \tau}$ derivatives. Since the mass and squashing couplings are introduced in a theory-independent way in the localization setup \cite{Hama:2012bg},
	we conclude that $\partial_\tau \partial_{\bar \tau} \partial_m^2\log Z\big\vert_{m=0,b=1}$ and $\partial_\tau \partial_{\bar \tau}  \partial_b^2\log Z\big\vert_{m=0,b=1}$ must be proportional up to a theory independent constant. A quick calculation in the abelian SYM theory which has partition function\footnote{Here we have included the abelian instanton contributions for completeness \cite{Wyllard:2009hg}, though they do not affect the physical observables that come from fourth derivatives of the SYM free energy.}
	\ie
	Z_{U(1)}=&\int da   \, {\Upsilon'_b(0)\over  \Upsilon_b(im+{b+1/b\over 2})}e^{- 2\pi \tau_2 a^2} \prod_{i=1}^\infty
	|1-q^i|^{2(Q^2/4-m^2-1)}
	\\
	=& {\Upsilon'_b(0)\over  \sqrt{2\tau_2}\Upsilon_b(im+{b+1/b\over 2})}\prod_{i=1}^\infty |1-q^i|^{2(Q^2/4-m^2-1)}\,,
	\label{U1PF}
	\fe
	confirms that this proportionality constant is one and thus the desired relation follows.

	\section{Solutions of inhomogeneous Laplace equations}
	\label{lapsol}
	In this appendix we will describe some properties of the generalised Eisenstein series that satisfy equations of the form \eqref{laplacewsource} that arise in the coefficients $\cH(q,\tau,\bar\tau)$ of even terms in the $1/N$ expansion \eqref{mderiv} up to order $1/N^3$.   The function $\cE(3,\threeh,\threeh,\tau,\bar\tau)$ (the coefficient of the $D^6R^4$ interaction in flat-space type IIB superstring theory) was completely determined in \cite{Green:2014yxa},    Certain properties of more general functions satisfying  \eqref{laplacewsource}  were presented in \cite{Green:2008bf} but these were mainly restricted to the perturbative terms, whereas we are here also interested in detailed properties of the D-instanton terms for the specific functions appearing in the $1/N$ expansion.
	
	\subsection{ $\cE(3,\threeh,\threeh,\tau,\bar\tau)$}
	\label{d6r4fun}
	
	We will first review the structure of this modular invariant based on the solution of the Laplace equation  \cite{Green:2005ba}\footnote{The overall normalisation of the source term on the right-hand side of this equation has been arbitrarily set equal to $-1$.  Since $ \cE(3,\threeh,\threeh,\tau,\bar\tau)$ is the coefficient of the $D^6R^4$ interaction in the low energy expansion of the flat-space type IIB superstring action its normalisation is simple to fix from the string theory scattering amplitude.}
	\es{d6eq}{ \left(\Delta_\tau - 12\right) \cE(3,\threeh,\threeh,\tau,\bar\tau)=-E(\threeh,\tau,\bar\tau)E( \threeh,\tau,\bar\tau)\,.}
	Following \cite{Green:2014yxa}, this equation may be solved in terms of its Fourier modes defined by
	\es{fourierm}{ \cE(3,\threeh,\threeh,\tau,\bar\tau) = \sum_k  \cF_k(\tau_2) e^{2\pi i k\tau_1}\,.
	}
	It is important to understand the boundary conditions imposed on the Fourier modes that are necessary in order for the complete function to be $SL(2,\Z)$ invariant.  According to a theorem proved in  \cite{Green:2014yxa}:
	{}
\ie
\parbox{.9\textwidth}{\it   $SL(2,\Z)$ invariance of a function that grows as $\tau_2^a$ as $\tau_2\to \infty$ implies that its Fourier  modes  are bounded by $\tau_2^{1-a}$ in the $\tau_2\to 0$ limit.}
\nonumber
\fe
	\noindent  Since $ \cE(r, s_1,s_2,\tau,\bar\tau)\underset{\tau_2\to \infty}{=} O(\tau_2^{s_1+s_2}$) the boundary conditions require $\cF_n(\tau_2) \underset{\tau_2\to 0}{=} O(\tau_2^{1-s_1-s_2})$.
	
	Writing the source term on the right-hand side of \eqref{d6eq}  as 
	\es{source}{-E(\threeh,\tau,\bar\tau)E( \threeh,\tau,\bar\tau)\ = \sum_{k_1,k_2} S_{k_1k_2} (\tau_2)e^{2\pi i (k_1+ k_2)\tau_1 } = \sum_k e^{2\pi i k \tau_1} \sum_{k_1} S_{k_1,k-k_1}(\tau_2)\,}
	the $k$th mode satisfies the equation 
	\es{modeq}{ (\tau_2^2\partial_{\tau_2}^2-12-4\pi^2k^2\tau_2^2)\, \cF_{k}(\tau_2)= S_k(\tau_2)\,, \qquad n\in \Z\,.  }
	
	The general solution to the above differential equation can be found in \cite{Green:2014yxa}, where it is expressed as the sum of a particular solution and a solution of the homogeneous equation. $ \cF_{k}(\tau_2) = \cF^P_{k}(\tau_2) +  \cF^H_{k}(\tau_2)$.  The coefficient of the homogeneous solution is uniquely determined by imposing the $\tau_2\to 0$ boundary condition described above.

	The particular solution that was determined in \cite{Green:2014yxa}, was written in the form $\cF^P_k(\tau_2) = \sum_{k_1} \sum_{k_2=k-k_1} f^P_{k_1,k_2}(\tau_2)$ where 
	\es{partint}{ (\tau_2^2\partial_{\tau_2}^2-12-4\pi^2 (k_1+k_2)^2\tau_2^2)\, f^P_{k_1,k_2}(\tau_2)= S_{k_1,k_2}(\tau_2)\,.}
	
	It is useful to consider the solutions in several sectors: (a) $k_1=-k_2$ (so $k=0)$; (b) $k_1\ne 0,k_2=0$ or $k_1= 0,k_2\ne0$;; (c)  $k_1\ne0,k_2\ne 0$.with $k=k_1+k_2\ne0$.
	
	\vskip 0.2cm
	{\bf (a) $k_1=-k_2$}
	
	These terms contribute to the $k=0$  mode.
	The term with $k_1=k_2=0$  is a sum of powers of $\tau_2$ that is given by 
	\ie \label{eq:perty}
	f_{0,0}(\tau_2) = \frac{2 \zeta (3)^2 \tau_2^3 }{3} +\frac{4  \zeta(2) \zeta (3)\tau_2 }{3} +\frac{4 \zeta(4)}{ \tau_2} + \beta \tau_2^{-3}\, .
	\fe
	The first three terms in this expression originate from $f^P_{0,0}(\tau_2)$ and are easily obtained by equating the coefficients of the powers of $\tau_2^3$, $\tau_2$ and $(\tau_2)^{-1}$ on both sides of   \eqref{partint}.
	The term $\beta \tau_2^{-3}$ is a solution of the homogenous equation, and its coefficient $\beta$ was determined  in \cite{Green:2005ba} by multiplying  both sides of   \eqref{partint}  by $E(4, \tau, \bar \tau)$ and integrating over a fundamental domain of  $SL(2, \mathbb{Z})$. A detailed analysis can be found in  \cite{Green:2005ba}, which leads to $\beta = 4 \zeta(6)/ 27 $ so that 
	\ie
	\label{betares}
	\cF^H_0 ( \tau_2) =  \frac{4 \zeta(6)} {27} \tau_2^{-3}\, .
	\fe
	
	In addition to the $(0,0)$ term, the $k=0$ mode receives contributions from a sum over an  infinite number of terms with $k_1=-k_2 \ne 0$, which represent D-instanton/anti D-instanton pairs. These terms are  bilinear in K-Bessel functions and are given by  
	\ie
	\label{zeromode}
	f^P_{k_1,-k_1}( \tau_2) = {32 \pi^2 \over 315 |k_1|^3} \sigma_2(|k_1|) \sigma_2(|k_1|) \sum^1_{i,j=0} q_3^{i,j}(\pi |k_1| \tau_2)K_i(2\pi |k_1| \tau_2) K_j(2\pi |k_1| \tau_2)\, ,
	\fe
	where the coefficients $q_3^{i,j}$ are given by
	\ie
	q_3^{0,0}(z) &= z \left(-512 z^4+48 z^2-15\right) \, , \cr
	q_3^{0,1}(z) &= q_3^{1, 0}(z) =  -128 z^4-12 z^2-15 \, , \cr
	q_3^{1,1}(z) &= 512 z^5+16 z^3+33 z-\frac{15}{z} \, .
	\fe
	Making use of the  weak coupling ($\tau_2\to \infty$) expansion of the $K$-Bessel functions,
	\es{asymk}{ K_{s-\half}(2\pi |k| \tau_2) \underset{\tau_2\to \infty}{=} \frac{ e^{-2\pi |k| \tau_2}}{2 (|k| \tau_2)^\half\,} \left(1+ \frac{s(s-1)}{4 |k| \pi \tau_2}+O(\tau_2^{-2})\right)\, \,,}
	we see that the expression \eqref{zeromode}  is suppressed by a factor proportional to $e^{-4\pi |k_1|\tau_2}$, which is characteristic of an instanton/anti-instanton pair.
	
	The complete zero mode is given by $ \cF^P_0 ( \tau_2)=  f^P_{0,0}( \tau_2)+  \sum_{k_1 \neq 0}   f^P_{k_1,-k_1}( \tau_2)$.  
	In order to check the small-$\tau_2$ boundary condition we note that in the small-$\tau_2$ limit
	\ie \label  {eq:kk-smally}
	\sum_{k_1 \neq 0}   f^P_{k_1,-k_1}( \tau_2) = \sum_{k_1 \neq 0} -\frac{8 \sigma _2(|k_1|){}^2}{21 \pi ^2 k_1^6 }  \tau_2^{-3} +O(\tau_2^{-2}) = -\frac{4 \zeta(6)} {27} \tau_2^{-3} +O(\tau_2^{-2})\, ,
	\fe
	where we have used the Ramanujan identity 
	\ie
	\sum_{n=1}^\infty   \frac{\sigma_p(n)\sigma_{p'}(n)}{n^r} =  \frac{\zeta (r) \,\zeta (r-p) \, \zeta (r-p' ) \, \zeta (-p-p' +r)}   {\zeta (-p-p' +2 r)} \, .
	\fe
	Using  \eqref {eq:perty}, \eqref{betares} and  \eqref{eq:kk-smally}  we see that $ \cF_0 ( \tau_2)= \cF^P_0 ( \tau_2)+ \cF^H_0 ( \tau_2)  \underset{\tau_2\to 0}{=} O(\tau_2^{-2})$, which is the required boundary condition. 
	
	The solution of the homogeneous equation for $k\ne 0$  has the form $\cF^H_{k}(\tau_2) = \alpha_k  \sqrt{\tau_2} K_{\sevenh}(2\pi |k| \tau_2)$.  This  depends only on the sum of the  source mode numbers, $k_1+k_2$, and $\alpha_k$  is determined by imposing the boundary condition at $\tau_2=0$. It turns out that $\alpha_k=0$ for $k\ne 0$ (although this is a property of the solution that was not noticed in  \cite{Green:2014yxa}).  Since $\cF^H_{k}(\tau_2)=0$ for $k\ne 0$ the  solution for mode $k$ is identified with the particular solution,  $\cF_{k}(\tau_2)=  \cF^P_{k}(\tau_2)$, so we  can drop the superscript $P$ in the following.

	\vskip 0.2cm
	{\bf (b) $k_1 = k\ne 0$, $k_2=0$ and $k_2 = k\ne 0$, $k_1=0$}
	
	These modes are solutions of \eqref{partint} given by
	\begin{multline}\label{exactfn0P}
	\widehat{f}_{k,0}(\tau_2)   =    \widehat{f}_{0,k}(\tau_2)   = 
	\frac{8 \,\sigma_2(|k|) }{9\,\pi\,|k|^3   }\times
	\Big(q_{k,0}^0(\pi |k| \tau_2)
	K_0(2 \pi |k| \tau_2)+ q_{k,0}^1(\pi |k| \tau_2)
	K_1(2 \pi|k|  \tau_2)\Big),
	\end{multline}
	where the coefficients are given by
	\begin{eqnarray}\label{coeffOP}
	q_{k,0}^0(z)={1\over z}\,
	\left(90\zeta(3)-k^2\pi^4+9z^2\zeta(3)\right)\,,  \qquad
	q_{k,0}^1(z)={1\over z^2}\,\left(
	90\zeta(3)-k^2\pi^4+54z^2\zeta(3)\right)\,.\nonumber\\
	\end{eqnarray}

	Using the weak coupling expression \eqref {asymk}
	we see that $f_{k,0}(\tau_2)$ has the form of a charge-$k$ D-instanton contribution with a characteristic $e^{-2\pi |k| \tau_2}$ suppression factor together with an unlimited number of perturbative corrections  (powers of $ \tau_2^{-1}$).  In order to find the complete expression for the $k$th mode we need to sum an infinite number of terms of the form  $\cF _k (\tau_2)= \sum_{k_1} f_{k-k_1, k_1}(\tau_2)$.

	\vskip 0.2cm
	{\bf (c) $k_1 \ne0$,  $k_2\ne 0$ with $k_1+k_2\ne 0$}
	
	In these cases the solution of \eqref{partint}
	was found in  \cite{Green:2014yxa}   to have the form
	\es{pricsol}{
		\widehat{f}_{k_1,k_2}(\tau_2)  = 
		\frac {32\,\pi\,\sigma_2(|k_1|)\,\sigma_2(|k_2|)}{3\,|k_1k_2|\,|k_1+k_2|^5} \sum_{ i,j=0, 1}
		q^{ i,j}_{k_1,k_2}(\pi |k_1+k_2|\tau_2) \,  K_i(2 \pi |k_1| \tau_2)  \, K_j(2 \pi |k_2| \tau_2)\,, }
	where $q^{i,j}_{k_1,k_2}(z)$ are specific  polynomials with powers of $z$ ranging from $z$ to $z^{-2}$ which we will not display here.  There are two distinct cases to consider:
	
	{\bf (i) $k_1 \, k_2 >0$} 
	
	The solution contains D-instantons of charge $k=k_1+k_2>0$, or anti D-instantons when $k<0$.  These are characterized by an exponentially suppressed behavior of the form $e^{-2\pi |k| \tau_2}$ in the $\tau_2\to \infty$ limit.
	
	{\bf (ii) $k_1 \, k_2 < 0$}
	
	This is again a contribution that has total D-instanton charge equal  to $k$, but the solution describes a D-instanton of charge $k_1$ together with an anti D-instanton of charge $k-k_1$ (when we assume that $k_1>0$).  The large-$\tau_2$ behavior is characterized by an exponential suppression factor of $e^{-2\pi( |k_1 + |k_2| )\tau_2}$.  Since $|k_1|+ |k_2|> |k|$ these terms do not contribute to the leading exponential behavior in the weak-coupling limit.
	
	Given the complete solution it is straightforward to compare  with our analysis of the $1/N$ terms in the large-$N$ expansion of the localized integrated correlator that are determined in  Section~\ref{sec:instantons} and Appendix~\ref{app:top}.    For example, among these terms there are certain perturbative contributions to the leading exponential dependence in the $k=1$  D-instanton term, which is the sum of the $(1,0)$ and $(0,1)$ components (the  $k=1$ components $(2,-1)$, $(3,-2)$, $\dots$
	are exponentially suppressed relative to the leading term).
	In particular, the first few terms in the perturbative expansion around the $k=1$ D-instanton contribution matches terms in the expansion of 
	$\cF_1(\tau_2) = f_{1,0} (\tau_2) + f_{0, 1} (\tau_2)$, which is given by 
	\ie
	e^{2\pi \tau_2} \cF_{1} (\tau_2) &=8 \zeta (3) \sqrt{\tau_2} +\frac{95 \zeta (3)}{2 \pi \sqrt{\tau_2} } + \left(\frac{5705 \zeta (3)}{64 \pi
		^2}-\frac{8 \pi ^2}{9}\right)  \frac{1}{\tau^{3/2} _2}  +
	\left(\frac{75285 \zeta (3)}{1024 \pi ^3}-\frac{5 \pi
	}{6}\right)  \frac{1}{\tau^{5/2}_2}  \cr
	&+\frac{35  \left( 95931
		\zeta (3)-1024 \pi ^4\right)}{196608 {\pi ^4}} \frac{1}{\tau_2^{7/2} } + \frac{35 \left(1024 \pi ^4-106821 \zeta
		(3)\right)}{1048576 \pi ^5} \frac{1}{\tau_2^{9/2} } + \cdots \,,
	\fe
	where $\tau_2^{-1} = g_{_{\rm YM}}^2/4\pi$. Similarly, 
	the leading exponential contribution to the $k=2$ mode gets contributions from the sum of  the $(1,1)$, $(2,0)$, and $(0,2)$ components.   The perturbative expansion around the leading exponential dependence of the  $k=2$ contribution should therefore match the expansion of  $\cF_2(\tau_2) =  f_{2,0}(\tau_2) + f_{0,2}(\tau_2) + f_{1,1}(\tau_2)$, which has the form 
	\ie
	\label{modes3}
	e^{4\pi \tau_2}   \cF_{2}(\tau_2)  &=
	\frac{10\zeta (3)  \sqrt{\tau_2} }{\sqrt{2}} +\frac{16 \pi ^2}{3}  +\frac{475 \zeta (3)}{16 \sqrt{2} \pi \sqrt{\tau_2} } +\frac{6 \pi
	}{\tau_2}    +  \left(\frac{28525 \zeta (3)}{1024 \sqrt{2} \pi ^2}-\frac{10 \pi ^2}{9
		\sqrt{2}}\right) \frac{1}{\tau_2^{3/2}  } \cr
	&+\frac{27}{8 \tau_2^2} +  \left(\frac{376425 \zeta (3)}{32768 \sqrt{2} \pi
		^3}-\frac{25 \pi }{48 \sqrt{2}}\right)  \frac{1}{\tau_2^{5/2} }  +\frac{55}{64 \pi  \tau_2^3}+ \cdots \, .  
	\fe
	An intriguing aspect of this expansion is that it contains a sum of  odd powers of $\sqrt{1/\tau_2}\sim g_{_{\rm YM}}$ that come from the expansion of $ f_{2,0}(\tau_2) + f_{0,2}(\tau_2) $ and even powers of $1/\sqrt{\tau_2}$ that come from the expansion of $ f_{1,1}(\tau_2)$.  
	
	\vskip 0.2cm
	\noindent{\bf A note on the r\^ole of the $\tau_2=0$ boundary condition.}\hfill\break\noindent 
	An expression such as \eqref{modes3} is uniquely determined by a large-$\tau_2$ expansions of the exact solution.  The uniqueness is associated with the fact that the solution is valid for all $\tau_2$ and builds in the $\tau_2=0$ boundary condition.    
	It is important to stress that simply solving  \eqref{partint} for the $(k_1,k_2)$ component  (with $k_1+k_2=k$) by means of  a perturbation expansion around the large-$\tau_2$ limit  does not determine the expansion coefficients uniquely since such an expansion does not address the boundary condition at $\tau_2=0$.  
	This is reflected  in the arbitrary coefficients of the solutions of the homogeneous equations for the $(k_1,k-k_1)$ sectors, $f^H_{k_1,k_2}(\tau_2)$. 
	It is enlightening to illustrate this by considering the perturbative solution around the $k=2$ D-instanton  that is defined by the solution of \eqref{partint} with $k_1+k_2=2$.   We have seen that  $e^{4\pi \tau_2}(f_{2,0}( \tau_2)+ f_{0,2}( \tau_2))$  has an expansion  in half-integer powers of $\tau_2^{-1}$.  However, the perturbative expansion of  $e^{4\pi \tau_2}\,f^H_{k_1,k_2}(\tau_2)$ (where $f^H_{k_1,k_2}(\tau_2)$ satisfies \eqref{partint} with the source term set to zero),  is  in integer powers of $\tau_2^{-1}$.  Therefore, there is no ambiguity in the perturbative solution in the $(2,0)+(0,2)$ sector.  However,  the perturbative expansion of $e^{4\pi \tau_2}\, f_{1,1}( \tau_2)$ (the $(1,1)$ sector) is in integer powers of $\tau_2^{-1}$, and so $f_{1,1}( \tau_2)$    mixes with the expansion of the solution of the homogeneous equation, $f^H_{1,1}(\tau_2)$.  More explicitly, we can extract the $k=2$ instanton power behavior by setting $k_1=k_2=1$ in  \eqref{partint} and writing 
	\ie
	\label{11sec}
	f_{1,1}( \tau_2) = e^{-4\pi \tau_2}\, C(\tau_2)\,, 
	\fe
	The function $C(\tau_2)$  satisfies a differential equation that is easily solved to any given order in  a perturbation expansion in $\tau_2^{-1}$, but the solution has one arbitrary constant that cannot be determined without additional information.   Since   \eqref{modes3} is an expansion of  the exact solution, it builds in the $\tau_2=0$ boundary condition, so there is no ambiguity involved in comparing  \eqref{modes3} with the results of the localization calculation in the main text.

	Other terms  of order $1/N$ were also determined from the localized correlator  in section~\ref{sec:instantons}  and appendix~\ref{app:top}. In particular,  the analysis of in appendix~\ref{app:top} determines the exact expression for the  $(1,-1)$ component of the $k=0$ mode rather than simply  its perturbative expansion.  
	In addition  section~\ref{sec:instantons}  and appendix~\ref{app:top} contain an  analysis of the expansion of the components   $(2,-2)$, $(3,-3)$, $(1,-2)$, $(1,-3)$ and $(2,-3)$ of the localized correlator  to the first few orders in powers of $g_{_{\rm YM}}\sim 1/\sqrt{\tau_2}$.   We have verified that these expansions match the components $f_{k,-k} (\tau_2)$ with $k \leq 3$, $f_{1,-2}(\tau_2)$, $f_{1,-3}(\tau_2)$ and $f_{2,-3}(\tau_2)$ of the solution of   \eqref{partint}.  
	
	\subsection{ $\cE(4,\threeh,\fiveh,\tau,\bar\tau)$ and  $\cE(6,\threeh,\fiveh,\tau,\bar\tau)$}
	
	The modular functions $\cE(4,\threeh,\fiveh,\tau,\bar\tau)$ and  $\cE(6,\threeh,\fiveh,\tau,\bar\tau )$ satisfy the inhomogeneous Laplace equations, 
	\es{d10eq}{ \left(\Delta_\tau - r(r+1)\right) \cE(r,\threeh,\fiveh,\tau,\bar\tau)=-E(\threeh,\tau,\bar\tau)E( \fiveh,\tau,\bar\tau)\,, \qquad\qquad r=4,6  \,. } 
	From here on we will write the Fourier expansion of  $\cE(r, s_1,s_2,\tau,\bar\tau)$ using the notation
	\es{fourier-d10}{ \cE(r,s_1,s_2,\tau,\bar\tau) = \sum_k \cF^{r,s_1,s_2}_k(\tau_2) e^{2\pi i k\tau_1}\  = \sum_{k_1} \sum_{k_2}
		f^{r,s_1,s_2}_{k_1, k_2}(\tau_2) e^{2\pi i (k_1+k_2)\tau_1}\,, }
	where we have introduced the  superscripts $r, s_1,s_2$ to indicate the eigenvalue and the source term.\footnote{In this notation the modes $f_k(\tau)$ in the previous subsection would be denoted $f_k^{3,\threeh,\threeh}(\tau_2)$.}
	
	We have not determined the complete solution in these cases, but we have determined many features that can be correlated with the results of the $1/N^2$ contribution to the large-$N$ expansion of the localized integrated correlator. Some of these properties of the solutions are summarised as follows.
	
	\vskip 0.2cm
	{\bf (i)  $k=0$, i.e. $k_1 =-k_2$, }
	
	We have determined the  complete zero-instanton sector, which again includes terms that are power-behaved in $\tau_2$, and $(k_1, -k_1)$ D-instanton/anti D-instanton terms.  
	
	The power-behaved  terms ($k_1=k_2=0$) with eigenvalues $r=4,6$  are given by
	\ie \label{eq:d10per}
	f^{4,\threeh,\fiveh}_{0,0}( \tau_2) &=\frac{\zeta (3) \zeta (5) \tau_2^4}{2}  +\frac{4 \zeta(2) \zeta(5) \tau_2^2}{9} +
	\frac{4 \zeta(4) \zeta (3)}{15}
	+  \frac{4 \zeta(6)}{3 \tau_2^2} +  {44 \zeta(8) \over 405 \tau_2^4 }  \, , \cr
	f^{6,\threeh,\fiveh}_{0,0}( \tau_2)  &= \frac{2  \zeta (3) \zeta (5) \tau_2^4}{15}+\frac{ \zeta(2) \zeta (5) \tau_2^2 }{5} +\frac{8\zeta(4) \zeta (3)}{63 }  + \frac{14 \zeta(6)}{27 \tau_2^2}  + \frac{88 \zeta (10)}{23625 \tau_2^6}   \,.
	\fe
	The $(k_1, -k_1)$ terms are zero-mode D-instanton/anti D-instanton contributions, that have the form 
	\ie
	f^{\,r,\threeh,\fiveh}_{k_1,-k_1}( \tau_2)  = {32 \pi^2 \over 315 k_1^4} \sigma_2(k_1) \sigma_4(k_1) \sum^3_{i,j=2} q_r^{i,j}(\pi k_1 \tau_2)K_i(2\pi k_1 \tau_2) K_j(2\pi k_1 \tau_2)\, ,
	\fe
	where the coefficients $q_r^{i,j}$ are symmetric, $q_r^{i,j}= q_r^{j,i}$.  We find that the expressions for $q_4^{i,j}(z)$ for $r=4$ are given by 
	\ie
	q_4^{2,2}(z) & = \frac{16384 z^6}{135}-\frac{3584 z^4}{27}+168 z^2-105 \,, \cr
	q_4^{2,3}(z) & =  \frac{4096 z^5}{27}-\frac{896 z^3}{9}+42 z  \,, \cr
	q_4^{3,3}(z) & =  -\frac{16384 z^6}{135}+\frac{512 z^4}{9}-\frac{56 z^2}{3} \,,
	\fe
	whereas for the case $r=6$, they are given by
	\ie
	q_6^{2,2}(z) & = \frac{524288 z^8}{12285}-\frac{16384 z^6}{351}+\frac{768 z^4}{13}-\frac{440 z^2}{13}-28 \,, \cr
	q_6^{2,3}(z) & =  \frac{131072 z^7}{2457}-\frac{4096 z^5}{117}+\frac{192 z^3}{13}+\frac{110 z}{13} \,, \cr
	q_6^{3,3}(z) & =  -\frac{524288 z^8}{12285}+\frac{16384 z^6}{819}-\frac{256 z^4}{39}-\frac{40
		z^2}{13}-\frac{21}{13} \,.
	\fe
	Once again it is important that the $\tau_2\to 0$ boundary condition is satisfied.  
	
	\vskip 0.2cm
	{\bf  (ii) $k\ne 0$} 
	
	We have determined perturbative expansions around the leading exponential behavior  in various instanton sectors that are needed in order to compare with the results obtained from the $1/N^2$ contribution to the  localized integrated correlator discussed in Section~\ref{sec:instantons}.    We will discuss the explicit perturbative expansions around the leading exponential behavior of the charge $k=1$ and $k=2$ D-instantons  to the first few orders  in powers of $1/\tau_2$.
	For the one-instanton contributions,  i.e. the $(1,0)+(0,1)$ components,   the expansions of  $\cF^{\,r,\threeh,\fiveh}_{1}( \tau_2)  = f^{\,r,\threeh,\fiveh}_{1,0}( \tau_2)  + f^{\,r,\threeh,\fiveh}_{0,1}( \tau_2)$  with $r=4,6$ have the form
	\ie
	\label{r4k1}
	e^{2\pi \tau_2}    \cF^{\,4,\threeh,\fiveh}_{1}( \tau_2)   & = \frac{4}{3}  \zeta (5) \tau_2^{3/2}+ 
	\left(\frac{8 \pi  \zeta (3)}{3}-\frac{145 \zeta (5)}{12 \pi }\right) \sqrt{\tau_2} +
	\left(\frac{49 \zeta (3)}{2}-\frac{15645 \zeta (5)}{128 \pi ^2}\right) \frac{1}{\sqrt{\tau_2}} \cr
	& +\frac{\left(4336416 \pi ^2
		\zeta (3)-21684915 \zeta (5)-16384 \pi ^6\right)}{55296 \pi ^3} \frac{1}{\tau^{3/2} _2} + \cdots \, , \cr
	e^{2\pi \tau_2}    \cF^{\,6,\threeh,\fiveh}_{1}( \tau_2)  & = \frac{4}{3}  \zeta (5) \tau_2^{3/2} +
	\left(\frac{8 \pi  \zeta (3)}{3}-\frac{107 \zeta (5)}{4 \pi }\right)   \sqrt{\tau_2} +
	\left(\frac{323 \zeta (3)}{6}-\frac{72317  \zeta (5)}{128 \pi ^2}\right) \frac{1}{\sqrt{\tau_2}} \cr
	& +\frac{ \left( 20455200 \pi ^2 \zeta (3)-214782435 \zeta (5)-16384 \pi ^6 \right)}{55296 \pi ^3} \frac{1}{\tau_2^{3/2} } + \cdots \,.
	\fe
	For the case of $k=2$, the leading exponential contributions to  $ \cF^{r ,\threeh,\fiveh}_{2}( \tau_2)  = f^{\,r,\threeh,\fiveh}_{2,0}( \tau_2)  + f^{\,r,\threeh,\fiveh}_{0,2}( \tau_2)  + f^{\,r,\threeh,\fiveh}_{1,1}( \tau_2)$  have expansions of the form
	\ie
	\label{r4k2}
	& e^{4\pi \tau_2}    \cF^{\,4,\threeh,\fiveh}_{2}( \tau_2)  
	= \frac{5 \zeta (5) \tau_2^{3/2}  }{3 \sqrt{2}} +\frac{\sqrt{\tau_2} \left(544 \pi ^2 \zeta (3)-725 \zeta
		(5)\right)}{96 \sqrt{2} \pi }  +\frac{32 \pi ^3}{15} + \frac{7 \left(7616 \pi ^2 \zeta (3)-11175 \zeta
		(5)\right)}{2048 \sqrt{2} \pi ^2 \sqrt{\tau_2} }  \cr
	& \,\,\, + \frac{4 \pi^2}{\tau_2}  -\frac{
		\left(-73719072 \pi ^2 \zeta (3)+108424575 \zeta (5)+1114112 \pi ^6\right)}{1769472
		\sqrt{2} \pi ^3 \tau_2^{3/2}}  +\frac{15 \pi }{4 \tau_2^2} + \cdots \, , \cr
	&   e^{4\pi \tau_2}   \cF^{\,6,\threeh,\fiveh}_{2}( \tau_2)  
	= \frac{5 \zeta (5) \tau_2^{3/2}  }{3 \sqrt{2}} +\frac{\sqrt{\tau_2} \left(544 \pi ^2 \zeta (3)-1605 \zeta
		(5)\right)}{96 \sqrt{2} \pi } -\frac{32 \pi ^3}{315} + \frac{\left(351424 \pi ^2 \zeta (3)-1084755 \zeta
		(5)\right)}{6144 \sqrt{2} \pi ^2 \sqrt{\tau_2} }  \cr
	& \,\,\, - \frac{28 \pi^2}{15\tau_2}  -\frac{ \left(544 \pi
		^2 \left(2048 \pi ^4-639225 \zeta (3)\right)+1073912175 \zeta (5)\right) }{1769472
		\sqrt{2} \pi ^3 \tau_2^{3/2}}  - \frac{65 \pi }{12 \tau_2^2} + \cdots \, ,
	\fe
	which have been obtained by expanding the exact solutions for these modes, which incorporates the $\tau_2=0$ boundary condition (although  in this case we have not displayed these solutions explicitly).   As we saw in the case of the perturbative expansion of the $k=2$ contribution to the  $1/N$ coefficient in \eqref{modes3}  this equation has half-integer powers of $\tau_2^{-1}$ that arise from the expansion of the $(2,0)+(0,2)$ sector and  integer powers from the $(1,1)$  sector.

	We finally note that we have also determined perturbative expansions of other contributions  such as $(1, -2), (1,-3), (2,-3)$. The computations of the perturbative expansions of these sectors are relatively straightforward.  They can be obtained by simply equating both sides of the differential equations without any subtlety.  We have checked that the results are all in agreement with localization results.
	
	\subsection{$\cE(r,\threeh,\threeh,\tau,\bar\tau)$,  $\cE(r,\fiveh,\fiveh,\tau,\bar\tau)$,  $\cE(r,\threeh,\sevenh,\tau,\bar\tau)$ with $r=5,7,9$}
	
	These functions enter into the description of the $1/N^3$ term in the large-$N$ expansion of the localized $\cN =4$ SYM correlation function.
	Here we will again list the coefficients of the zero Fourier mode  ($k=0$), that include terms that are power behaved in $\tau_2$ as well as the sequence of $(k_1, -k_1)$ (D-instanton/anti D-instanton) contributions. We have also evaluated many terms in the perturbative expansion  in powers of $\sqrt{1/\tau_2}$ around D-instanton contributions with $k\ne 0$, that include the instanton sectors of $(k_1, k_2)=(2,0), (1,1), (1,-2), (1,-3), (2,-3)$.  We will not present them explicitly here. These $1/N^3$ contributions take similar forms to the $1/N^2$ contributions in  \eqref{r4k1} and \eqref{r4k2}. We find that all of coefficients match perfectly with the localization computation.

	\subsubsection{$\cE(r,\threeh,\threeh,\tau,\bar\tau)$}
	
	We described this function with $r=3$ in detail in Section~\ref{d6r4fun}  in order to compare with the coefficient of the $1/N$ term in the large-$N$ expansion of the localized correlator. In the $r=5,7,9$  cases the perturbative terms (the terms power behaved in $\tau_2$) are given by 
	\ie
	f_{0,0}^{5,\threeh,\threeh}(\tau_2)  &= \frac{\tau_2^3 \zeta (3)^2}{6}+\frac{8 \zeta(2) \zeta (3) \tau_2 }{15} +\frac{10 \zeta(4)}{7 \tau_2}  + \frac{104 \zeta(8) }{31095 \tau_2^5} \, , \cr
	f_{0,0}^{7,\threeh,\threeh}(\tau_2) &=\frac{2 \tau_2^3 \zeta (3)^2}{25}+\frac{2 \zeta(2)  \zeta (3) \tau_2}{7} +\frac{20 \zeta(4) }{27 \tau_2} +  \frac{77792 \zeta(10)}{199387125 \tau_2^7} \, ,  \cr
	f_{0,0}^{9,\threeh,\threeh}(\tau_2) &= \frac{\tau_2^3 \zeta (3)^2}{21}+\frac{8 \zeta(2) \zeta(3) \tau_2}{45}  +\frac{5 \zeta(4)}{11 \tau_2} + \frac{70720  \zeta(12) }{693031059 \tau_2^9}  \, .
	\fe
	The non-perturbative contribution with $k=0$ comes from the sum of the  $(k_1, -k_1)$  D-instanton/anti D-instanton  contributions of the form
	\ie
	f_{k_1,-k_1}^{r,\threeh,\threeh}(\tau_2) = {32 \pi^2 \over 315 k_1^3} \sigma_2(k_1) \sigma_2(k_1) \sum^3_{i,j=2} q_r^{i,j}(\pi k_1 \tau_2)K_i(2\pi k_1 \tau_2) K_j(2\pi k_1 \tau_2)\,.
	\fe
	The coefficients are given by the following polynomials:
	
	\begin{itemize}
		\item $r=5$
		\ie
		q_5^{2,2}(z) &=-\frac{32768 z^7}{165 \pi }+\frac{7168 z^5}{33 \pi }-\frac{3024 z^3}{11 \pi }+\frac{147
			z}{\pi }+\frac{105}{\pi  z} \, , \cr
		q_5^{2,3}(z) &= -\frac{8192 z^6}{33 \pi }+\frac{1792 z^4}{11 \pi }-\frac{756 z^2}{11 \pi }-\frac{105}{2
			\pi } \,, \cr
		q_5^{3,3}(z) &= \frac{32768 z^7}{165 \pi }-\frac{1024 z^5}{11 \pi }+\frac{336 z^3}{11 \pi }+\frac{273
			z}{11 \pi } \,. 
		\fe
		\item   $r=7$
		\ie
		q_7^{2,2}(z) &=-\frac{1048576 z^9}{23625 \pi }+\frac{32768 z^7}{675 \pi }-\frac{1536 z^5}{25 \pi
		}+\frac{176 z^3}{5 \pi }+\frac{97 z}{5 \pi }+\frac{252}{5 \pi  z}  \, , \cr
		q_7^{2,3}(z) &= -\frac{262144 z^8}{4725 \pi }+\frac{8192 z^6}{225 \pi }-\frac{384 z^4}{25 \pi }-\frac{44
			z^2}{5 \pi }-\frac{129}{5 \pi } \,, \cr
		q_7^{3,3}(z) &= \frac{1048576 z^9}{23625 \pi }-\frac{32768 z^7}{1575 \pi }+\frac{512 z^5}{75 \pi
		}+\frac{16 z^3}{5 \pi }+\frac{57 z}{5 \pi }-\frac{9}{5 \pi  z} \,. 
		\fe
		\item  $r=9$
		\ie
		q_9^{2,2}(z) &= -\frac{134217728 z^{11}}{20738025 \pi }+\frac{4194304 z^9}{592515 \pi }-\frac{65536
			z^7}{7315 \pi }+\frac{2048 z^5}{399 \pi }+\frac{208 z^3}{57 \pi }+\frac{109 z}{19 \pi
		}+\frac{560}{19 \pi  z} \, , \cr
		q_9^{2,3}(z) &= -\frac{33554432 z^{10}}{4147605 \pi }+\frac{1048576 z^8}{197505 \pi }-\frac{16384
			z^6}{7315 \pi }-\frac{512 z^4}{399 \pi }-\frac{52 z^2}{19 \pi }-\frac{30}{19 \pi 
			z^2}-\frac{605}{38 \pi } \,, \cr
		q_9^{3,3}(z) &= \frac{134217728 z^{11}}{20738025 \pi }-\frac{4194304 z^9}{1382535 \pi }+\frac{65536
			z^7}{65835 \pi }+\frac{2048 z^5}{4389 \pi }+\frac{16 z^3}{19 \pi }-\frac{90}{19 \pi 
			z^3}+\frac{125 z}{19 \pi }-\frac{60}{19 \pi  z} \,. 
		\fe
	\end{itemize}
	
	\subsubsection{$\cE(r,\fiveh,\fiveh,\tau,\bar\tau)$}
	
	The $k=0$ terms with $r=5,7,9$ that are power-behaved in $\tau_2$ are given by 
	\ie
	f_{0,0}^{5,\fiveh,\fiveh}(\tau_2)  &= \frac{2 \tau_2^5 \zeta (5)^2}{5} +\frac{16 \zeta(4) \zeta(5) \tau_2}{45}+\frac{112 \zeta(8) }{243 \tau_2^3} +  \frac{1664 \zeta(10)}{18657 \tau_2^5}  \, , \cr
	f_{0,0}^{7,\fiveh,\fiveh}(\tau_2)  &= \frac{\tau_2^5 \zeta (5)^2}{9}+\frac{4 \zeta(4) \zeta(5) \tau_2}{21} +\frac{56 \zeta(8) }{297 \tau_2^3} + \frac{77792 \zeta(12)}{17090325 \tau_2^7}\, , \cr
	f_{0,0}^{9,\fiveh,\fiveh}(\tau_2)  &= \frac{2 \tau_2^5 \zeta (5)^2}{35}+\frac{16 \zeta(4) \zeta(5) \tau_2}{135} +\frac{112 \zeta(8) }{1053 \tau_2^3}+ \frac{5657600 \zeta(14)}{6237279531 \tau_2^9}  \, .
	\fe
	The  non-perturbative $k=0$ terms are  given by the sum of $(k_1, -k_1)$ D-instanton/anti D-instanton contributions  that takes the following form 
	\ie
	f_{k_1,-k_1}^{r,\fiveh,\fiveh}(\tau_2)  =   {32 \pi^2 \over 315 k_1^5} \sigma_4(k_1) \sigma_4(k_1) \sum^3_{i,j=2} q_r^{i,j}(\pi k_1 \tau_2)K_i(2\pi k_1 \tau_2) K_j(2\pi k_1 \tau_2)\, . 
	\fe
	The coefficients  in this equation are given by, the following polynomials.
	\begin{itemize}
		\item $r=5$
		\ie
		q_5^{2,2}(z) &=-\frac{131072 \pi  z^7}{4455}+\frac{28672 \pi  z^5}{891}-\frac{448 \pi  z^3}{11}+28 \pi 
		z \, , \cr
		q_5^{2,3}(z) &=-\frac{32768 \pi  z^6}{891}+\frac{7168 \pi  z^4}{297}-\frac{112 \pi  z^2}{11}  \,, \cr
		q_5^{3,3}(z) &= \frac{131072 \pi  z^7}{4455}-\frac{4096 \pi  z^5}{297}+\frac{448 \pi  z^3}{99}-\frac{28
			\pi  z}{11} \,. 
		\fe
		\item $r=7$
		\ie
		q_7^{2,2}(z) &=-\frac{1048576 z^9}{23625 \pi }+\frac{32768 z^7}{675 \pi }-\frac{1536 z^5}{25 \pi
		}+\frac{176 z^3}{5 \pi }+\frac{97 z}{5 \pi }+\frac{252}{5 \pi  z}  \, , \cr
		q_7^{2,3}(z) &= -\frac{262144 z^8}{4725 \pi }+\frac{8192 z^6}{225 \pi }-\frac{384 z^4}{25 \pi }-\frac{44
			z^2}{5 \pi }-\frac{129}{5 \pi } \,, \cr
		q_7^{3,3}(z) &= \frac{1048576 z^9}{23625 \pi }-\frac{32768 z^7}{1575 \pi }+\frac{512 z^5}{75 \pi
		}+\frac{16 z^3}{5 \pi }+\frac{57 z}{5 \pi }-\frac{9}{5 \pi  z} \,. 
		\fe
		\item $r=9$
		\ie
		q_9^{2,2}(z) &= \pi( -\frac{536870912   z^{11}}{346621275}+\frac{16777216  z^9}{9903465}-\frac{262144
			z^7}{122265}+\frac{8192 z^5}{6669}+\frac{448 z^3}{513}+\frac{644 
			z}{171}-\frac{80 }{171 z}  )\, , \cr
		q_9^{2,3}(z) &=\pi  (-\frac{134217728  z^{10}}{69324255}+\frac{4194304  z^8}{3301155}-\frac{65536 
			z^6}{122265}-\frac{2048  z^4}{6669}-\frac{112  z^2}{171}-\frac{80 }{57
			z^2}-\frac{140 }{171} ) \,, \cr
		q_9^{3,3}(z) &=\pi  ( \frac{536870912   z^{11}}{346621275}-\frac{16777216  z^9}{23108085}+\frac{262144
			z^7}{1100385}+\frac{8192   z^5}{73359}+\frac{448   z^3}{2223}-\frac{80 
		}{19 z^3}-\frac{140  z}{171}-\frac{160 }{57 z})\\  
		\fe
	\end{itemize}
	
	\subsubsection{$\cE(r,\threeh,\sevenh,\tau,\bar\tau)$}
	
	The $k=0$ terms with $r=5,7,9$ that are power-behaved in $\tau_2$ are given by 
	\ie
	f_{0,0}^{5,\threeh,\sevenh}(\tau_2)  &= \frac{2 \zeta (3) \zeta (7) \tau_2^5 }{5}  +\frac{ \zeta(2) \zeta (7) \tau_2^3 }{3} 
	+\frac{16 \zeta(6) \zeta (3)}{105 \tau_2}+\frac{64 \zeta(8)}{81 \tau_2^3} + \frac{1456 \zeta(10)}{17275 \tau_2^5} \, ,\cr
	f_{0,0}^{7,\threeh,\sevenh}(\tau_2)   &= \frac{ \zeta (3) \zeta (7) \tau_2^5}{9} +\frac{4 \zeta(2)  \zeta (7) \tau_2^3}{25}  +\frac{32 \zeta(6) \zeta (3)}{405 \tau_2}  +\frac{32 \zeta(8) }{99 \tau_2^3}+ \frac{113152 \zeta(12)}{30762585 \tau_2^7}  \, , \cr
	f_{0,0}^{9,\threeh,\sevenh}(\tau_2)   &= \frac{2  \zeta (3) \zeta (7) \tau_2^5}{35} +\frac{2 \zeta(2) \zeta(7)  \tau_2^3 }{21}  +\frac{8 \zeta(6) \zeta (3)}{165 \tau_2} +\frac{64 \zeta(8) }{351 \tau_2^3} + \frac{16539776 \zeta(14)}{24256087065 \tau_2^9} \, .
	\fe
	The non-perturbative contribution with $k=0$ comes from the sum of the  $(k_1, -k_1)$  D-instanton/anti D-instanton  contributions of the form
	\ie
	f_{k_1,-k_1}^{r,\threeh,\sevenh}(\tau_2) =   {32 \pi^2 \over 315 k_1^5} \sigma_2(k_1) \sigma_6(k_1) \sum^3_{i,j=2} q^{i,j}(\pi k_1 \tau_2)K_i(2\pi k_1 \tau_2) K_j(2\pi k_1 \tau_2) \, . 
	\fe
	The coefficients are given by the following polynomials.
	\begin{itemize}
		\item $r=5$
		\ie
		q_5^{2,2}(z) &=-\frac{131072 \pi  z^7}{10395}+\frac{4096 \pi  z^5}{297}-\frac{192 \pi  z^3}{11}+\frac{36
			\pi  z}{5}  \, , \cr
		q_5^{2,3}(z) &=-\frac{32768 \pi  z^6}{2079}+\frac{1024 \pi  z^4}{99}-\frac{48 \pi  z^2}{11}-6 \pi  \,, \cr
		q_5^{3,3}(z) &= \frac{131072 \pi  z^7}{10395}-\frac{4096 \pi  z^5}{693}+\frac{64 \pi  z^3}{33}+\frac{204
			\pi  z}{55} \,. 
		\fe
		\item  
		$r=7$
		\ie
		q_7^{2,2}(z) &= -\frac{4194304 \pi  z^9}{1002375}+\frac{917504 \pi  z^7}{200475}-\frac{14336 \pi 
			z^5}{2475}+\frac{448 \pi  z^3}{135}+\frac{116 \pi  z}{135} \, , \cr
		q_7^{2,3}(z) &= -\frac{1048576 \pi  z^8}{200475}+\frac{229376 \pi  z^6}{66825}-\frac{3584 \pi 
			z^4}{2475}-\frac{112 \pi  z^2}{135}-\frac{164 \pi }{45} \,, \cr
		q_7^{3,3}(z) &=\frac{4194304 \pi  z^9}{1002375}-\frac{131072 \pi  z^7}{66825}+\frac{14336 \pi 
			z^5}{22275}+\frac{448 \pi  z^3}{1485}+\frac{92 \pi  z}{45}-\frac{8 \pi }{5 z} \,. 
		\fe
		\item
		$r=9$
		\ie
		q_9^{2,2}(z) &=\pi \,  \Big( -\frac{536870912   z^{11}}{706080375}+\frac{16777216   z^9}{20173725}-\frac{786432
			z^7}{747175}+\frac{8192  z^5}{13585}+\frac{448  z^3}{1045}+\frac{476 
			z}{3135}-\frac{20 }{57 z} \Big)\, , \cr
		q_9^{2,3}(z) &=\pi \,  \Big( -\frac{134217728  z^{10}}{141216075}+\frac{4194304   z^8}{6724575}-\frac{196608
			z^6}{747175}-\frac{2048  z^4}{13585}-\frac{336  z^2}{1045}-\frac{20 
		}{19 z^2}-\frac{1582 }{627}\Big)\,  \cr
		q_9^{3,3}(z) &= \pi \,  \Big( \frac{536870912  z^{11}}{706080375}-\frac{16777216 z^9}{47072025}+\frac{262144
			\pi  z^7}{2241525}+\frac{8192  z^5}{149435}+\frac{1344   z^3}{13585}-\frac{60
		}{19 z^3}  +\frac{812 \pi  z}{627}-\frac{40  }{19 z}\Big)\,  
		\fe
	\end{itemize}

	%%%%%%%%%
	%%%%%%%%%
	\section{Topological recursion}
	\label{app:top}
	
	In this appendix we will show the details of the localization calculations whose results were discussed in the main text. All of these calculations involve computing expectation values with respect to the $m=0,b=1$ free gaussian matrix model in \eqref{Zfree}. In fact, as explained in \cite{Chester:2019pvm}, if an expectation value only depends on the difference of eigenvalues $a_{ij}$, as all the ones we consider do, then we can equivalently take the expectation value with respect to the $U(N)$ $\cN=4$ SYM matrix model
	\es{Zfree2}{
		Z\big \vert_{m=0,b=1}
		=  \int d^{N} a\, e^{-\frac{8 \pi^2  }{ g_{_{\rm YM}} ^2} \sum_i a_i^2}  \prod_{i < j} a_{ij}^2\,,
	}
	where we now integrate over $N$ eigenvalues with no constraint, unlike the $SU(N)$ matrix model in \eqref{Zfree}. In the following we will for simplicity take all expectation values with respect to \eqref{Zfree2}. We will then compute these expectation values using topological recursion, which we will briefly review following \cite{Chester:2019pvm}.
	
	Let us begin by defining the $n$-point operator 
	\es{RnDef}{
		R^n(y_1, \dots, y_n) \equiv  \sum_{i_1} \frac{1}{y_1 -a_{i_1} }\cdots \sum_{i_n} \frac{1}{y_n -a_{i_n} } \,.
	}
	The expectation value of this operator with respect to \eqref{Zfree2} can be shown to obey recursion relations in $n$ and $1/N$, which are called topological recursion. It is customary to write down these recursion relations in terms of the connected correlators 
	\es{W}{
		W^n(y_1,\dots, y_n)\equiv
		N^{n-2}\langle R^n(y_1, \dots, y_n)\rangle_\text{conn} =  N^{n-2}\left\langle\sum_{i_1} \frac{1}{y_1 -a_{i_1} }\cdots \sum_{i_n} \frac{1}{y_n -a_{i_n} }\right\rangle_\text{conn.}\,,
	}
	which in a slight abuse of notation we will refer to as resolvents. These resolvents can then be expanded in $1/N^2$ as 
	\es{W2}{
		W^n(y_1,\dots, y_n)\equiv\sum_{m=0}^\infty\frac{1}{N^{2m}} W^n_m(y_1,\dots, y_n)\,,
	}
	and each genus-$m$ term $W^n_m$ can be computed for finite $\lambda$ using a recursion formula in $n,m$ \cite{Eynard:2004mh,Eynard:2008we} starting with the base case $W^1_0$, as described e.g. in \cite{Chester:2019pvm}. We use resolvents up to $n+m\leq 5$, which were given in  \texttt{Mathematica} files attached to \cite{Chester:2019pvm,Chester:2020dja}, except one should set $\sqrt{y_i^2-\lambda/(4\pi^2)} \to y_i \sqrt{1-\lambda/(4\pi^2 y_i^2)}$ in all expressions given there, so that the resolvents have the correct properties as $y\to\infty$. In the following subsections, we will relate the expectation values we are interested in to these resolvents, which allows us to compute their $1/N$ expansion.
	
	\subsection{Details of perturbative calculation}
	\label{app:pert}
	
	The goal of this subsection is to compute \eqref{treeExpect2} starting from the expectation values in \eqref{pertmain}. We start by reviewing the calculation of \cite{Chester:2019pvm,Chester:2020dja}, where the former computed the two-body operator $\mathcal{I}(\omega)$, and the latter computed the four-body operator $\mathcal{J}(\omega,w)$.
	
	Define the inverse Laplace transform of a function $f$ by 
	\es{L}{
		& f(b_1,\dots, b_n)\equiv \frac{1} {(2\pi i)^n} \left[\prod_{i=1}^n\int_{\gamma_i-i\infty}^{\gamma_i+i\infty}dy_i e^{b_i y_i} \right]f(y_1,\dots,y_n)\,,
	}
	with $\gamma_{i}$ chosen so that the contour lies to the right of all singularities in the integrand. We then write the expectation values in \eqref{exp2} as
	\es{expToW}{
		\mathcal{I}(\omega)=&N^2 \widehat W^1(2i\omega)\; \widehat W^1(-2i\omega)+ \widehat W^2(2i\omega,-2i\omega)\,,\\
		\mathcal{J}(\omega,w)=&N^2\mathcal{J}^0(\omega,w)+\mathcal{J}^1(\omega,w)+N^{-2}\mathcal{J}^2(\omega,w)\,,\\
	}
	where we define
	\es{expToW3}{
		\mathcal{J}^0(\omega,w) &\equiv \widehat W^1(2i\omega)\;\widehat W^1(2iw)\;\widehat W^2(-2i\omega,-2iw)\\
		&{}+\widehat W^1(2i\omega)\;\widehat W^1(-2iw)\;\widehat W^2(-2i\omega,2iw)\\
		&{}+\widehat W^1(-2i\omega)\;\widehat W^1(2iw)\;\widehat W^2(2i\omega,-2iw)\\
		&{}+\widehat W^1(-2i\omega)\;\widehat W^1(-2iw)\;\widehat W^2(2i\omega,2iw)\,,\\
	}
	\es{expToW4}{
		\mathcal{J}^1(\omega,w) &\equiv \widehat W^2(2i\omega,2iw)\;\widehat W^2(-2i\omega,-2iw)\\
		&{}+\widehat W^2(2i\omega,-2iw)\;\widehat W^2(-2i\omega,2iw)\\
		&{}+\widehat W^1(2i\omega)\;\widehat W^3(-2i\omega,-2iw,2iw)\\
		&{}+\widehat W^1(-2i\omega)\;\widehat W^3(2i\omega,-2iw,2iw)\\
		&{}+\widehat W^1(2iw)\;\widehat W^3(-2i\omega,-2iw,2i\omega)\\
		&{}+\widehat W^1(-2iw)\;\widehat W^3(2i\omega,-2i\omega,2iw)\,,\\
	}
	\es{expToW5}{
		\mathcal{J}^2(\omega,w) &\equiv \widehat W^4(2i\omega,-2i\omega,2iw,-2iw)\,.\\
	}
	We then take the inverse Laplace transform in \eqref{expToW3} of the explicit resolvents to get the $1/N^2$ expansion at finite $\lambda$ for $\mathcal{I}(\omega)$ and $\mathcal{J}(\omega,w)$ in terms of integrals over the Fourier variables $w,\omega$ shown in \eqref{pertmain}. For instance, at leading order in $1/N^2$ we need only consider the genus-zero resolvents in $\mathcal{J}^0(\omega,w)$ and $\mathcal{I}^0(\omega)$, which give
	\es{tree221}{
		\mathcal{J}^0(\omega,w)\big\vert_{N^2}=&\frac{8 \pi  
			J_1(\frac{\sqrt{\lambda } \omega }{\pi })
			J_1(\frac{w \sqrt{\lambda }}{\pi })}{\sqrt{\lambda }
			(w^2-\omega^2 ) }\left[\textstyle\omega  J_0\left(\frac{\sqrt{\lambda } \omega }{\pi }\right)
		J_1\left(\frac{w \sqrt{\lambda }}{\pi }\right)-w
		J_1\left(\frac{\sqrt{\lambda } \omega }{\pi }\right)
		J_0\left(\frac{w \sqrt{\lambda }}{\pi }\right)\right]\,,\\
		\mathcal{I}^0(\omega)\big\vert_{N^2}=&\frac{4 \pi ^2  J_1(\frac{ \sqrt{\lambda }\omega}{\pi
			}){}^2}{\omega^2\lambda }\,.
	}
	We can then plug these expressions into \eqref{pertmain} to get the leading order in $N^2$ result at finite $\lambda$:
	\es{4mLead}{
		& \partial_m^4\log Z^\text{pert}\vert_{m=0,b=1}=N^2\Bigg[\int_0^\infty d\omega\frac{32\omega \pi ^2  J_1(\frac{ \sqrt{\lambda }\omega}{\pi
			}){}^2}{w^2\lambda\sinh^2\omega }\\
		&+\int_0^\infty d\omega \int_0^\infty dw\frac{96w\omega \pi  
			J_1(\frac{\sqrt{\lambda } \omega }{\pi })
			J_1(\frac{w \sqrt{\lambda }}{\pi })}{\sinh^2w\sinh^2\omega\sqrt{\lambda }
			(w^2-\omega^2 ) }\left[\textstyle\omega  J_0\left(\frac{\sqrt{\lambda } \omega }{\pi }\right)
		J_1\left(\frac{w \sqrt{\lambda }}{\pi }\right)-w
		J_1\left(\frac{\sqrt{\lambda } \omega }{\pi }\right)
		J_0\left(\frac{w \sqrt{\lambda }}{\pi }\right)\right]\Bigg] \\
		&{}+O(N^0)\,,
	}
	and the higher order in $1/N^2$ terms take a similar form of integrals of two Bessel functions for the 2-body terms, and four Bessel functions for the 4-body terms.  We need to take the large $\lambda$ expansion of these results, which will correspond to the large $N$ expansion after we set $\lambda=g_{_{\rm YM}} ^2N$. As described in Appendix D of \cite{Binder:2019jwn}, the first step is to express products of Bessel functions in their Mellin-Barnes form
	\es{mbbessel}{
		J_\mu(x)J_\nu(x) = \frac 1 {2\pi i}\int_{c-\infty i}^{c+\infty i}ds\frac{\Gamma(-s)\Gamma(2s+\mu+\nu+1)\left(\frac12 x\right)^{\mu+\nu+2s}}{\Gamma(s+\mu+1)\Gamma(s+\nu+1)\Gamma(s+\mu+\nu+1)} \,.
	}
	For the two-body terms, we can them perform the resulting integrals over $\omega$ in \eqref{pertmain} using the identity
	\es{id}{
		\int_0^\infty d\omega\ \frac{\omega^{a}}{\sinh^2\omega} =& \frac{1}{2^{a-1}}\Gamma(a+1)\zeta(a) \,.\\
	}
	After doing these $\omega$ integrals, we can then do the $s$ integral in \eqref{mbbessel} by closing the contour to the left, which gives an expansion in $1/\lambda$. 
	
	For the 4-body term in \eqref{4mLead}, we can now apply \eqref{mbbessel} twice to get 
	\es{tree221Again}{
		&\int dsdt\frac{3\ (2\pi)^{-2 (s+t-1)}  w^{2 s+3} \Gamma (-s)
			\omega ^{2 t+3} \Gamma (-t) \lambda ^{s+t+1} }{\Gamma (s+1) \Gamma
			(s+2)^2 \Gamma (s+3) \Gamma (t+1) \Gamma (t+2)^2 \Gamma (t+3)
			(w^2-\omega^2 ) \sinh^2w\sinh^2\omega}\\
		&\times (\Gamma (s+1)
		\Gamma (2 s+3) \Gamma (t+3) \Gamma (2 t+2)-\Gamma (s+3) \Gamma
		(2 s+2) \Gamma (t+1) \Gamma (2 t+3))\,,\\
	}
	where note that the $w,\omega$ dependence does not factorize due to the $w^2-\omega^2$ in the denominator. While in general it is difficult to compute the $s,t$ integrals by closing the contour to the left, since there are likely poles that can only be seen after doing the $w,\omega$ integrals, for the  poles at $s,t=-\frac32,-\frac52,\dots$ we find that the residues at each order in $\lambda$ factorize in $w,\omega$. The $w,\omega$ integrals can then be computed with \eqref{id} analytically continued to negative even integers (recall that this quantity is only divergent for $a=1,-1,-3,-5,\dots$). These poles correspond to the $N^0\lambda^{-\text{integer}}$ terms discussed in the main text, which is why we can compute all of them analytically. Unfortunately, this factorization after taking poles does not apply to all the expected large $\lambda$ terms, such as the $N^0\lambda^{-\frac{\text{integer}}{2}}$ terms that we know to exist from the numerical results of \cite{Chester:2020dja}, nor does it apply to any terms at higher orders in $1/N^2$.
	
	\subsection{$\tau_2^2 \partial_\tau\partial_{\bar\tau} \partial_m^2 \log Z\big\vert_{m=0,b=1} $ at large $N$ and finite $g_{_{\rm YM}} $}
	\label{oldnew}
	
	Before we discuss the instanton sector contribution to the relations in \eqref{relation1} and $\partial_m^4 \log Z\big\vert_{m=0,b=1}$ in \eqref{newpaper}, we first introduce a new large $N$ and finite $g_\text{YM}$ method that we will use for these calculations, by demonstrating it in the simpler case of $\tau_2^2 \partial_\tau\partial_{\bar\tau} \partial_m^2 \log Z\big\vert_{m=0,b=1} $ in \eqref{oldpaper}. This result was previously computed in \cite{Chester:2019jas} to the first couple orders in $1/N$ at finite $g_{_{\rm YM}} $, and at subsequent orders in $1/N$ in a small $g_{_{\rm YM}} $ expansion. Here, we complete this derivation by computing all orders in $1/N$ at finite $g_{_{\rm YM}} $.
	
	\subsubsection{One-instanton sector}
	
	We start by considering the one-instanton contribution to $\tau_2^2 \partial_\tau\partial_{\bar\tau} \partial_m^2 \log Z\big\vert_{m=0,b=1} $, and for simplicity we will consider just $\partial_m^2 \log Z\big\vert_{m=0,b=1}$, since the $\tau$ derivatives can be trivially applied to the result.
	
	For this calculation, it is useful to express $ Z_\text{inst}^{(1)}(m,b,a_{ij})$ in \eqref{I1ExpressionM} as a contour integral
	\es{I1ExpressionMa}{
		Z_\text{inst}^{(1)}(m,b,a_{ij})&  =(b+1/b)\frac{m^2+\frac14(b-1/b)^2}{m^2+\frac14(b+1/b)^2} \int \frac{dz}{2 \pi} \left[\exp\left({\sum_{j=1}^N \log\frac{(z-a_j)^2-m^2}{(z-a_j)^2+\frac14(b+1/b)^2}} \right)- 1  \right] \,, 
	}
	where the integration contour is counter-clockwise around the poles at $z=a_j+i$, and the subtraction of $1$ from the integrand does not contribute to the final result, but makes the integrand decay as $1/z^2$ at $|z|\to\infty$, so that the contour can be taken to be the real line. We can then take the $m$ derivatives to get
	\es{I1Expression}{
		\partial_m^2 Z_\text{inst}^{(1)}(m, b,a_{ij})  \big|_{m=0,b=1} = 4 \int \frac{dz}{2 \pi} \left[e^{Q(z)} - 1  \right]  \,,
	}
	where we define 
	\es{Q}{
		Q(z) \equiv - \sum_j \log \left[1 + \frac{1}{(z - a_j)^2}\right] = \sum_{k=0}^\infty \frac{(-1)^n}{(n+1) (2n+1)!} \frac{\partial^{2n+1} R(z)}{\partial z^{2n+1}} 
	}
	in terms of the resolvent operator $R$ given in \eqref{RnDef}.
	We now take the expectation value, and use the cumulant expansion
	\es{cum}{
		\langle e^{A}\rangle=e^{\sum_{m=1}^\infty \frac{\langle A^m\rangle_\text{conn}}{m!}}
	}
	to get
	\es{ToCalc2}{
		\langle  \partial_m^2 Z_\text{inst}^{(1)}(m, b,a_{ij})   \rangle \big|_{m=0,b=1} 
		= 4 \int \frac{dz}{2 \pi} \left[ \exp \left(\sum_{j=1}^\infty\frac{\langle Q(z)^j \rangle_\text{conn}}{j!} \right) - 1\right] \,.
	}
	This can then be written in terms of the connected resolvents $W^n$ defined in \eqref{W} as
	\es{ToCalc3}{
		& \langle  \partial_m^2 Z_\text{inst}^{(1)}(m, b,a_{ij})   \rangle \big|_{m=0,b=1} =
		4 \int \frac{dz}{2 \pi} \, \Biggl[-1 +  \exp \biggl[  N \left(\partial_z - \frac{\partial_z^3}{12} + \frac{\partial_z^5}{360} + \cdots \right)  W^1(z)  \\
		&{}+ \frac{1}{2} 
		\left(\partial_{z_1} - \frac{\partial_{z_1}^3}{12} + \frac{\partial_{z_1}^5}{360} + \cdots \right)
		\left(\partial_{z_2} - \frac{\partial_{z_2}^3}{12} + \frac{\partial_{z_2}^5}{360} + \cdots \right) W^2(z_1 ,z_2) \big|_{z_i  = z} + \cdots 
		\biggr]  \Biggr]\,,
	} 
	where each $W^n$ can then be expanded to any order in large $N$ using topological recursion in terms of the $W^n_m$ defined in \eqref{W2}. We will have to evaluate $W^n_m$ at values of $z$ of order $\sqrt{\lambda}$, where at fixed $g_\text{YM}$, we have
	\es{WnmScaling}{
		N^{2-n} \partial_z^k W^n_m(z_1, \ldots, z_n) \propto N^{2 -\frac{3n + k}{2}  - 2m} \,.
	}
	From this scaling, we see that only the first term in the exponent of \eqref{ToCalc3} gives a contribution of order $N^0$ (namely the term $N \partial_z W^1_0$), while the rest are all suppressed in $1/N$, so one can further expand the exponential of these terms. We can thus write \eqref{ToCalc2} as
	\es{ToCalc4}{
		& \langle  \partial_m^2 Z_\text{inst}^{(1)}(m, b,a_{ij})   \rangle \big|_{m=0,b=1} =
		4 \int_{-\infty}^{\infty} \frac{dz}{2 \pi} \, \Biggl[-1 +  e^{  N \partial_z W^1_0(z)} \Biggl(1  - \frac{N}{12} \partial_z^3 W^1_0(z) \\
		&{}  +  \frac{N^2 \left( \partial_z^3 W^1_0(z)  \right)^2}{288}  + \frac{\partial_{z_1} \partial_{z_2} W^2_0(z_1 ,z_2) }{2} \bigg|_{z_i  = z}   + \frac{N}{360} \partial_z^5 W^1_0(z)  + \frac{1}{N} \partial_z W^1_1(z) + \cdots 
		\Biggr) \Biggr]\,,
	}
	Writing $z = x \sqrt{\lambda} / (2 \pi)$, and using the explicit expressions for the $W^n_m$ we can write \eqref{ToCalc4} as
	\es{ToCalc5}{
		& \langle  \partial_m^2 Z_\text{inst}^{(1)}(m, b,a_{ij})   \rangle \big|_{m=0,b=1} 
		=  \frac{2 g_{_{\rm YM}}  \sqrt{N}}{\pi^2} 
		\int dx \, \Biggl[
		- 1 +\theta(x-1) e^{\frac{8 \pi^2}{g_{_{\rm YM}} ^2}
			\left( 1 - \frac{x}{\sqrt{x^2 - 1}} \right) } 
		\Biggl( 1
		+ \frac{8 \pi^4 x}{(x^2 - 1)^{\frac 52} g_{_{\rm YM}} ^4 N} \\
		&{}+
		\frac{1}{N^2} \left( \frac{32 \pi^2 x^2}{g_{_{\rm YM}} ^8 (x^2 - 1)^5} 
		+ \frac{\pi^4 (8 x^2 + 1)}{g_{_{\rm YM}} ^4 (x^2 - 1)^4} 
		- \frac{5 \pi^2 x}{4 g_{_{\rm YM}} ^2 (x^2 - 1)^{\frac 72}}
		- \frac{16 \pi^6 x (4x^2 + 3)}{3 g_{_{\rm YM}} ^6 (x^2 - 1)^{\frac 92}}\right) + \cdots \Biggr) 
		\Biggr] \,.
	}
	These integrals can then be performed as described in \cite{Chester:2019jas} to get 
	\es{Expectation1}{
		& \langle  \partial_m^2 Z_\text{inst}^{(1)}(m, b,a_{ij})   \rangle \big|_{m=0,b=1} =e^{ \frac{8 \pi^2 }{g_{_{\rm YM}} ^2} }\Big[ - \sqrt{N} \frac{16 K_1 (8 \pi^2 / g_{_{\rm YM}} ^2)}{g_\text{YM}}  +  \frac{2 K_2 (8 \pi^2 / g_{_{\rm YM}} ^2)}{\sqrt{N}g_{_{\rm YM}} } \\
		&\quad+\frac{1}{32 g_{_{\rm YM}}  N^{\frac32}}\left[  -13 K_1 (8 \pi^2 / g_{_{\rm YM}} ^2)+9 K_3 (8 \pi^2 / g_{_{\rm YM}} ^2) \right] \\
		&\quad+\frac{1}{128 g_{_{\rm YM}}  N^{\frac52}}\left[  -25 K_2 (8 \pi^2 / g_{_{\rm YM}} ^2)+15 K_4 (8 \pi^2 / g_\text{YM}^2) \right] \\
		&\quad +\frac{1}{g_{_{\rm YM}} N^{\frac72}}\left[\frac{1533 K_1\left(\frac{8 \pi ^2}{g_{_{\rm YM}} ^2}\right)}{16384 }-\frac{5355 K_3\left(\frac{8 \pi ^2}{g_{_{\rm YM}} ^2}\right)}{32768 }+\frac{2625
			K_5\left(\frac{8 \pi ^2}{g_{_{\rm YM}} ^2}\right)}{32768 }\right]+ O(N^{-\frac92})\Big] \,.\\
	} 
	We can then take the $\tau$ derivatives and compare to the one-instanton term in \eqref{oldpaper}.
	
	\subsubsection{Higher instanton sector}
	
	We can similarly compute the $k>1$ instanton terms. As described in \cite{Chester:2019jas}, these instantons are described by rectangular Young diagrams of height $p$ and length $q$, which will correspond to the partition of unity in the divisor sum that defines the Eisenstein series. Following \cite{Chester:2019jas}, we thus define
	\es{pq}{
		\langle \partial_m^2Z_\text{inst}^{(k)}(m,b,a_{ij})\rangle\big\vert_{m=0,b=1}\equiv\sum_{p,q}I_{p \times q} \,,
	}
	for integers $p,q$ such that $k=pq$. This $I_{p \times q}$ was given in \cite{Chester:2019jas} as
	\es{eq:pq-tab}{
		I_{p \times q} &=  \oint  {dz \over 2\pi}
		\prod_{k_a}
		\prod_{j=1}^N {(z-a_j + k_a i )^2\over (z-a_j + k_a i)^2+1 } \times
		\left[
		{4 \over 1 + \delta_{pq}} \left({1\over p^2}+{1\over q^2} \right) \right.
		\\
		& \left. + \sum_{j=1}^N
		{i f(p, q) \over (z-a_j+(p+q -1) i) (z-a_j+ (q-1) i) (z-a_j+(p-1)i)}
		\right] \\
		&=  \oint  {dz \over 2\pi}
		\prod_{k_a}
		\prod_{j=1}^N {(z-a_j + k_a i )^2\over (z-a_j + k_a i)^2+1 } \times
		\left[
		{4 \over 1 + \delta_{pq}} \left({1\over p^2}+{1\over q^2} \right) \right.
		\\
		& \left. + \sum_{j=1}^N\left(
		-\frac{2 i (p+q) (p-q)^2}{p^2 q^2 (z- a_j+i (p+ q-1))}-\frac{2 i (p+q) (p-q)}{p^2 q (z-a_j+i(q-1))}+\frac{2 i (p+q) (p-q)}{p q^2 (z-a_j+i(p-1) )}\right)
		\right] \,.
	}
	We can write this in terms of resolvents as
	\es{eq:pq-tab2}{
		I_{p \times q} &=  \oint  {dz \over 2\pi}
		e^{\sum_{k_a} Q(z+k_a i)}\times
		\left[
		{4 \over 1 + \delta_{pq}} \left({1\over p^2}+{1\over q^2} \right) \right.
		\\
		& \left. +2i(p^2-q^2)\left(\frac{1}{p q^2 }R(z+i(p-1))-\frac{ (p-q)}{p^2 q^2 }R(z+i(p+q-1))-\frac{1}{p^2 q }R(z+i(q-1))\right)
		\right] \,.
	}
	We can then put all resolvents in the exponential by
	\es{eq:pq-tab3}{
		I_{p \times q} &= \left(\frac{4} {1 + \delta_{pq}} \left({1\over p^2}+{1\over q^2} \right)\right) \mathcal{S}
		+\frac{2i(p^2-q^2)}{p q^2 }\mathcal{Q}(i(p-1))-\frac{2i(p^2-q^2)(p-q)}{p^2 q^2 }\mathcal{Q}(i(p+q-1))\\
		&-\frac{2i(p^2-q^2)}{p^2 q }\mathcal{Q}(i(q-1))\,,\\
		\mathcal{Q}(x)&\equiv\int  {dz \over 2\pi}
		\partial_se^{s R(z-x)+\sum_{k_a} Q(z+k_a i)}\big\vert_{s=0}\,,\qquad \mathcal{S}\equiv\int  {dz \over 2\pi}
		e^{\sum_{k_a} Q(z+k_a i)}\,.
	}
	Finally, we can compute the expectation value using the cumulant expansion \eqref{cum} to get
	\es{cumulant}{
		\langle I_{p \times q} \rangle &= \left(\frac{4} {1 + \delta_{pq}} \left({1\over p^2}+{1\over q^2} \right)\right) \langle\mathcal{S}\rangle
		+\frac{2i(p^2-q^2)}{p q^2 }\langle\mathcal{Q}(i(p-1))\rangle-\frac{2i(p^2-q^2)(p-q)}{p^2 q^2 }\langle\mathcal{Q}(i(p+q-1))\rangle\\
		&-\frac{2i(p^2-q^2)}{p^2 q }\langle\mathcal{Q}(i(q-1))\rangle\,,\\
		\langle\mathcal{Q}(x)\rangle&=\int  {dz \over 2\pi}
		\partial_se^{\sum_{n=1}^\infty\frac{1}{n!}\langle (s R(z-x)+\sum_{k_a} Q(z+k_a i))^n\rangle_\text{conn} }\big\vert_{s=0}  \,,\qquad \langle\mathcal{S}\rangle=\int  {dz \over 2\pi}
		e^{\sum_{n=1}^\infty\frac{1}{n!}\langle(\sum_{k_a} Q(z+k_a i))^n\rangle_\text{conn}}\,.
	}
	We can then expand at large $N$ and perform the integrals similarly to the one-instanton case to get
	\es{ExpectationFinal}{
		&\langle I_{p \times q} \rangle =\frac{e^{ \frac{8pq \pi^2 }{ g_{_{\rm YM}} ^2} }}{1+\delta_{p,q}}\Big[ - \sqrt{N} \frac{16 K_1 (\frac{8pq \pi^2 }{ g_{_{\rm YM}} ^2})}{g_{_{\rm YM}}  }\left({\frac{p}{q}+\frac{q}{p}}\right)  +  \frac{2 K_2 (\frac{8pq \pi^2 }{ g_{_{\rm YM}} ^2})}{ g_{_{\rm YM}}  \sqrt{N}}\left({\frac{p^2}{q^2}+\frac{q^2}{p^2}}\right)\\
		&+\frac{1}{32 g_{_{\rm YM}} N^{\frac32}}\left[  -13 K_1 \left(\frac{8pq \pi^2 }{ g_{_{\rm YM}} ^2}\right)\left({\frac{p}{q}+\frac{q}{p}}\right)+9 K_3 \left(\frac{8pq \pi^2 }{ g_\text{YM}^2}\right)\left({\frac{p^3}{q^3}+\frac{q^3}{p^3}}\right) \right] \\
		&+\frac{1}{128 g_{_{\rm YM}} N^{\frac52}}\left[  -25  K_2 \left(\frac{8pq \pi^2 }{ g_{_{\rm YM}} ^2}\right)\left({\frac{p^2}{q^2}+\frac{q^2}{p^2}}\right)+15  K_4 \left(\frac{8pq \pi^2 }{ g_{_{\rm YM}} ^2}\right)\left({\frac{p^4}{q^4}+\frac{q^4}{p^4}}\right) \right] \\
		& +\frac{1}{g_{_{\rm YM}} N^{\frac72}}\left[\frac{1533 K_1\left(\frac{8pq \pi ^2}{g_{_{\rm YM}} ^2}\right)}{16384 }\left({\frac{p}{q}+\frac{q}{p}}\right)-\frac{5355 K_3\left(\frac{8pq \pi ^2}{g_{_{\rm YM}} ^2}\right)}{32768 }\left({\frac{p^3}{q^3}+\frac{q^3}{p^3}}\right)+\frac{2625
			K_5\left(\frac{8pq \pi ^2}{g_{_{\rm YM}} ^2}\right)}{32768 }\left({\frac{p^5}{q^5}+\frac{q^5}{p^5}}\right)\right]\\
		&+ O(N^{-\frac92})\Big] \,.\\
	} 
	We can then take the $\tau$ derivatives, take the $p,q$ sum in \eqref{pq}, and compare to the relevant instanton term in \eqref{oldpaper}, which is the complete finite $g_\text{YM}$ derivation of this result to any order in $1/N$.

	\subsection{Details of instanton calculation}
	\label{app:inst}
	
	We now continue with the calculation of the expectation values that show up in the relations \eqref{relation1} and $\partial_m^4 \log Z\big\vert_{m=0,b=1}$ in \eqref{newpaper}, and address the instanton terms. For some of these calculations, we will use the the large $N$ and small $g_{_{\rm YM}}$ method introduced in \cite{Chester:2019jas}, while for others we will use the new large $N$ and finite $g_{_{\rm YM}}$ method that we demonstrated in the previous section. We follow the main text and discuss the one-instanton sector, then the two-instanton sector, and finally the mixed instanton/anti-instanton sector.
	
	\subsubsection{One-instanton sector}
	
	We start by detailing the large $N$ and finite $g_\text{YM}$ calculation of \eqref{top1mmbb}. Consider the contour integral representation of $ Z_\text{inst}^{(1)}(m,b,a_{ij})$ given in \eqref{I1ExpressionMa}. We can then take derivatives in $m,b$ to get
	\es{I1Expression4}{
		\partial_m^4 Z_\text{inst}^{(1)}\big\vert_{m=0,b=1} = &\frac{24}{\pi} \int {dz} \left[e^{Q(z)}(\partial_z R(z)-1)  - 1  \right]\,, \\
		\partial_m^2\partial_b^2 Z_\text{inst}^{(1)} \big\vert_{m=0,b=1}  = &\frac{2}{\pi} \int {dz} \left[e^{Q(z)}(2\partial_z R(z)-3+2\sum_{j=0}^\infty(-1)^j\frac{\partial_z^{2j+1}R(z)}{(2j+1)!})  - 1  \right]\,, \\
	}
	where $R$ is the resolvent operator given in \eqref{RnDef}, and $Q$ was defined in \eqref{Q}. We then take the expectation value and use the cumulant expansion \eqref{cum} to get
	\es{I1Expression42}{
		\langle\partial_m^4 Z_\text{inst}^{(1)}\rangle\big\vert_{m=0,b=1}&= \frac{24}{\pi}\int  {dz }\left[
		\partial_se^{\sum_{n=1}^\infty\frac{1}{n!}\langle (s \partial_zR(z)+Q(z))^n\rangle_\text{conn} }\big\vert_{s=0}-e^{\sum_{n=1}^\infty\frac{1}{n!}\langle ( Q(z))^n\rangle_\text{conn} }-1\right]\,,\\
		\langle\partial_m^2\partial_b^2 Z_\text{inst}^{(1)} \big\vert_{m=0,b=1}  \rangle&= \frac{2}{\pi}\int  {dz }\left[2
		\partial_se^{\sum_{n=1}^\infty\frac{1}{n!}\langle (s \sum_{j=1}^\infty\frac{\partial_z^jR(z)(1+\delta_{j,1})}{j!}+Q(z))^n\rangle_\text{conn} }\big\vert_{s=0}-3e^{\sum_{n=1}^\infty\frac{1}{n!}\langle ( Q(z))^n\rangle_\text{conn} }-1\right]\,,\\
	}
	where we introduced the derivatives of $s$ to put all terms in \eqref{I1Expression4} into the exponential. From \eqref{Q}, we see that this expression is written in terms of connected correlators of $R$, i.e. resolvents with the known $1/N^2$ expansion described in previous sections. We can then expand \eqref{I1Expression42} at large $N$ and perform the integrals, just as in Section \ref{oldnew}, to get 
	\es{mandbresult}{
		&\langle\partial_m^4 Z_\text{inst}^{(1)}\rangle\big\vert_{m=0,b=1}=e^{\frac{8\pi^2}{ g_{_{\rm YM}} ^2}}\Bigg[\sqrt{N}\frac{768 \pi ^2 \left(K_0\left(\frac{8 \pi ^2}{g_{_{\rm YM}} ^2}\right)-2 K_1\left(\frac{8 \pi
				^2}{g_{_{\rm YM}} ^2}\right)+K_2\left(\frac{8 \pi ^2}{g_{_{\rm YM}} ^2}\right)\right)}{g_{_{\rm YM}} ^3}\\
		&+\frac{1}{\sqrt{N}}\frac{24}{g_{_{\rm YM}} ^3} \left(-4 \left(g_{_{\rm YM}} ^2-2 \pi ^2\right) K_0\left(\frac{8 \pi ^2}{g_{_{\rm YM}} ^2}\right)-\frac{\left(g_{_{\rm YM}} ^4-2 \pi
			^2 g_{_{\rm YM}} ^2+8 \pi ^4\right) K_1\left(\frac{8 \pi ^2}{g_{_{\rm YM}} ^2}\right)}{\pi ^2}\right)\\
		&+ \frac{3 \left(4 \pi ^2 \left(-75 g_{_{\rm YM}} ^4+72 \pi ^2 g_{_{\rm YM}} ^2+64 \pi ^4\right) K_0\left(\frac{8 \pi
				^2}{g_{_{\rm YM}} ^2}\right)-\left(75 g_{_{\rm YM}} ^6-72 \pi ^2 g_{_{\rm YM}} ^4+640 \pi ^4 g_{_{\rm YM}} ^2+256 \pi ^6\right) K_1\left(\frac{8 \pi
				^2}{g_{_{\rm YM}} ^2}\right)\right)}{64 \pi ^4 N^{\frac32}g_{_{\rm YM}} ^3}\\
		&+\frac{15 \left(2 \pi ^2 \left(-63 g_{_{\rm YM}} ^4+36 \pi ^2 g_{_{\rm YM}} ^2+32 \pi ^4\right) K_1\left(\frac{8 \pi
				^2}{g_{_{\rm YM}} ^2}\right)-\left(63 g_{_{\rm YM}} ^6-36 \pi ^2 g_{_{\rm YM}} ^4+136 \pi ^4 g_{_{\rm YM}} ^2+64 \pi ^6\right) K_2\left(\frac{8 \pi
				^2}{g_{_{\rm YM}} ^2}\right)\right)}{128 \pi ^4N^{\frac52} g_{_{\rm YM}} ^3 }\\
		&+O(N^{-\frac72})\Bigg]\,,\\
	}
	and
	\es{mandbresult2}{
		&\langle\partial_m^2\partial_b^2 Z_\text{inst}^{(1)} \big\vert_{m=0,b=1}  \rangle=e^{\frac{8\pi^2}{ g_{_{\rm YM}} ^2}}\Bigg[\sqrt{N}\frac{512 \pi ^2 K_0\left(\frac{8 \pi ^2}{g_{_{\rm YM}} ^2}\right)+16 \left(3 g_{_{\rm YM}} ^2-32 \pi ^2\right) K_1\left(\frac{8
				\pi ^2}{ g_{_{\rm YM}} ^2}\right)}{g_{_{\rm YM}} ^3}\\
		&\!\!\!\!\!\!\!  +\frac{1}{\sqrt{N}}\frac{4 \left(32 \pi ^2-11 g_{_{\rm YM}} ^2\right) K_0\left(\frac{8 \pi ^2}{g_{_{\rm YM}} ^2}\right)+\frac{\left(-11 g_{_{\rm YM}} ^4+32 \pi
				^2 g_{_{\rm YM}} ^2-128 \pi ^4\right) K_1\left(\frac{8 \pi ^2}{g_{_{\rm YM}} ^2}\right)}{\pi ^2}}{2 g_{_{\rm YM}} ^3}\\
		&\!\!\!\!\!\!\! + \frac{4 \pi ^2 \left(-195 g_{_{\rm YM}} ^4+288 \pi ^2 g_{_{\rm YM}} ^2+256 \pi ^4\right) K_0\left(\frac{8 \pi
				^2}{g_{_{\rm YM}} ^2}\right)-\left(195 g_{_{\rm YM}} ^6-288 \pi ^2 g_{_{\rm YM}} ^4+1824 \pi ^4 g_{_{\rm YM}} ^2+1024 \pi ^6\right) K_1\left(\frac{8
				\pi ^2}{g_{_{\rm YM}} ^2}\right)}{256 \pi ^4 N^{3/2} g_{_{\rm YM}} ^3}\\
		&\!\!\!\!\!\!\!  +\frac{5 \left(2 \pi ^2 \left(-315 g_{_{\rm YM}} ^4+288 \pi ^2 g_{_{\rm YM}} ^2+256 \pi ^4\right) K_1\left(\frac{8 \pi
				^2}{ g_{_{\rm YM}} ^2}\right)-\left(315 g_{_{\rm YM}} ^6-288 \pi ^2 g_{_{\rm YM}} ^4+784 \pi ^4 g_{_{\rm YM}} ^2+512 \pi ^6\right) K_2\left(\frac{8 \pi
				^2}{ g_{_{\rm YM}} ^2}\right)\right)}{1024 \pi ^4 g_{_{\rm YM}} ^3 N^{5/2}}  \\
		&\!\!\!\!\!\!\!  +O(N^{-\frac52})\Bigg]\,,\\
	}
	which we combine to get \eqref{top1mmbb}.
	
	Next, we compute $\partial_m^4  Z\big\vert^\text{NP}_{m=0,b=1}$, which consists of the two-body term $\langle\partial_m^4 Z_\text{inst}^{(1)}\rangle\big\vert_{m=0,b=1}$ computed above as well as the higher-body term $\cZ^{(1)}$ in \eqref{ZZ1}. For $\cZ^{(1)}$, it is difficult to perform the large $N$ and finite $g_{_{\rm YM}}$ calculation due to the $e^{2\omega a_{ij}}$ terms and the Fourier integral over $\omega$. Instead, we will perform a large $N$ and small $g_{_{\rm YM}}$ expansion by expanding $\partial Z^{(1)}_\text{inst}\big\vert_{m=0,b=1}$ at small eigenvalue, which corresponds to small $g_{_{\rm YM}}^2$, to get an infinite series of $n$-body terms. We will then compute their expectation value with $e^{2\omega a_{ij}}$ in a large $N$ expansion at finite $\lambda$ using topological recursion, and then do the large $\lambda$ expansion as we did with the perturbative terms of Section \ref{app:pert}. After setting $\lambda=g_{_{\rm YM}}^2N$, these steps give a consistent large $N$ and small $g_{_{\rm YM}}$ expansion.
	
	We start by expanding $\partial_m^2 Z^{(1)}_\text{inst}\big\vert_{m=0,b=1}$ in \eqref{I1ExpressionM} at small eigenvalue to get
	\es{expansion}{
		&  \partial_m^2 Z^{(1)}_\text{inst}\big\vert_{m=0,b=1} = f_0(N) + f_1(N) C_2
		+ \left( f_2^{(1)}(N) C_2^2 + f_2^{(2)}(N) C_4 \right) \\
		&+ \left(  f_3^{(1)}(N) C_2^3 + f_3^{(2)}(N) C_2 C_4 + f_3^{(3)}(N) C_6
		+ f_3^{(4)}(N) D_{2,2,2}  \right) \\
		& +\left(  f_4^{(1)}(N) C_2C_6 + f_4^{(2)}(N) C_2 D_{2,2,2} + f_4^{(3)}(N) C_4^2
		+ f_4^{(4)}(N)C_2^2C_4+f_4^{(5)}C_2^4+f_4^{(6)}D_{4,2,2}+f_4^{(7)}D_{4,4,0}  \right)+\dots\,,
	}
	where we defined the invariants
	\es{invars}{
		C_p =& \sum_{j, k} (a_j - a_k)^p  \,, \qquad
		D_{p,q,r} = \sum_{j, k, l} (a_j - a_k)^p (a_k - a_l)^q (a_l - a_j)^r \,,\\
	}
	and the coefficients are
	\es{Gotf}{
		f_0 &=- \frac{4 \Gamma(N + \frac 12)}{\sqrt{\pi} \Gamma(N)}\,, \qquad f_1 =- \frac{3 \Gamma(N - \frac 12)}{2\sqrt{\pi} \Gamma(N + 2)} \,, \qquad f_2^{(1)} = - \frac{315 \Gamma(N - \frac 32)}{64 \sqrt{\pi} \Gamma(N+4)} \,, \\
		f_2^{(2)} &=- \frac{15 (3 - N + 4 N^2) \Gamma(N - \frac 32)}{16 \sqrt{\pi} \Gamma(N+4)}  \,, \qquad
		f_3^{(1)} = -\frac{45045 \Gamma(N -\frac 52)}{256 \sqrt{\pi} \Gamma(N + 6)} \,, \\
		f_3^{(2)} &= \frac{3465 (4 N^2 - 3N + 15) \Gamma(N -\frac 52)}{128 \sqrt{\pi} \Gamma(N + 6)}  \,, \qquad
		f_3^{(3)} = - \frac{105 (4 N^4 - 10 N^3 + 53 N^2 - 11 N + 18) \Gamma( N - \frac 52)}{32 \sqrt{\pi} \Gamma(N + 6)}  \,, \\
		f_3^{(4)} &= \frac{105 (8 N^3  - 36 N^2 +189 N - 15) \Gamma( N - \frac 52)}{32 \sqrt{\pi} \Gamma(N + 6)}    \,, \\
		f_4^{(1)} &= \frac{45045 \left(N \left(N \left(4 N^2-22 N+181\right)-123\right)+252\right) \Gamma
			\left(N-\frac{7}{2}\right)}{256 \sqrt{\pi } \Gamma (N+8)}\,,\\
		f_4^{(2)} &= -\frac{45045 \left(N \left(8 N^2-60 N+465\right)-105\right) \Gamma \left(N-\frac{7}{2}\right)}{256
			\sqrt{\pi } \Gamma (N+8)}\,,\\
		f_4^{(3)} &=\frac{45045 (N (N (8 N (2 N+1)-149)+192)+105) \Gamma \left(N-\frac{7}{2}\right)}{2048 \sqrt{\pi }
			\Gamma (N+8)}\,,\\
		f_4^{(4)} &=-\frac{2297295 (N (4 N-5)+35) \Gamma \left(N-\frac{7}{2}\right)}{2048 \sqrt{\pi } \Gamma (N+8)}\,,\qquad  f_4^{(5)} = \frac{43648605 \Gamma \left(N-\frac{7}{2}\right)}{8192 \sqrt{\pi } \Gamma (N+8)}\,,\\
		f_4^{(6)} &=-\frac{315 (N (N (N (N (2 N (4 N+53)-1105)+2100)-5049)-2766)+7650) \Gamma
			\left(N-\frac{7}{2}\right)}{64 \sqrt{\pi } N \Gamma (N+8)}\,,\\
		f_4^{(7)} &= -\frac{945 (N (N (N (8 N (N (4 N-35)+322)-4063)+11462)-2077)+2550) \Gamma
			\left(N-\frac{7}{2}\right)}{256 \sqrt{\pi } N \Gamma (N+8)}\,.\\
	}
	Each $n$-body operator in \eqref{expansion} will give an $(n+2)$-body operator in \eqref{ZZ1} when combined with $e^{2\omega a_{ij}}$, whose expectation value can be computed using topological recursion as in Section \ref{app:pert} by applying the inverse Laplace transform to a resolvent. For instance, the leading term from topological recursion is
	\es{4bodtree}{
		\langle\partial_m^2 Z^{(1)}_\text{inst}\rangle\big\vert_{m=0,b=1}=-N^2\left[\frac{18 \Gamma \left(N-\frac{1}{2}\right)}{4 \sqrt{\pi } \Gamma (N+2)} \int_0^\infty d\omega\frac{1}{\sinh^2\omega} \left(\frac{ J_1\left(\frac{\omega \sqrt{\lambda }}{\pi
			}\right){}^2}{\omega}-\frac{ \sqrt{\lambda } J_0\left(\frac{\omega \sqrt{\lambda }}{\pi
			}\right) J_1\left(\frac{\omega \sqrt{\lambda }}{\pi }\right)}{\pi }\right)\right]+\dots\,,
	}
	which we can then expand at large $\lambda$ using \eqref{mbbessel} and \eqref{id} and convert $\lambda=g_{_{\rm YM}}^2N$ to get
	\es{firstg}{
		\langle\partial_m^2 Z^{(1)}_\text{inst}\rangle\big\vert_{m=0,b=1}=-\frac{9 g_{_{\rm YM}} ^2}{4\pi^{\frac52}}\sqrt{N}+\dots\,.
	}
	We can systematically include more terms in large $N$ and small $g_{_{\rm YM}}$ by including more terms in the eigenvalue expansion \eqref{expansion}, the topological recursion expansion of the $(n+2)$-body operators, and the large $\lambda$ expansion of the result from topological recursion. After combining these terms with the small $g_{_{\rm YM}}$ expansion of the two-body term $\langle\partial_m^2 Z^{(1)}_\text{inst}\rangle\big\vert_{m=0,b=1}$ in \eqref{mandbresult}, we get the result \eqref{top1Fin}.

	\subsubsection{Two-instanton sector}
	
	The calculation in the two-instanton sector is similar to the one-instanton sector, except all the expressions are much more complicated, so we only do calculations in the large $N$ and small $g_{_{\rm YM}}$ expansions. For the two-body terms $\partial_m^4 Z_\text{inst}^{(2)}\big\vert_{m=0,b=1}$ and $\partial_m^2\partial_b^2 Z_\text{inst}^{(2)}\big\vert_{m=0,b=1}$, we expand to leading order in small eigenvalue to get $N$-dependent coefficients that satisfy complicated recursion relations, similar those found at $k>1$ instantons in the small eigenvalue expansion of $\partial_m^2 Z_\text{inst}^{(2)}\big\vert_{m=0,b=1}$ in \cite{Chester:2019jas}. We can then expand these recursion relations at large $N$ and perform the trivial expectation value (since their is no eigenvalue dependence to leading order) to get
	\es{2instsmalle}{
		\langle\partial_m^4 Z_\text{inst}^{(2)}\rangle\big\vert_{m=0,b=1}=&\Bigg[\frac{48 N}{\pi }+30 \sqrt{\frac{2}{\pi }} \sqrt{N}-\frac{12}{\pi }-\frac{219 \sqrt{\frac{1}{N}}}{4
			\sqrt{2 \pi }}-\frac{33}{2 \pi  N}-\frac{11685 \left(\frac{1}{N}\right)^{3/2}}{512 \sqrt{2 \pi
		}}\\
		&-\frac{249}{8 \pi  N^2}-\frac{150285 \left(\frac{1}{N}\right)^{5/2}}{8192 \sqrt{2 \pi
		}}-\frac{8943}{128 \pi  N^3}+O(N^{-\frac72})\Bigg]+O(g_{_{\rm YM}}^2)\,,\\
		\langle\partial_m^2\partial_b^2 Z_\text{inst}^{(2)}\rangle\big\vert_{m=0,b=1}=&\bigg[\frac{16 N}{\pi }+5 \sqrt{\frac{2}{\pi }} \sqrt{N}-\frac{4}{\pi }-\frac{61 \sqrt{\frac{1}{N}}}{8
			\sqrt{2 \pi }}-\frac{11}{2 \pi  N}-\frac{1225 \left(\frac{1}{N}\right)^{3/2}}{1024 \sqrt{2 \pi
		}}\\
		&-\frac{83}{8 \pi  N^2}+\frac{2795 \left(\frac{1}{N}\right)^{5/2}}{1024 \sqrt{2 \pi
		}}-\frac{2981}{128 \pi  N^3}+O(N^{-\frac72})\Bigg]+O(g_{_{\rm YM}}^2)\,,
	}
	which satisfies the second relation in \eqref{relation1} for the two-instanton sector. For the higher body term $\cZ^{(2)}$, we note that the first term in \eqref{ZZ2} can be computed to leading order in $g_{_{\rm YM}}^2$ by simply squaring the leading order expression in \eqref{expansion} to get
	\es{Z2simple}{
		-\langle\partial_m^2Z_\text{inst}^{(1)}\rangle^2&=- \frac{16 \Gamma(N + \frac 12)^2}{{\pi} \Gamma(N)^2}+O(g_{_{\rm YM}}^2)\\
		&=\Big[-\frac{16 N}{\pi }+\frac{4}{\pi }-\frac{1}{2 \pi  N}-\frac{1}{8 \pi  N^2}+\frac{5}{128 \pi 
			N^3}+O(N^{-4})\Big]+O(g_{_{\rm YM}}^2)\,.
	}
	Note that the $N^1$ terms cancel between \eqref{2instsmalle} and \eqref{Z2simple} (after including the factor of 3 in \eqref{explicitDers}), so the combined expansion begins at order $\sqrt{N}$ as expected. For the other terms in $\cZ^{(2)}$ in \eqref{ZZ2}, the calculation is very similar to the one-instanton case in the previous section except the $N$-dependent coefficients small eigenvalue expansion of $ \partial_m^2 Z^{(2)}_\text{inst}\big\vert_{m=0,b=1} $ are expressed by complicated recursion relations given in \cite{Chester:2019jas}. We can combine the results of this calculation with \eqref{2instsmalle} and \eqref{Z2simple} to get \eqref{top2Fin}
	
	\subsubsection{Instanton/anti-instanton sector}
	\label{oneminone}
	
	Finally, we consider the mixed instanton/anti-instanton sector. For $\partial_m^4  \log Z\big\vert^{\text{NP},(1,-1)}_{m=0,b=1}$ we perform this calculation at large $N$ and finite $g_{_{\rm YM}}$. We combine \eqref{pqinst} with \eqref{I1Expression} to get
	\es{finiteG1-1}{
		\partial_m^4  \log Z\big\vert^{\text{NP},(1,-1)}_{m=0,b=1}=&\frac{24}{\pi^2}\int d z_1 dz_2(\langle e^{Q(z_1)+Q(z_2)} \rangle-\langle e^{Q(z_1)}\rangle\langle e^{Q(z_2)} \rangle)\\
		&\frac{24}{\pi^2}\int d z_1 dz_2( e^{\sum_{n=1}\frac{1}{n!}\langle[Q(z_1)+Q(z_2)]^n\rangle_\text{conn}} -e^{\sum_{n=1}\frac{1}{n!}\langle Q(z_1)^n+Q(z_2)^n\rangle_\text{conn}} )\,,
	}
	where in the second equality we did the usual cumulant expansion in \eqref{cum}. We then collect large $N$ terms as in the similar one-instanton calculation in Section \ref{oldnew}, and transform to  $z_i=x_i\sqrt{\lambda}/(2\pi)$, to get the leading large $N$ term
	\es{finiteG1-1largeN}{
		\frac{6 g_{_{\rm YM}}^2}{\pi^4N}e^{\frac{16\pi^2}{ g_{_{\rm YM}}^2}}\int d x_1 dx_2 \theta(|x_1|-1) \theta(|x_2|-1)e^{-\frac{8\pi^2}{g_{_{\rm YM}}^2}\frac{|x_1|}{\sqrt{x_1^2-1}}}e^{-\frac{8\pi^2}{g_{_{\rm YM}}^2}\frac{|x_2|}{\sqrt{x_2^2-1}}}\partial_{x_1}\partial_{x_2}W_0^2(x_1,x_2)\,,
	}
	where we have
	\es{W20}{
		W_0^2(x_1,x_2)=-\frac{2 \pi ^2 \left(x_1 x_2 \left(\sqrt{1-\frac{1}{x_1^2}}
			\sqrt{1-\frac{1}{x_2^2}}-1\right)+1\right)}{g_{_{\rm YM}}^2 \sqrt{1-\frac{1}{x_1^2}} x_1
			\sqrt{1-\frac{1}{x_2^2}} x_2 (x_1-x_2)^2}\,.
	}
	While we do not know how to compute this integral analytically, it can be checked numerically for many values of $g_\text{YM}$ that \eqref{finiteG1-1largeN} matches
	\es{expected}{
		&\frac{6 e^{\frac{16 \pi ^2}{g_{_{\rm YM}} ^2}}}{35 g_{_{\rm YM}} ^{10}N} \left(\frac{8 \left(15 g_{_{\rm YM}} ^8+192 \pi ^4 g_{_{\rm YM}} ^4+32768 \pi ^8\right) g_{_{\rm YM}} ^2
			K_0\left(\frac{8 \pi ^2}{g_{_{\rm YM}} ^2}\right) K_1\left(\frac{8 \pi ^2}{g_{_{\rm YM}} ^2}\right)}{\pi ^2}\right.\\
		&\left.+16 \left(15
		g_{_{\rm YM}} ^8-768 \pi ^4 g_{_{\rm YM}} ^4+131072 \pi ^8\right) K_0\left(\frac{8 \pi ^2}{g_{_{\rm YM}} ^2}\right){}^2\right.\\
		&\left.-\frac{\left(-15
			g_{_{\rm YM}} ^{12}+528 \pi ^4 g_{_{\rm YM}} ^8+4096 \pi ^8 g_{_{\rm YM}} ^4+2097152 \pi ^{12}\right) K_1\left(\frac{8 \pi
				^2}{g_{_{\rm YM}} ^2}\right){}^2}{\pi ^4}\right)\,,
	}
	which is the expected $(1,-1)$ sector term in \eqref{newpaper}. The sub-leading terms in $1/N$ take a similar form and can be similarly compared numerically to the terms listed in \eqref{newpaper}  using the properties of the $\cE$ functions given in appendix~\ref{lapsol}.  We have verified this up to $O(N^{-3})$.
	
	For the other mixed instanton terms $\partial_m^4  \log Z\big\vert^{\text{NP},(p,-q)}_{m=0,b=1}$ with $p,q\leq3$, we performed the calculation in a large $N$ and small $g_\text{YM}$ expansion. For this calculation, we simply plug in the small eigenvalue expansion of $\partial_m^2 Z_\text{inst}^{(k)}$ into \eqref{pqinst}, where the $k=1$ value was given in \eqref{expansion} and $k=2,3$ values are given in \cite{Chester:2019jas}. We can then easily perform the expectation values of the resulting polynomial in eigenvalue operators using Wick contractions in the gaussian matrix model, which yields $\partial_m^4  \log Z\big\vert^{\text{NP},(2,-2)}_{m=0,b=1}$ as given in \eqref{pq2-2}, as well as the other cases
	\es{pq3-3}{
		\partial_m^4  \log Z\big\vert^{\text{NP},(3,-3)}_{m=0,b=1}&=\frac{1}{N}\left[\frac{25 g_{_{\rm YM}} ^4}{144 \pi ^5}-\frac{125 g_{_{\rm YM}} ^6}{13824 \pi ^7}+O(g_{_{\rm YM}} ^{8})\right]\\
		&-\frac{1}{N^2}\left[-\frac{1025 g_{_{\rm YM}} ^4}{1728 \pi ^5}-\frac{1025 g_{_{\rm YM}} ^6}{165888 \pi ^7} +O(g_{_{\rm YM}} ^{8})\right]\\
		&+\frac{1}{N^3}\left[\frac{4625 g_{_{\rm YM}} ^4}{41472 \pi ^5}-\frac{154225 g_{_{\rm YM}} ^6}{3981312 \pi ^7}+O(g_{_{\rm YM}} ^{8})\right]+O(N^{-4})\,,
	}
	\es{pq1-2}{
		\partial_m^4  \log Z\big\vert^{\text{NP},(1,-2)}_{m=0,b=1}&=\frac{1}{N}\left[\frac{135 g_{_{\rm YM}} ^4}{256 \sqrt{2} \pi ^5}-\frac{2025 g_{_{\rm YM}} ^6}{32768 \left(\sqrt{2} \pi ^7\right)}+\frac{192375
			g_{_{\rm YM}} ^8}{8388608 \sqrt{2} \pi ^9}+O(g_{_{\rm YM}} ^{10})\right]\\
		&-\frac{1}{N^2}\left[-\frac{3645 g_{_{\rm YM}} ^4}{4096 \left(\sqrt{2} \pi ^5\right)}-\frac{5265 g_{_{\rm YM}} ^6}{524288 \left(\sqrt{2} \pi
			^7\right)}-\frac{20522565 g_{_{\rm YM}} ^8}{134217728 \left(\sqrt{2} \pi ^9\right)}+O(g_{_{\rm YM}} ^{10})\right]\\
		&+\frac{1}{N^3}\left[ \frac{233145 g_{_{\rm YM}} ^4}{524288 \sqrt{2} \pi ^5}-\frac{6413175 g_{_{\rm YM}} ^6}{67108864 \left(\sqrt{2} \pi
			^7\right)}+\frac{13478466825 g_{_{\rm YM}} ^8}{17179869184 \sqrt{2} \pi ^9}+O(g_{_{\rm YM}} ^{10})\right]+O(N^{-4})\,,
	}
	\es{pq1-3}{
		\partial_m^4  \log Z\big\vert^{\text{NP},(1,-3)}_{m=0,b=1}&=\frac{1}{N}\left[\frac{5 \sqrt{3} g_{_{\rm YM}} ^4}{32 \pi ^5}-\frac{25 g_{_{\rm YM}} ^6}{512 \left(\sqrt{3} \pi ^7\right)}+O(g_{_{\rm YM}} ^{8})\right]\\
		&-\frac{1}{N^2}\left[-\frac{35 g_{_{\rm YM}} ^4}{32 \left(\sqrt{3} \pi ^5\right)}+\frac{55 g_{_{\rm YM}} ^6}{6144 \sqrt{3} \pi ^7}+O(g_{_{\rm YM}} ^{8})\right]\\
		&+\frac{1}{N^3}\left[\frac{5 \sqrt{3} g_{_{\rm YM}} ^4}{128 \pi ^5}-\frac{25 g_{_{\rm YM}} ^6}{384 \left(\sqrt{3} \pi ^7\right)}+O(g_{_{\rm YM}} ^{8})\right]+O(N^{-4})\,,
	}
	\es{pq2-3}{
		\partial_m^4  \log Z\big\vert^{\text{NP},(2,-3)}_{m=0,b=1}&=\frac{1}{N}\left[\frac{25 \sqrt{\frac{3}{2}} g_{_{\rm YM}} ^4}{128 \pi ^5}-\frac{625 g_{_{\rm YM}} ^6}{16384 \left(\sqrt{6} \pi
			^7\right)}+O(g_{_{\rm YM}} ^{8})\right]\\
		&-\frac{1}{N^2}\left[-\frac{3325 g_{_{\rm YM}} ^4}{2048 \left(\sqrt{6} \pi ^5\right)}-\frac{11975 g_{_{\rm YM}} ^6}{786432 \left(\sqrt{6} \pi
			^7\right)}+O(g_{_{\rm YM}} ^{8})\right]\\
		&+\frac{1}{N^3}\left[\frac{82325 g_{_{\rm YM}} ^4}{262144 \sqrt{6} \pi ^5}-\frac{11510125 g_{_{\rm YM}} ^6}{100663296 \left(\sqrt{6} \pi
			^7\right)}+O(g_{_{\rm YM}} ^{8})\right]+O(N^{-4})\,.
	}

	\bibliographystyle{ssg}
	\bibliography{instantonNew}

\end{document}